\documentclass{aa}  

\usepackage{graphicx}
\usepackage{siunitx}
\usepackage{natbib}
\usepackage{amsmath}
\usepackage{tabularx}
\usepackage{float}
\usepackage{placeins}
\usepackage{txfonts}
\usepackage[dvipsnames]{xcolor}
\usepackage[breaklinks, colorlinks, citecolor=Cerulean]{hyperref}

\newcommand{\Hb}{\hbox{H$\beta$}}
\newcommand{\Ha}{\hbox{H$\alpha$}}
\newcommand{\oi}{[O~\textsc{i}]$\lambda6300$}
\newcommand{\oiii}{[O~\textsc{iii}]$\lambda5007$}
\newcommand{\nii}{[N~\textsc{ii}]$\lambda6584$}
\newcommand{\siia}{[S~\textsc{ii}]$\lambda6717$}
\newcommand{\siib}{[S~\textsc{ii}]$\lambda6731$}
\newcommand{\sii}{[S~\textsc{ii}]$\lambda\lambda6717,6731$}
\newcommand{\siii}{[S~\textsc{iii}]$\lambda9068$}
\newcommand{\ngc}{NGC~}
\newcommand{\Msun}{$\mbox{M}_{\sun}$}

\newcommand{\hii}{H~{\sc ii}}
\newcommand{\sn}{S$/$N}
\newcommand{\ebv}{$E(B-V)$}
\newcommand{\izi}{{\tt IZI}}
\DeclareSIUnit\angstrom{\text {\AA}}

\defcitealias{Santoro22}{S22}

\begin{document}

\title{PHANGS-MUSE: Detection and Bayesian classification of $\sim$40000 ionised nebulae in nearby spiral galaxies\thanks{The catalog of nebulae is available only in electronic form at the CDS via anonymous ftp to cdsarc.cds.unistra.fr (130.79.128.5) or via \url{https://cdsarc.cds.unistra.fr/cgi-bin/qcat?J/A+A/}. The catalog, together with the segmentation maps, is also available through the CADC via \url{http://dx.doi.org/10.11570/23.0006}}}

\titlerunning{PHANGS-MUSE: classification of ionised nebulae}
\authorrunning{Congiu et al.} 


\author{
Enrico Congiu\inst{1,2}\thanks{econgiu@das.uchile.cl}, 
Guillermo A. Blanc\inst{3,1},
Francesco Belfiore\inst{4},
Francesco Santoro\inst{5},
Fabian Scheuermann\inst{6},
Kathryn Kreckel\inst{6},
Eric Emsellem\inst{7,8},
Brent Groves\inst{9}, 
Hsi-An Pan\inst{10},
Frank Bigiel\inst{11},
Daniel~A.~Dale\inst{12},
Simon~C.~O.~Glover\inst{13},
Kathryn Grasha\inst{14,15},
Oleg~V.~Egorov\inst{6},
Adam Leroy\inst{16,17},
Eva Schinnerer\inst{5},
Elizabeth J. Watkins\inst{6},
Thomas G. Williams\inst{5}.
}

\institute{
Departamento de Astronom\'{i}a, Universidad de Chile, Camino del Observatorio 1515, Las Condes, Santiago, Chile; \and 
European Southern Observatory (ESO), Alonso de Córdova 3107, Casilla 19, Santiago 19001, Chile; \and
Observatories of the Carnegie Institution for Science, 813 Santa Barbara Street, Pasadena, CA 91101, USA; \and
INAF -- Osservatorio Astrofisico di Arcetri, Largo E. Fermi 5, I-50157 Firenze, Italy; \and
Max-Planck-Institut f\"{u}r Astronomie, K\"{o}nigstuhl 17, D-69117, Heidelberg, Germany; \and
Astronomisches Rechen-Institut, Zentrum f\"{u}r Astronomie der Universit\"{a}t Heidelberg, M\"{o}nchhofstra\ss e 12-14, 69120 Heidelberg, Germany; \and
European Southern Observatory, Karl-Schwarzschild Stra{\ss}e 2, D-85748 Garching bei M\"{u}nchen, Germany; \and
Univ Lyon, Univ Lyon1, ENS de Lyon, CNRS, Centre de Recherche Astrophysique de Lyon UMR5574, F-69230 Saint-Genis-Laval France; \and
International Centre for Radio Astronomy Research, University of Western Australia, 7 Fairway, Crawley, 6009 WA, Australia; \and
Department of Physics, Tamkang University, No.151, Yingzhuan Road, Tamsui District, New Taipei City 251301, Taiwan; \and
Argelander-Institut f\"{u}r Astronomie, Universit\"{a}t Bonn, Auf dem H\"{u}gel 71, 53121 Bonn, Germany; \and
Department of Physics \& Astronomy, University of Wyoming, Laramie, WY 82070, USA; \and
Instit\"ut  f\"{u}r Theoretische Astrophysik, Zentrum f\"{u}r Astronomie der Universit\"{a}t Heidelberg, Albert-Ueberle-Strasse 2, 69120 Heidelberg, Germany; \and
Research School of Astronomy and Astrophysics, Australian National University, Canberra, ACT 2611, Australia; \and
2 ARC Centre of Excellence for All Sky Astrophysics in 3 Dimensions (ASTRO 3D), Australia; \and
Department of Astronomy, The Ohio State University, 140 West 18th Avenue, Columbus, OH 43210, USA; \and
Center for Cosmology and Astroparticle Physics, 191 West Woodruff Avenue, Columbus, OH 43210, USA.
}

\date{Received XXX; accepted}

\abstract{In this work, we present a new catalogue of $>40000$ ionised nebulae distributed across the 19 galaxies observed by the PHANGS-MUSE survey.
The nebulae have been classified using a new model-comparison-based algorithm that exploits the odds ratio principle to assign a probabilistic classification to each nebula in the sample.
The resulting catalogue is the largest catalogue containing complete spectral and spatial information for a variety of ionised nebulae available so far in the literature.
We developed this new algorithm to address some of the main limitations of the traditional classification criteria, such as their binarity, the sharpness of the involved limits, and the limited amount of data they rely on for the classification.
The analysis of the catalogue shows that the algorithm performs well when selecting \hii\ regions.
In fact, we can recover their luminosity function, and its properties are in line with what is available in the literature.
We also identify a rather significant population of shock-ionised regions (mostly composed of supernova remnants), which is an order of magnitude larger than any other homogeneous catalogue of supernova remnants currently available in the literature.
The number of supernova remnants we identify per galaxy is in line with results in our Galaxy and in other very nearby sources.
However, limitations in the source detection algorithm result in an incomplete sample of planetary nebulae, even though their classification seems robust.
Finally, we demonstrate how applying a correction for the contribution of the diffuse ionised gas to the nebulae's spectra is essential to obtain a robust classification of the objects and how a correct measurement of the extinction using diffuse-ionised-gas-corrected line fluxes prompts the use of a higher theoretical \Ha$/$\Hb\ ratio (3.03) than what is commonly used when recovering the \ebv\ via the Balmer decrement technique in massive star-forming galaxies. 
}


\maketitle

\section{Introduction}
\label{sec:intro}

The interstellar medium (ISM) is one of the main components of galaxies, and its study is critical to understanding their formation and evolution.
The vast majority of the ISM is in the form of gas, whose properties span a wide range of physical conditions, mostly depending on the gas phase (atomic, molecular, ionised) \citep[e.g.][]{Ferriere01, Haffner05, Cox05}.
Each gas phase is related to specific properties of the host galaxy, and has a different role in its structure and evolution.
In particular, the warm phase of the ISM, composed of partially ionised gas with temperatures around $8000$--$10000\,\si{K}$, \citep[e.g.][]{Osterbrock06}, is probably one of the most studied components of the ISM since it produces an emission-line-rich spectrum in the optical band.
While only a relatively small fraction of the mass of the ISM is in this form \citep[$\sim 15$--$20$\% in the Milky Way;][]{Ferriere01}, its study is crucial in understanding many properties of galaxies (e.g. kinematics, chemical composition and enrichment, and massive star formation).
Part of this ionised gas is concentrated in nebulae connected to specific ionisation mechanisms, which are processes that provide energy to the gas and ionise it.
The vast majority of these nebulae can be classified into three classes: \hii\ regions, planetary nebulae (PNe), and supernova remnants (SNRs).
The first clas of nebulae, \hii\ regions, are clouds of gas associated with regions of recent massive star formation.
The young, massive and hot stars produced in these regions emit large quantities of ultraviolet photons with energies capable of ionising the surrounding hydrogen ($13.6$ eV).
These nebulae are typically characterised by an electron temperature of the order of $\sim 10000\,\si{K}$, relatively low electron density ($10$--$10^3\,\si{cm^{-3}}$), and sizes varying between $\sim1$ and $100\,\si{pc}$ depending on the type, number, and distribution of the stars powering the nebula \citep[e.g.][]{Maciel13}. 
Since they are strictly connected with the star formation process, \hii\ regions are primarily found in star-forming galaxies, where they account for the majority of the observed nebulae.
They are a precious tool for investigating many aspects of the star formation process and the properties of the ionised gas, such as the initial abundances of the gas from which stars are currently being produced \citep[e.g.][]{Stasinska04}.

The second class of sources, PNe, are again clouds of gas photo-ionised by a central ionising source.
This time, however, each nebula is ionised by a single star, typically an extremely hot \citep[$50000$--$300000$ K;][]{Maciel13} white dwarf. 
Those harder spectral sources produce high ionisation lines, which are significantly stronger than in \hii\ regions.
In particular, PNe are bright in the \oiii\ line, which is used to identify them in extragalactic environments.
On the other hand, the low luminosity of the white dwarfs leads to relatively small ionised regions \citep[e.g. $< 1$ pc for bright PNe;][]{Acker92}.
Planetary nebulae play an essential role in studying galaxies. 
For example, their luminosity function can be used to measure the distance of relatively nearby galaxies \citep[e.g.][]{Jacoby89, Ford96, Ciardullo02, Rekola05, Herrmann08, Kreckel17, Scheuermann22} and it is, therefore, a key step of the distance ladder.
They are also widely used as test particles to trace galaxies' gravitational potential and dark matter content via dynamical modelling \citep[e.g.][]{Hui95, Arnaboldi96, Arnaboldi98,Douglas02}.
Their physics is also similar to that of \hii\ regions, so it is relatively easy to use their spectra to study, for example, the temperature, density and chemical composition of the gas \citep{Stasinska04}.
However, PNe trace gas that has been processed by the central star in its previous life stages, while \hii\ regions typically trace the unprocessed gas from which they were recently formed.

Finally, SNRs, as their name says, are the result of a supernova explosion interacting with the ISM.
During a supernova explosion, a large amount of material processed by the exploding star is ejected into the ISM, where it starts expanding at velocities of the order of one to ten thousands of km s$^{-1}$.
This creates powerful shock waves that heat up the gas to extremely high temperatures ($\sim 10^{7}$--$10^{8}\,\si{K}$).
In these nebulae, the main ionisation mechanism of the gas is not photo-ionisation from a central source, as in \hii\ regions and PNe, but collisional ionisation by radiative shocks.
Supernova remnants are characterised by strong synchrotron continuum emission in the radio band, but their optical spectra are relatively similar to those of other nebulae.
The main differences reside in the line ratios that involve low ionisation lines (i.e. \sii, \nii), which are typical shock tracers, and the kinematics of the gas since the high expansion velocities are reflected in the profile and properties of the observed emission lines \citep[e.g.][]{Maciel13}.
These nebulae mostly trace the positions where supernova explosions occurred in the past.
Knowing where, how and when supernovae exploded is fundamental to understanding the energetics and chemical evolution of the ISM since these episodes are one of its main sources of feedback and chemical enrichment.

In this work, we take advantage of the data from one of the main surveys carried out by the Physics at High Angular Resolution in Nearby GalaxieS\footnote{\url{www.phangs.org}} (PHANGS) collaboration, the PHANGS-MUSE survey \citep{Emsellem22}, to develop a new automatic algorithm that satisfies all these criteria.
In Sec.~\ref{sec:intro_class} we give an overview of the traditional methods used to classify ionised nebulae in the literature.
In Sec.~\ref{sec:data} we describe the data we used for this work.
Section~\ref{sec:creation} focuses on the creation of the nebula catalogue we use to develop the classification algorithm.
In Sec.~\ref{sec:class} we describe the classification process extensively, while in Sec.~\ref{sec:analysis} we compare its result with the classification obtained by applying traditional classification criteria.
In Section~\ref{sec:disc} we discuss some of the main properties of the catalogue and how it performs in some common applications.
Finally, in Sec.~\ref{sec:conc} we provide a final summary of this work.

\section{Traditional classification of nebulae}
\label{sec:intro_class}

Identifying and correctly classifying nebulae is fundamental to measuring their properties and understanding their nature and origin.
In our Galaxy and in a few nearby sources like the Magellanic Clouds, classifying these objects is relatively easy. 
Most of the nebulae we observe in these systems can be easily resolved in multiple different bands (e.g. optical, radio) with the current instrumentation. 
Thus, we can identify the ionising source powering them with reasonable confidence and study their spatially resolved properties.
When we move to other galaxies, even to the closest ones (e.g. M31, M33), we soon lose both the ability to resolve most nebulae (except for the largest star-forming regions) and to perform their multiwavelength characterisation (most of them can be detected only in the optical and infrared bands).
This makes the classification of single nebulae extremely challenging.
Nevertheless, several classification methods have been developed in the literature that leverage the main differences in the optical spectra of the nebulae (e.g. line ratios or luminosities) to define simple criteria that can be applied efficiently to large samples of nebulae.

Traditionally, the search for ionised nebulae has been performed using narrow-band images, that is, images acquired with filters characterised by a narrow pass-band ($<100\,\si{\angstrom}$) centred on specific emission lines \citep[e.g.][]{Jacoby89, Ciardullo02, Kennicutt03, Kennicutt08}.
This method allows one to cover a large field of view, arcminutes or even degrees in diameter depending on the instrument, and analyse many nebulae simultaneously.
However, it is highly time-consuming because of the inherent long exposure time needed to get good signal-to-noise in narrow-band images, and the large overheads caused by the acquisition of off-band images needed to remove the contribution underlying stellar continuum.
Consequently, only a small subset of lines is typically observed when studying ionised nebulae, the two or three lines where the specific type of nebulae the authors are interested in is brighter.
This strongly influenced the characteristics of many classification criteria commonly used in the literature.

\subsection{Planetary nebulae}
\label{sec:intro_pne}

Originally, extragalactic PNe were identified by blinking photographic plates acquired with different filters.
One of these filters was centred on one or more lines (`on-band'), typically \oiii, while the other one was observing a nearby region of the spectrum free of emission lines (`off-band').
Planetary nebulae would appear as unresolved objects in the on-band images but would be undetected in the off-band images \citep[e.g.][]{Baade55,Ford73}.
One of the first attempts to create simple, objective, criteria to classify ionised nebulae using their spectral properties is the one from \citet{Sabbadin77}.
Using data from galactic nebulae, they define several regions in two different diagnostic diagrams that could be used to identify the nature of the considered nebulae (their fig. 1 and 2).
This criterion, revised recently by \citet{Riesgo06}, has been used for decades, particularly for identifying PNe.
These diagrams, however, require detecting a relatively large number of emission lines (\sii, \Ha, \nii) which makes this method more suitable for studying bright galactic nebulae than faint extragalactic sources, especially when using narrow-band images.
Therefore, the blinking technique continued to be used for many years until \citet{Ciardullo02} finally developed a more objective way to separate PNe from other contaminants.
While they still use the blinking technique to identify the first sample of nebulae, the new method could be used independently as long as precise measurements of the \oiii\ and \Ha\ lines are available for all the objects.
Specifically, they consider as PNe all those sources that have $\rm R = [O~{\textsc{iii}}]\lambda5007/ \left(H\alpha+[N~{\textsc{ii}}]\lambda6584\right) > 1.6$\footnote{The authors argue that the proper ratio should be $\rm R > 2$ but their \Ha\ images were not deep enough to reach such high ratios.}.
Later, \citet{Herrmann08} defined a new relation that identifies the parameter space in the R vs. \oiii\ absolute magnitude diagram where PNe are supposed to be located based on the work of \citet{Ciardullo02}.
Although, it must be noted that this criterion is less restrictive than Ciardullo et al.'s one, and it could produce samples with rather high contamination.
Despite this, it is one of the most commonly used criteria to identify PNe \citep[e.g.][]{Kreckel17, Roth21, Galan21, Scheuermann22} together with a more refined version of the on- and off-band blinking \citep[e.g][]{Bhattacharya, Hartke20}.
Other classification criteria based on the BPT diagrams have been developed in the literature, for example \citet{Frew10}, but even though they require flux measurements for more emission lines to be applied (which makes them more suitable for bright, galactic nebulae) they still do not ensure an exact classification \citep[e.g.][]{Roth21}.

\subsection{Supernova remnants}
\label{sec:intro_snr}

The vast majority of Galactic SNRs were historically detected via their radio emission \citep[e.g,][]{Bolton49, Mills52, Bennet63}.
At optical wavelengths, they appear as faint and diffuse objects, and the heavy dust extinction of the Galactic disk makes them even more difficult to observe \citep{Green84, Magnier95}. 
However, the first searches for SNRs outside the Milky Way still focus on the radio emission to identify them.
\citet{Mathewson72} were the first who tried to define an optical criterion to isolate SNRs from \hii\ regions, by noticing that, typically, SNRs have a \sii$/$\Ha\ ratio which is relatively close to 1, while the \Ha\ line observed in \hii\ regions is more or less an order of magnitude brighter than the \sii\ doublet.
While this is a poor quantitative criterion, it provided the framework for the development of more refined criteria in future works.
The first quantitative definition of the \sii$/$\Ha\ criterion comes from \citet{Dodorico78}.
Based on the analysis of line ratios reported in the literature for both SNRs and \hii\ regions, they classify as SNRs those nebulae with \sii$/$\Ha$\geq 0.6$ and as candidate SNRs other nebulae with \sii$/$\Ha$\geq 0.4$.
A few years later, \citet{Dodorico80} finally set \sii$/$\Ha$\geq 0.4$ as the criterion to distinguish between SNRs and \hii\ regions in extragalactic environments.

Despite being more than forty years old, this classification criterion is still widely used for identifying extragalactic SNRs \citep[e.g.][]{Long10, Lee14, Moumen19}, but it is often associated with other criteria requiring, for example, a shell-like morphology of the nebulae or the absence of an obvious ionising source (i.e. a blue, hot star).
The already mentioned \citet{Frew10} and \citet{Sabbadin77} diagrams can be used to identify also SNRs, but since they require spectroscopic observations or more narrow-band imaging to produce the required diagnostic ratios, most of the search for SNRs performed in extragalactic environments still prefer to use the simpler \sii$/$\Ha\ criterion.

Other narrow-band friendly classification criteria have been proposed in the literature.
For example, \citet{Fesen85} proposed new criteria based on the \oi$/$\Hb\ or [O~\textsc{II}$\lambda3727$]$/$\Hb\ ratios while \citet{Boeshaar80} one based on the \sii$/$ [Ar~\textsc{III}]$\lambda7136$ line.
More recently, \citet{Kopsacheili20} used shock models from \citet{Allen08} and starburst models from \citet{Kewley01} and \citet{Levesque10} to study the best approach to the identification of SNRs.
They show that while a multi-ratio approach outperforms single-ratio approaches, the \oi$/$\Ha\ is the single-ratio criterion that best discriminates between SNRs and other nebulae, while the \sii$/$\Ha\ generally does not perform as well.
Therefore, while some of these criteria seem to be more effective in discriminating SNRs from other regions than the \sii$/$\Ha\ ratio, probably the faintness of the required lines (\oi, [Ar~\textsc{III}]) or the need for precise extinction correction has prevented them from becoming particularly popular.

With the diffusion of integral field spectroscopy, however, it is becoming more clear that simple narrow-band imaging is not sufficient to correctly identify SNRs but that only a careful evaluation of the spectrum can produce reliable results.
For example, \citet{Long22} observationally confirm \citet{Kopsacheili20} results, showing that while objects with \sii$/$\Ha $>0.4$ are typically SNRs, a relatively large fraction of objects with properties compatible with those of SNRs (e.g. enhanced \oi$/$\Ha, \nii$/$\Ha, \siii$/$\sii, broadened lines) are characterised by a \sii$/$\Ha $<0.4$ and would be missed by this traditional criterion.
On the other hand, some more complex diagrams involving several lines and the line velocity dispersion have been developed in the literature \citep[e.g.][]{Davies17, Dagostino19, Dagostino19b}, and confirmed that indeed, the velocity dispersion is critical in correctly identifying shock-ionised regions.
However, most of these methods focus on identifying shock-ionised regions connected to galactic outflows or to the interaction between AGN jets and the ISM. 
Consequently, some of the criteria used are unsuitable for identifying SNRs. 
In particular, SNRs are expected to have enhanced gas velocity dispersion with respect to PNe and \hii\ regions, but there is no consensus around an exact cut that can be used to distinguish between the different classes of nebulae because the observed velocity dispersion depends on the SNR age.
The older the SNR is, the lower the expected velocity dispersion.
High velocity dispersion objects are rarely observed in nearby galaxies \citep[e.g.][]{Long22, Winkler21}, so applying cuts at velocity dispersion $\sim \SI{100}{\km \per \s}$ as in \citet{Davies17} to identify SNRs would return only this young population of objects.

\subsection{\hii\ regions}
\label{sec:intro_hii}

\hii\ regions are the most common nebulae in star-forming galaxies, with hundreds or thousands of nebulae detected in single galaxies (e.g. \citealt{Knapen98, Bradley06, Azimlu11, Rousseau18, Santoro22}, hereafter \citetalias{Santoro22}).
They can span a wide range of sizes and luminosities, and they are typically both brighter and larger than most PNe and SNRs.
Their size and brightness make them relatively easy to observe even in an extragalactic environment, and since their blended emission still appears as an \hii\ region, they can also be observed in relatively distant objects where it is not possible to resolve the single nebulae \citep[e.g.][]{Espinosa20}.

An emission-line-rich spectrum is characteristic of \hii\ regions, with \Ha\ typically being the brightest line \citep{Osterbrock06}.
As a consequence, \Ha\ emission became one of the main \hii\ region tracers, both in our galaxy \citep[e.g.][]{Sharpless53,Gum55,Sharpless59} and outside \citep[e.g.][]{Baade64, Hodge69, Pellet78}.
Other detection techniques based on their radio or infrared emission exist and have been applied to search for \hii\ regions in the Milky Way \citep[e.g.][]{Reifenstein70, Lockman89, Kuchar97, Anderson14}, but \Ha\ is still the main \hii\ region tracer in other galaxies (e.g. \citealt{Azimlu11,Espinosa20}; \citetalias{Santoro22}).

While \hii\ regions are bright and easy to detect, for several reasons, no simple criterion based on a single line ratio has been defined yet for their identification.
First of all, if we consider data with similar depth, \hii\ regions dominate in numbers over PNe and SNRs in star-forming galaxies.
For example, \citet{Hodge99} found 2338 \hii\ regions in M~33, while \citet{Lee14b} found only 199 SNR and \citet{Magrini00} 134 PNe.
Secondly, PNe emit most of their radiation through the \oiii\ line, with \Ha\ a factor of $\sim2$ fainter than \oiii, so they are relatively difficult to detect in the \Ha\ narrow-band images typically used to identify \hii\ regions.
Finally, both PNe and SNRs are typically fainter than most \hii\ regions \citep[see the different luminosity functions, e.g.][]{Lee14, Ciardullo02, Kennicutt89}.
They contaminate only the faint end of the \hii\ region luminosity function (which is typically ignored when fitting and studying this quantity; \citetalias{Santoro22}), and they are easily excluded from spectroscopically characterised samples of \hii\ regions.
So for most applications, cleaning \hii\ region catalogues is not as essential as cleaning samples of SNRs or PNe.

Since \hii\ regions can be relatively bright, it is possible to acquire high-quality spectra of these nebulae also in more distant galaxies.
Therefore, it is possible to characterise the properties of these regions rather easily and confirm their classification, for example, by building traditional diagnostic diagrams \citep[also known as BPT diagrams;][]{Baldwin81, Veilleux87} to identify the ionising mechanism of the nebulae.
The relations from \citet{Kewley01} and \citet{Kauffmann03} are used to separate star-forming regions from other types of nebulae in these diagrams.
These relations can be applied to spectra of both integrated sources (i.e. galaxies) and single nebulae.
With the advent of integral field spectroscopy, this technique has been used to identify \hii\ regions in several integral field surveys of galaxies (\citealt{Lopez13, Sanchez18, Kreckel19, Dellabruna20, Espinosa20, Grasha22}; \citetalias{Santoro22}).

\subsection{The need for a new classification scheme}

As emphasised in the previous section, while such classification criteria have been used for decades, including in recent surveys (e.g. \citealt{Moumen19, Scheuermann22}; \citetalias{Santoro22}), they suffer from significant limitations.

The first thing to notice is that most of the criteria are defined to select a specific class of nebulae.
This is a consequence of using narrow-band images to identify nebulae.
Acquiring these images can be time-consuming and with large amounts of overheads, and it encourages the definition of criteria aimed at selecting a specific type of nebulae (requiring the observation of a small number of emission lines), more than classifying them comparatively.
Some criteria that try to classify all the three classes of nebulae simultaneously exist \citep[e.g.][]{Sabbadin77, Frew10}, but the need for a relatively large number of narrow-band images or spectroscopic follow-ups probably limited their popularity and detailed quantification of their performance.

Another characteristic of all criteria is that they are based on sharp limits, which can fail to correctly classify nebulae with properties close to the limits.
For example, \citet{Kopsacheili20} showed how there is an overlap between shock-ionised and photo-ionised regions and that the traditional \sii$/$\Ha\ criterion can both misclassify photo-ionised regions as shocks and vice versa.
Similarly, the PNe classification criterion from \citet{Herrmann08} and \citet{Ciardullo02} identifies the region of a specific parameter space where PNe lives, but it does not exclude the presence of other types of nebulae.
This also creates degeneracies between different classifications that must be solved in other ways.
Also, faint regions with low \sn\ detections in the important lines can be easily misclassified 
by these sharp criteria, since relatively small errors in the flux measurements can result in different classifications.
For this reasons, these criteria often require dedicated complex analysis or human interaction (e.g. morphological classification of barely resolved regions, association with faint ionising sources, etc.) to perform as expected, and are not suitable for the automatic analysis of a large number of sources such as the ones produced by modern integral field surveys \citep[thousands or tens of thousands, e.g.][]{Rousseau18, Sanchez18, Kreckel19}.
While, for obvious reason, there is no overlap in the criteria that can classify different types of nebulae at the same time \citep[e.g.][]{Sabbadin77, Frew10}, they still suffer from the problem of having sharp limits, and not being able to correctly classify nebulae at the edge of the classification criteria \citep{Roth21}.

Finally, in the last decade, the advent of new integral field instruments capable of acquiring spatially resolved spectra of large fields of view, such as the Multi Unit Spectroscopic Explorer \citep[MUSE,][]{Bacon10} and the Spectromètre Imageur à Transformée de Fourier pour l'Etude en Long et en Large de raies d'Emission \citep[SITELLE,][]{Grandmont12} changed our view of the ISM in nearby galaxies completely.
With these instruments, it is possible to simultaneously observe properties of the nebulae that would require both imaging (position, size, morphology) and spectroscopic (multiple line fluxes, kinematics) observations.
Despite this wealth of information, the classification of the nebulae is still primarily performed using the traditional classification methods (e.g. \citealt{Sanchez18, Moumen19, Espinosa20, Dellabruna20, Rhea21, Scheuermann22}; \citetalias{Santoro22}) focusing on a single class of nebulae at a time.

An ideal classification algorithm should have a few key characteristics.
First, it should be probabilistic.
That is, it should return a probability associated with each classification.
It should then take advantage of the wealth of information recovered from integral field units (IFU) data while maintaining a certain degree of flexibility (e.g. it should work with incomplete datasets).
Finally, the algorithm should be automated and objective.
This is essential to work on the large influx of data that is being produced by modern IFU surveys without the need for any human intervention.
\section{Data}
\label{sec:data}

In this paper, we analyse 19 galaxies observed with MUSE \citep[][]{Bacon10} as part of the PHANGS-MUSE ESO Large Program \citep[P.I. Schinnerer;][]{Emsellem22}.
The galaxies in the sample are all nearby ($\rm D < 20\,\si{Mpc}$), star-forming and mildly inclined (i $< \ang{75}$), to minimise the effect of extinction and blending.
Each galaxy was observed with a variable number of pointings (from 3 to 15), to map a significant fraction of its star-forming disk.
All the targets were observed in wide field mode (WFM).
All the data have been reduced using \verb!pymusepipe!\footnote{\url{https://github.com/emsellem/pymusepipe}}, a \verb!python! wrapper of the ESO data reduction pipeline \citep{Weilbacher20}, specifically developed for the reduction of PHANGS data.
Then, the reduced data have been combined, to produce a single mosaicked datacube for each target galaxy.
In this work, we use a homogenised version of the mosaics, with constant point-spread-function both spatially and as a function of wavelength.
All the details about the sample selection, the data reduction, convolution, and much more can be found in \citet{Emsellem22}.

The fully reduced data are then processed by the data analysis pipeline (DAP) to extract high-level data products, such as continuum subtracted cubes, two-dimensional maps of line properties (flux, velocity, velocity dispersion), stellar mass, age, etc.
In particular, the DAP is a \verb!python! software based on the \verb!gist! code \citep{Bittner19} that allows us to extract high-level data products (e.g. fluxes, kinematics) from the emission lines and stellar continuum observed in our data.
The data analysis takes place in several steps, which are extensively described in \citet{Emsellem22}.

To compile our new catalogue of nebulae, however, we did not use the DAP emission-line maps because, by design, the DAP cannot return negative flux values when fitting emission lines. 
As a consequence, performing the fitting of a line where no emission is present results in positively biased fluxes which can simulate the emission of faint regions.
These artefacts then can be identified by the segmentation algorithm when trying to identify faint objects.
To minimise this issue we integrated the flux over a fixed wavelength window in a continuum subtracted cube to create channel-integrated flux maps.
While the continuum subtracted cubes are still somewhat connected with the Gaussian line fitting, especially in the Balmer lines where both the absorption and emission features are fitted simultaneously, this approach should depend much less on the results of the actual fit especially for lines not overlapping with strong absorption features like the \oiii\ line and the \sii\ doublet.

We used the \Ha\ velocity maps produced by the DAP to estimate the reference wavelength of each line in every single spaxel.
Since \Ha\ is detected in the vast majority of the observed area \citep{Emsellem22}, the kinematics of the line is reliable enough for our goal, particularly considering that we define a large extraction window ($\SI{500}{\km\per\s}$) centred on each line to allow for eventual differences in the line kinematics.
In the case of the \sii\ doublet, we position the extraction window at the average wavelength of the two lines, and we increase its size to extract both of them simultaneously to avoid cross-contamination of the line maps that may arise by their vicinity.
Finally, for each line channel-integrated flux map, we also create a error map by summing in quadrature the errors extracted from the original datacubes in the considered wavelength range.
In Sec.~\ref{sec:creation} we describe how we used these maps to produce the detection map we use to create the catalogue.

\section{Catalogue creation}
\label{sec:creation}

The first step towards the creation of our catalogue of classified nebulae is identifying all the nebulae we detect in our data.
We use CLUMPFIND \citep{Williams94} to perform the first identification and to create a tentative segmentation map, which is then processed by a custom algorithm that rejects spurious detections and refines the segmentation.

\subsection{CLUMPFIND analysis}
\label{sec:clumpfind}

CLUMPFIND is an algorithm developed by \citet{Williams94} to detect emission regions in a variety of astronomical data.
It uses a contour-based approach first to identify peaks, and then to connect pixels inside the same contour to the nearest peak in a similar fashion as what is done by the human eye.
The algorithm is efficient at identifying peaks, especially in crowded areas, but it tends to overestimate sizes and to have a significant fraction of spurious detections caused by noise spikes.
This motivates a post-processing of the CLUMPFIND output via a dedicated algorithm (see Sec.~\ref{sec:post}).

To proceed with the detection, CLUMPFIND needs two inputs: an image, and a configuration file containing all the segmentation parameters, including the definition of the contours that will be used during the analysis.
To maximise the number of detected regions, we run the algorithm on a detection image, created by combining the \oiii, \Ha\ and \sii\ line channel-integrated flux maps described in Sec.\,\ref{sec:data}. 
This particular combination of lines accounts for the fact that some of the nebulae we are looking for can be particularly bright in some lines while undetected in the others.
The final detection image is a weighted average of the three channel-integrated flux maps described above, where the weight of each pixel in each line flux map is defined by the square of its signal-to-noise ratio (\sn), that is the ratio between the line flux and its error according to the channel-integrated flux and error maps produced in the previous step.
This allows us to boost the detection of regions emitting only in a sub-sample of the considered lines.
We also extract the error map associated with the detection image by summing in quadrature the emission-line flux error maps weighted by the same weights used for the creation of the detection image, as indicated by standard error propagation techniques.

The second step is to define the contours that CLUMPFIND will consider for the segmentation process.
To do so, we measure the average background level and the standard deviation of the detection image in areas without the evident presence of structured emission (e.g. nebulae, filaments). 
This is used to set up a lower limit (background + standard deviation) that defines the faintest contour considered by the algorithm.
From this lower limit (reported in Tab.~\ref{tab:sample}), we define logarithmically and evenly spaced contours to allow a good sampling of the faintest regions, without crowding the brightest peaks with too many contours.
At this point, we run CLUMPFIND on all the galaxies of the sample.
We use the version of the algorithm implemented in the CUPID\footnote{\url{http://starlink.eao.hawaii.edu/starlink/CUPID}} software package \citep{Currie14,Berry07}, in legacy mode \citep[to behave like the original CLUMPFIND IDL algorithm from][]{Williams94}, and with default options\footnote{For the default values see the \href{http://starlink.eao.hawaii.edu/docs/sun255.htx/sun255ss5.html\#x11-260000}{CLUMPFIND documentation}}.

\subsection{Post-processing and catalogue creation}
\label{sec:post}

\begin{figure*}[ht]
\centering
\includegraphics[width=0.9\textwidth]{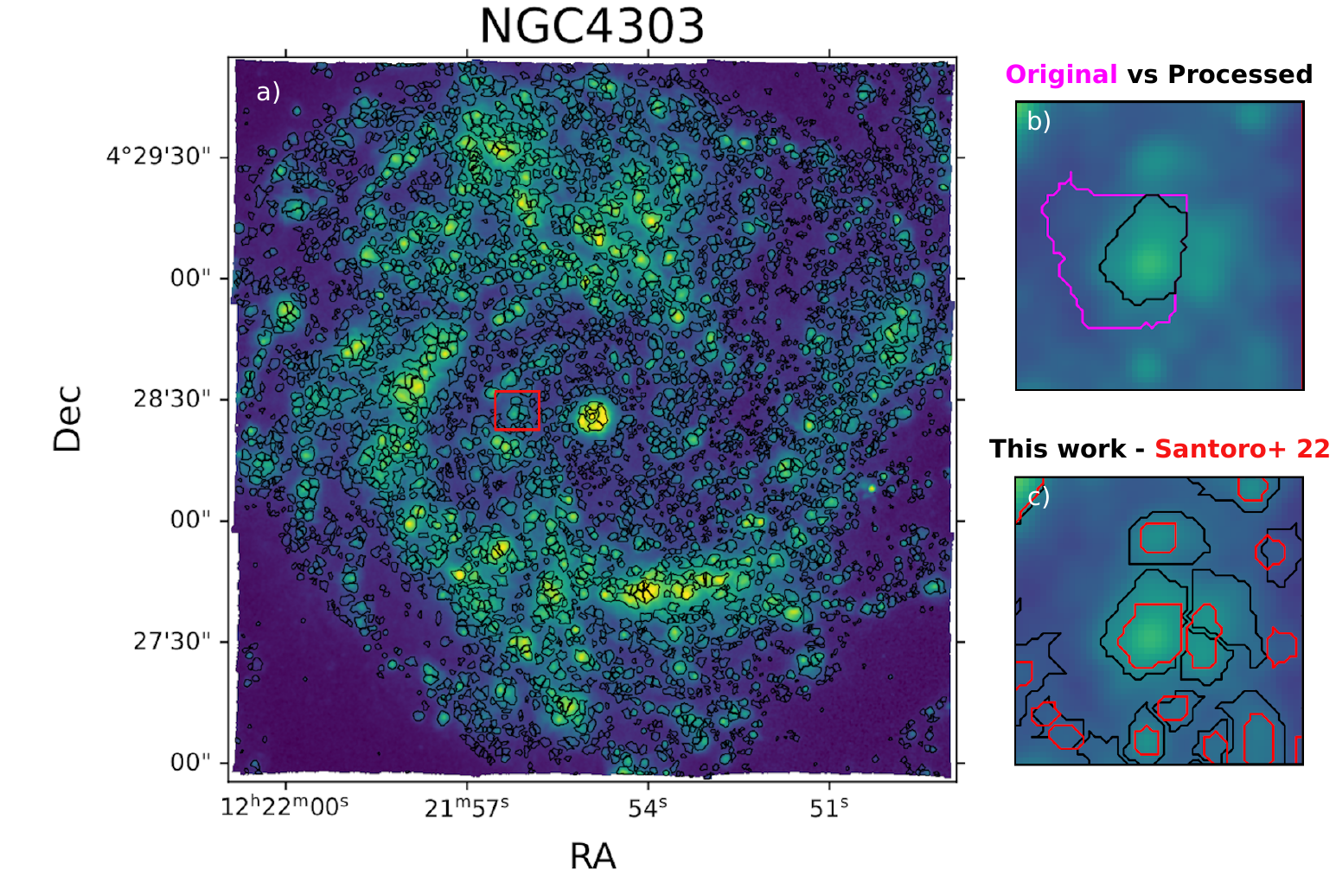}
\caption{NGC4303 segmentation map. \textbf{Panel a)} Map of the S/N-weighted combination of the \oiii, \Ha\ and \sii\ lines (our detection image) of \ngc4303 used for the identification of the nebulae with the final segmentation map (after the post-processing), shown in black contours. The red box highlights the area shown in panels b) and c) \textbf{Panel b)} Zoom-in on a small region of the galaxy showing the difference between the original CLUMPFIND region boundaries (in magenta) and those that result after the post-processing described in Sec.~\ref{sec:post} (in black, only one region is shown). \textbf{Panel c)} Zoom-in on the same area of panel b) showing the difference between our new segmentation map (black) and that produced by \citetalias{Santoro22} (red). As is clear in this panel we find that our regions are larger on average, but also the segmentation for some regions is different.}
\label{fig:segmentation}
\end{figure*}

We run CLUMPFIND on all the 19 galaxies of the sample, producing the first tentative segmentation maps and region catalogues.
However, because of the aggressive CLUMPFIND settings, this first version of the catalogues still contains a significant number of spurious detections.
Moreover, most of the detected regions are overgrown, because, by construction, the algorithm assigns every pixel inside the lowest contour to one of the regions.
To address these issues, we developed a post-processing algorithm that cleans the spurious detections from the CLUMPFIND catalogues, and assigns a more realistic size to the final regions.

The algorithm uses CLUMPFIND catalogues and segmentation maps to create one-dimensional surface-brightness profiles for each region centred on their CLUMPFIND peak, with these profiles then fitted with a Gaussian superimposed on a constant background.
Most regions do not have a Gaussian profile, since they are resolved.
However, a Gaussian fit allows us to extract information like a first guess for its FWHM that will be used later in the rejection criteria.
The background is an estimate of the local contribution of the diffuse ionised gas (DIG) and the contamination from nearby nebulae.

At this point, the algorithm estimates the new boundaries of the regions.
It first subtracts from each region's pixels the constant background estimated by the fit.
Then, it computes the cumulative sum of the remaining flux for each region and identifies the isophote containing $90\%$ of this flux of the considered nebula, that then becomes the new region boundary (Fig.~\ref{fig:segmentation}, panel b).
We also pass the new regions through an algorithm that fills ``holes'' in the regions created by the new definition of the boundaries.

Finally, several criteria are evaluated to reduce the number of false detections included in the final catalogue.
First, the region's total \sn\ measured from the detection image and associated error map is evaluated, and regions with \sn\ $< 5$ are rejected.
Then, all regions with an area $< 12\,\si{pix}$ or FWHM $< 1\,\si{pix}$ are rejected as spurious detections. 
Regions where the surface brightness cumulative distribution does not grow monotonically as a function of distance from the peak are also considered spurious detections and rejected.
Finally, we reject all regions whose peak flux is $< 5\times rms$, where the $rms$ is the standard deviation of the detection image background as computed in Sec.~\ref{sec:clumpfind}, all regions with a FWHM $> 50\%$ of the size of the one-dimensional profile (typically diffuse, filamentary regions) and all regions that overlap with the stellar masks as defined in \citet{Emsellem22}. 

All these criteria, including the $90\%$ flux limit used to redefine the regions boundaries, have been carefully tuned by eye, by comparing the final segmentation map of each galaxy with the detection image and with a colour image obtained associating each channel-integrated line map to a different colour (blue for the \oiii\ line, green for \Ha, red for the \sii\ doublet).
The colour information significantly simplifies distinguishing real emission from spurious detections.
All the applied limits are the result of a compromise between rejecting the fake detections returned by CLUMPFIND, while retaining as many regions that the human eye can pick up as possible.
While these criteria to define the boundaries of the regions are somewhat arbitrary, currently no clear agreement exists in the literature on how best to define the boundary of an ionised gas nebula in external galaxies.
The flattening of the \Ha\ surface brightness profile of each nebula is used to define the boundary of the regions by common segmentation algorithms like HIIPhot \citep{Thilker00}, while other works experiment with using the \sii$/$\Ha\ ratio \citep[e.g.][]{Kreckel16, Dellabruna20} or the \sii$/$\oiii\ ratio \citep{Pellegrini12} to distinguish between \hii\ regions and the DIG.
Moreover, for density-bounded regions a clear boundary does not simply exist.

The output of the detection code is a segmentation map (see Fig.~\ref{fig:segmentation}, panel a) and a catalogue with basic information, such as the position of the region peak, the flux-weighted centre, the area (in pixels) covered by the region, its maximum radius, and its circularised radius.
The total number of detected nebulae across our sample of 19 galaxies is 40920.
On average, we detect $\sim$2150 nebulae per galaxy, with a minimum of 853 in \ngc7496, which is one of the less massive targets and the galaxy with the smallest number of observed pointings (3) and a maximum of 4101 in \ngc628, one of the galaxies with the largest covered area.
In Table~\ref{tab:sample} we report, in detail, the number of detected nebulae for all the observed galaxies.

\begin{table*}
\caption{Properties of the examined galaxies and the results of the detection algorithm.}
 \label{tab:sample}
\centering
\small
\begin{tabularx}{\textwidth}{lccScSSSSXXXX}
\hline\hline
Name& RA(J2000) & Dec(J2000) & {Distance\tablefootmark{a}} & {Scale} & {Stellar Mass\tablefootmark{b}} & {SFR\tablefootmark{b}} & {Area\tablefootmark{c}}&{Min\tablefootmark{d}}&{All\tablefootmark{e}}&{H$\alpha$\tablefootmark{e}}&{OIII\tablefootmark{e}}& {BPT\tablefootmark{e}}\\
    & {hh:mm:ss.s} & {dd:mm:dd} & {Mpc} & {pc/arcsec} & {$\log($\Msun$)$} & {\Msun$/$yr} & {kpc$^2$} & & & & \\
\hline
IC~5332  & 23:34:27.5 & $-$36:06:04 & 8.18  & 40 & 9.45  & 0.31  & 34& 38 &972 & 862 & 569 & 407\\
\ngc628  & 01:36:41.7 & +15:47:01 & 9.84  & 48 & 10.24 & 1.73  & 89 &62&3502 & 2813 & 1594 & 1067\\
\ngc1087 & 02:46:25.2 & $-$00:29:55 & 15.85 & 77 & 9.88  & 1.26  & 126& 22 &1168 & 965 & 740 & 590\\
\ngc1300 & 03:19:41.0 & $-$19:24:40 & 18.99 & 92 & 10.57 & 1.07  & 356& 21 &1707 & 1566 & 1061 & 806\\
\ngc1365 & 03:33:36.4 & $-$36:08:25 & 19.57 & 95 & 10.82 & 16.35 & 409& 20 &2351 & 1784 & 1341 & 901\\
\ngc1385 & 03:37:28.6 & $-$24:30:04 & 17.22 & 83 & 9.94  & 2.01  & 101& 29 &1138 & 905 & 726 & 577\\
\ngc1433 & 03:42:01.5 & $-$47:13:19 & 12.11 & 59 & 10.41 & 0.42  & 426& 17 &2768 & 2451 & 1651 & 939\\
\ngc1512 & 04:03:54.1 & $-$43:20:55 & 17.13 & 83 & 10.58 & 0.79  & 266& 11 &1137 & 947 & 743 & 437\\
\ngc1566 & 04:20:00.4 & $-$54:56:17 & 17.69 & 85 & 10.66 & 4.36  & 208& 46&3274 & 2444 & 1765 & 1145\\
\ngc1672 & 04:45:42.5 & $-$59:14:50 & 19.40 & 94 & 10.64 & 7.39  & 250& 41 &1953 & 1470 & 1030 & 768\\
\ngc2835 & 09:17:52.9 & $-$22:21:17 & 12.38 & 60 & 9.85  & 1.18  & 87& 45 &1489 & 1122 & 858 & 666\\
\ngc3351 & 10:43:57.8 & +11:42:13 & 9.96  & 48 & 10.27 & 1.17  & 73&31&1820 & 1565 & 973 & 526\\
\ngc3627 & 11:20:15.0 & +12:59:29 & 11.32 & 55 & 10.73 & 3.66  & 85&39&2105 & 1483 & 1024 & 590\\
\ngc4254 & 12:18:49.6 & +14:24:59 & 13.0  & 63 & 10.30 & 2.97  & 169&21&3536 & 2635 & 2028 & 1466\\
\ngc4303 & 12:21:54.9 & +04:28:25 & 16.99 & 82 & 10.64 & 5.14  & 214&40&3756 & 2697 & 1683 & 1231\\
\ngc4321 & 12:22:54.9 & +15:49:20 & 15.21 & 74 & 10.71 & 3.48  & 191&27&2858 & 2016 & 1265 & 814\\
\ngc4535 & 12:34:20.3 & +08:11:53 & 15.77 & 76 & 10.49 & 2.04  & 124&52 &2551 & 2145 & 1158 & 715\\
\ngc5068 & 13:18:54.7 & $-$21:02:19 & 5.16  & 25 & 9.36  & 0.28  & 23&55&1974 & 1612 & 1219 & 955\\
\ngc7496 & 23:09:47.3 & $-$43:25:40 & 18.72 & 91 & 9.84  & 2.08  & 92&33&861 & 728 & 468 & 377\\
\hline
 &  &  &  &  &  &  & &{\textbf{Total}} & 40920 & 32210 & 21896 & 14977\\
\hline
\end{tabularx}
\tablefoot{
\tablefoottext{a}{Distances from \citet{Anand21},}
\tablefoottext{b}{From \citet{Leroy21},}
\tablefoottext{c}{Projected area covered by the survey from \citet{Belfiore22},} 
\tablefoottext{d}{Threshold used by CLUMPFIND to create the first segmentation map in $10^{-20}\,\si{erg/(cm^2.s.\angstrom)}$,} 
\tablefoottext{e}{Samples as defined in Sec.\,\ref{sec:analysis}.}
}
\end{table*}

We use the segmentation map to extract the one-dimensional spectrum of each nebula, \textbf{which we then process with the DAP to recover the fluxes of the brightest lines} that are available within the MUSE wavelength range and the gas kinematics.
The lines included in the fit, together with their wavelength and the ID used in the catalogue, are listed in Table~\ref{tab:lineid}. 
The emission lines are divided into two different groups.
The first group includes the hydrogen emission lines (\Ha\ and \Hb), and the second group includes all the remaining lines.
The kinematics of the lines is tied within each group (to the \Ha\ line and the \nii\ line, respectively) to improve the fit of the faintest ones, as described in \citet{Emsellem22}.
Differently from what was done in that work, we do not separate the high and low ionisation metal lines but include them in the same group.
In particular, we noticed that leaving free the kinematics of the \oiii\ line results in unrealistic velocity dispersions, too large with respect to the profile of the line observed in the spectra, for a significant number of faint regions. 
Fixing the kinematic of the \oiii\ to the other metal lines does not significantly influence the fit in high \sn\ spectra, but it improves the fit in low \sn\ spectra.

\begin{table}
\caption{Emission lines included in the catalogue. Wavelengths are recovered from the National Institute of Standards and Technology (NIST\tablefootmark{a}).}
 \label{tab:lineid}
\centering
\begin{tabular}{lcl}
\hline\hline
Name& Wavelength (\AA)& ID\\
\hline
\Hb\   & 4861.35& \verb!HB4861!\\
\oiii\ & 5006.84& \verb!OIII5006!\\
\Ha\   & 6562.79& \verb!HA6562!\\
\nii\  & 6583.45& \verb!NII6584!\\
\siia\ & 6716.44& \verb!SII6716!\\
\siib\ & 6730.82& \verb!SII6730!\\
\oi\   & 6300.30& \verb!OI6300!\\
\siii\ & 9068.60& \verb!SIII9068!\\
\hline
\end{tabular}
\tablefoot{
\tablefoottext{a}{\url{https://physics.nist.gov/PhysRefData/ASD/lines_form.html}}
}
\end{table}

\subsection{DIG correction}
\label{sec:dig}

\begin{figure*}
\centering
\includegraphics[width=0.9\textwidth]{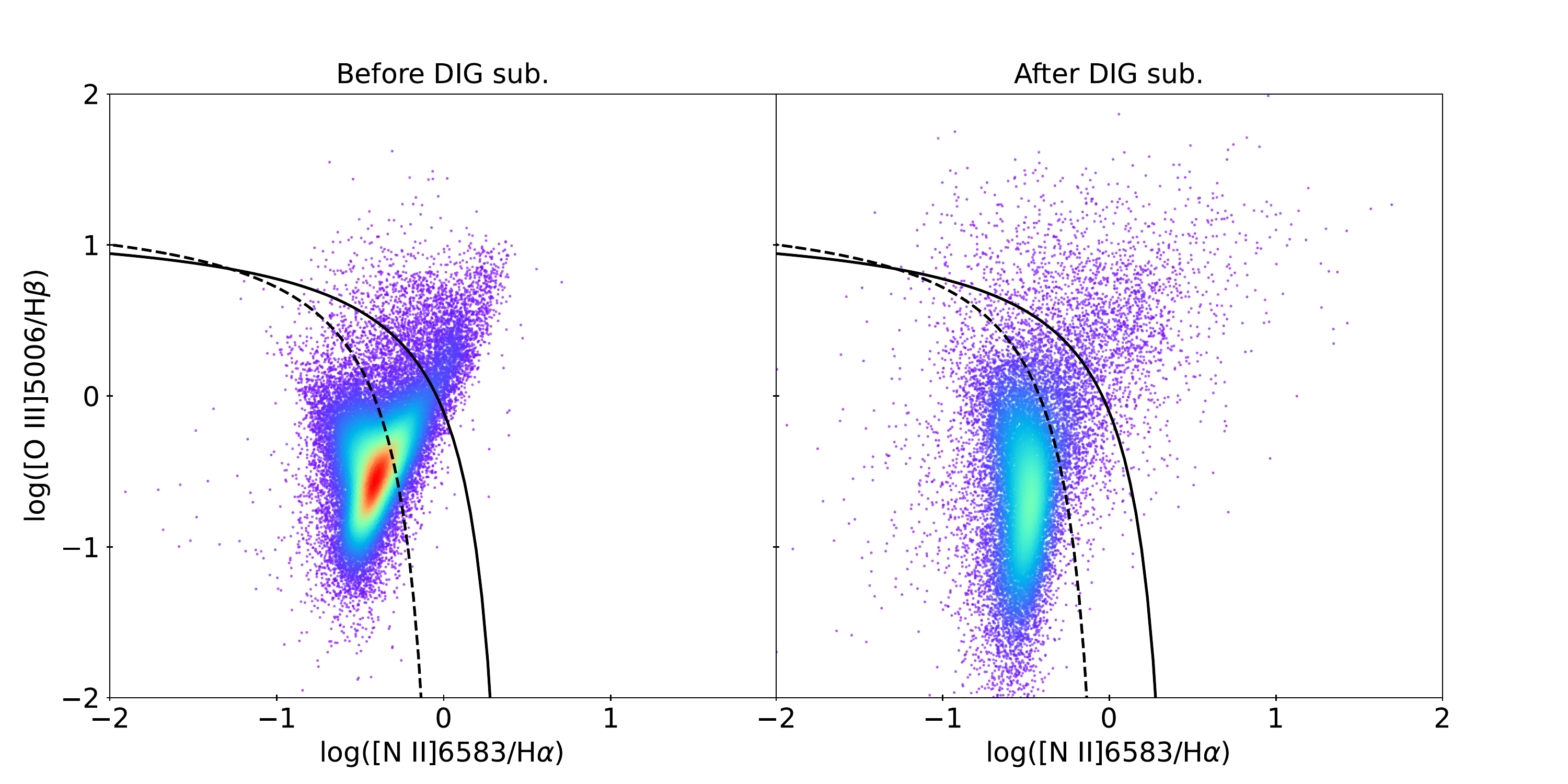}
\caption{$\rm \log([N \textsc{ii}]/\Ha)$ vs $\rm \log([O \textsc{iii}]/\Hb)$ diagram of the nebulae before (\textbf{left}) and after (\textbf{right}) the correction for DIG contamination. Only nebulae where \Ha, \Hb, \oiii\ and \nii\ are detected at $3\sigma$ after the DIG correction are plotted in the right panel. The colour map qualitatively shows how the points are concentrated. The solid and dashed black lines are the relation from \citet{Kewley01} and \citet{Kauffmann03} which separate star-forming regions from regions ionised by other ionising sources.} 
\label{fig:bpt}
\end{figure*}

In all galaxies, there is a warm, low-density component of ionised gas, called DIG, which is distributed across the whole galactic disk.
While the ionisation source of this gas is still a matter of debate \citep[e.g.][]{Belfiore22}, it has been proven that the DIG can contribute a significant fraction to the total line emission in some galaxies \citep[between 10 and 60\% of the total \Ha\ emission of a galaxy,][]{Thilker02, Oey07, Blanc09}.

This diffuse emission has a larger scale height with respect to the distribution of ionised nebulae \citep{Reynolds97} and it can contaminate our integrated regions due to projection effects.
Consequently, when measuring the emission lines flux in nebulae such as \hii\ regions, they will be contaminated by the DIG emission.
The level of contamination depends on several factors, for example, the relative strength of the nebula emission with respect to the DIG. 
The brighter the nebula is with respect to the DIG, the less important the contamination is, and vice versa.

The main problem is that the DIG spectrum is characterised by enhanced low-ionisation line emission \citep[e.g.][]{Reynolds85, Hoopes96, Otte99} with respect to the spectrum of other nebulae like \hii\ regions and PNe.
As a consequence, the DIG emission can significantly alter the line ratios of the observed nebulae resulting in the likely misclassification of DIG-dominated regions if its contamination is not taken into account during the classification procedure.

To correct the spectra of our sources for DIG emission, we use the following procedure.
First, for each region, we use a Gaussian filter with a standard deviation equal to the circularised radius of the nebula to increase the size of the nebula itself.
Then, the area covered by the original nebula is masked, and we recover an annulus whose size depends on the circularised radius of the original nebula. 
We further mask any overlap between this annulus and the neighbouring regions.
The algorithm extracts the DIG spectrum associated with the region using the annulus and feeds it to the same spectral extracting algorithm used to measure the line fluxes of the nebulae.
Finally, the correction is applied by subtracting the estimated DIG flux of each line from the regions' flux, after applying a rescaling factor to consider the difference in size between the nebula and the associated annulus.
In some cases, the estimated DIG emission is larger than the flux emitted by the considered region.
This happens mostly for faint regions in crowded locations, where the DIG varies strongly.
These lines are considered undetected, and when needed in the following analysis their flux is substituted by an upper limit after conducting a formal error propagation of the nebula and DIG emission line fluxes.

Figure~\ref{fig:bpt} shows a comparison between the \oiii$/$\Hb\ vs \nii$/$\Ha\ diagnostic diagram before and after applying the DIG correction.
It is possible to see how the distribution of the points significantly changes after applying a DIG correction to our catalogue of nebulae.
Before subtracting the DIG contribution, the points are relatively clustered close to the centre of the plot, following the distribution typically observed in diagnostic diagrams when comparing the properties of different types of sources \citep[e.g.][]{Kewley01}, with a branch of nebulae extending from the core of the distribution towards the upper right (LINER-like) region of the diagram.
After the DIG correction, the distribution of points appears to be less concentrated, with a larger scatter. 
The bulk of the points seems to follow a sequence that runs parallel to the delimitation line and is consistent with a dominant \hii\ region population.
Only regions where all lines necessary to compute the diagnostic ratios (\Ha, \Hb, \oiii, \nii) are detected at a $3\sigma$ level after the DIG correction ($\sim 46\%$ of the sample) are plotted in the right panel, so the scatter is real, and not only a consequence of low \sn\ regions.
Something interesting to notice is that the prominent structure of points connecting the star-forming region (on the bottom$/$left of the \citealt{Kauffmann03} line) to the region of the diagram typically associated with AGN, LINER and shock emission is not as prominent after the DIG correction.
This is the typical location where DIG-dominated regions are found \citep[e.g. figure 15 of][]{Belfiore22}.
Its absence, therefore, means that we are successfully removing DIG contamination from the spectra of the nebulae, and it highlights the importance of this step prior to the classification of individual nebulae.

\section{Classification}
\label{sec:class}

Here, we develop a new classification algorithm based on comparing the regions' morphological and spectral characteristics with models of nebular properties using the principle of the odds ratio.
This new algorithm is designed to overcome most of the issues of the traditional classification criteria: 1) it does a comparative classification, 2) it associates a probability to each classification, 3) it consistently considers non-detections and upper limits, 4) it can take advantage of the plethora of morphological and spectroscopic information that can be recovered from modern high spatial resolution integral field spectroscopic datasets. 

\subsection{Odds ratio}
\label{sec:odds}

Our classification algorithm is based on the concept of the \emph{odds ratio}.
According to Bayesian statistics, the odds ratio is defined as:
\begin{equation}
    \label{eq:odds}
    O_{ij} = \frac{p(M_i,\lvert I)}{p(M_j,\lvert I)}\frac{p(D,\lvert M_i, I)}{p(D,\lvert M_j, I)},
\end{equation}
where $p(M_{i},\lvert I)$ are the priors on the model $i$  and $p(D,\lvert M_{i}, I)$ is the global likelihood of the data $D$ given the model $i$ \citep{Gregory05}, also called the `evidence'.
When comparing two models to the same data and given some priors on them, if we can estimate the evidence of each model, Eq.\,\ref{eq:odds} gives us a way to determine which one is better describing the data.
The odds ratio also works when comparing more than two models.
In this case, the probability of each model can be recovered with the following equation:
\begin{equation}
\label{eq:oddsprob}
    p(M_i, \lvert D, I) = \frac{O_{i1}}{\Sigma_{i=1}^{N_{mod}} O_{i1}},
\end{equation}
where $p(M_i, \lvert D, I)$ is the probability of model $i$ given the data and the priors on the data, and $O_{i1}$ is the odds ratio between model $i$ and model $1$, which is assumed as a reference.

The main problem with this approach is the computation of the evidence. 
Standard model fitting techniques like Markov-Chain Monte Carlo (MCMC) methods are extremely powerful when used for parameter estimation, but they are not designed for evidence estimation.
An alternative method, that has been specifically developed to perform well both for parameter estimation and for evidence estimation, is `nested sampling' \citep{Skilling04,Skilling06}.
Based on the concept of `typical sets', this algorithm estimates the evidence and the posterior simultaneously by integrating the prior in nested shells of constant likelihood.

\subsection{Models}
\label{sec:models}

To classify our sources, we want to compare their spectra to different families of models, each one of them associated with a different class of nebulae.
As explained in Sec.~\ref{sec:intro}, almost all the emission-line regions belong to one of three classes of sources: \hii\ regions, PNe and SNRs.
To perform our classification, we need to define models corresponding to each class.

We use two-parameter grids of standard photo-ionisation or shock models, that adequately represent different families of objects. 
While these models suffer from limitations associated with the simplistic assumptions they adopt (e.g. simple geometries, homogeneity in the gas distribution, reduction of extra secondary parameters), they are sufficient for our purposes as we are not interested in inferring the physical properties of individual nebulae at this stage, but instead just assigning a classification.
Moreover, the larger the model grid, the more computationally expensive is the fitting procedure.

To represent \hii\ regions, we use the models from \citet{PerezMontero14}.
These models are a small grid developed to study chemical abundance in \hii\ regions using direct methods.
The grid depends on a total of three parameters (ionisation parameter, metallicity, N$/$O ratio), and there are two versions of it, one where the C abundance is linked to the O abundance and the second where the C is connected to the N one.
In this work, we use the grid with the C abundance tied to the O abundance, and a fixed $\log(\mathrm{N/O}) = -0.75$.
We selected these values by comparing the grids with the BPT diagrams \citep{Baldwin81} and identifying the grid that was best covering the \hii\ regions area of the BPT diagrams.
Table~\ref{tab:models} reports a summary of the selected grid parameters.

Planetary nebulae are represented by the models from \citet{Delgado14}.
These are a large grid of models that have been produced to study the ionisation correction factor of several ions in PNe.
The grid depends on eight parameters (spectral energy distribution shape, density law, metallicity, dust depletion, temperature of the ionising source, hydrogen density, stellar luminosity and inner radius of the nebula).
As done for \hii\ regions, we consider only a sub-sample of full grid of models.
The details of the parameters of the selected grid are reported in Table\,\ref{tab:models}.
We selected the more realistic models produced with stellar atmospheres from \citet{Rauch03} and including dust depletion.
Abundances are the default CLOUDY solar abundances since they are the only ones with which the models are provided.
For the inner radius of the nebula and the gas density, we selected the intermediate values between the set provided in the grids.
This left as our free parameters the temperature and luminosity of the ionising star.

Finally, we select the fast shock models by \citet{Allen08} to represent SNRs.
These are a set of simulations of shocks produced using the MAPPINGS photo-ionisation and shock modelling code typically used to study SNRs \citep[e. g.][]{Dopita10,Sabin13,Blair14,Micelotta16,Kopsacheili20}.
As we did for the \hii\ regions and PNe grids, we selected a sub-sample of the available models representing standard conditions found in SNRs \citep[e.g.][]{Dopita10}.
One problem with these models is that they have been extensively used in the literature to also describe the emission of the narrow-line region and extended narrow-line region of AGN \citep[e. g.][and many others]{Allen08,Schlesinger09,Holt09,Sarzi10}.
While most of our galaxies are purely star-forming galaxies, a few objects are known to host an AGN in their nucleus (e.g. \ngc1365).
So it might be possible that some of the regions classified using these models are AGN ionised regions and not real SNRs.
For this reason, we decided to identify the nebulae classified by these models as shocks or shock-ionised nebulae, and we leave to Sec.~\ref{sec:shocks_dis} a discussion of how other types of nebulae are contaminating our sample of SNRs.

\begin{figure*}
\centering
\includegraphics[width=0.9\textwidth]{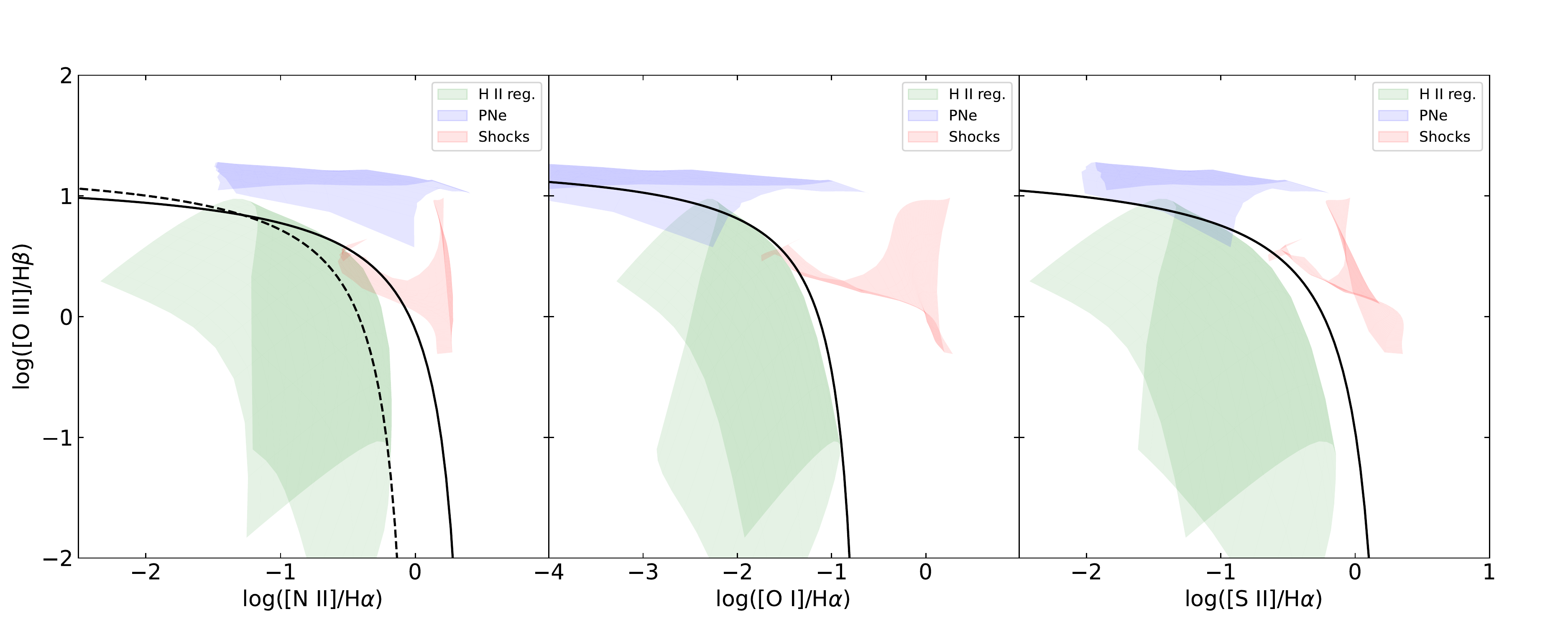}
\caption{Line ratios from the models used for classifying the nebulae in the catalogue plotted in the traditional diagnostic diagrams from \citet{Baldwin81}. Black solid lines represent the relations from \citet{Kewley01} which separates star-forming regions from other kinds of sources. The black dashed line represents the limit for pure star formation identified by \citet{Kauffmann03}.}
\label{fig:bpt_models}
\end{figure*}

The \citet{Delgado14} and \citet{PerezMontero14} models have been downloaded from the Mexican Million Model database\footnote{\url{https://sites.google.com/site/mexicanmillionmodels}} (3MdB).
They are a rerun of the original grids performed with Cloudy v13 \citep{Ferland13} and Cloudy v17.01 \citep{Ferland17} respectively.
The \citet{Allen08} models are downloaded from the shock section of the 3MdB\footnote{\url{3mdb.astro.unam.mx:3686}}, and they are a rerun of the original grid with MAPPINGS V \citep{Sutherland18}.
Figure \ref{fig:bpt_models} shows how the models cover the traditional diagnostic diagrams for classifying ionised nebulae \citep{Baldwin81}.

\begin{table}
\caption{Summary of the model grids used by the classification algorithm. For each grid, we report the reference, the free parameters defining the grids, and the fixed parameter that we used to reduce the dimension of the grid. }
 \label{tab:models}
\centering
\small
\begin{tabular}{lc}
\hline\hline
\multicolumn{2}{c}{\hii{} region}\\
\hline
Reference& \citealt{PerezMontero14}\\
Free parameters& Ionisation par., $12+\log(\mathrm{O}/\mathrm{H})$\\
C tied to & O\\
$\log(\mathrm{N}/\mathrm{O})$ & $-0.75$\\
\hline\hline
\multicolumn{2}{c}{Planetary Nebulae}\\
\hline
Reference& \citealt{Delgado14}\\
Free parameters& Luminosity, Temperature\\
Spectral energy distribution& Rauch (TR)\\
Dust depletion& Yes\\
Density& $3000\,\si{cm^{-3}}$\\
Inner radius& $\SI{1e17}{\cm}$\\
Model set& Matter bounded (M40)\\
Abundances & $8.94$ $(12+\log(\mathrm{O}/\mathrm{H}))$\\
\hline\hline
\multicolumn{2}{c}{Shock}\\
\hline
Reference & \citealt{Allen08}\\
Free parameters& Magnetic field, Shock velocity\\
Model set& shock w$/$o precursor\\
Pre-shock density& $1 \,\si{cm^{-3}}$ \\
Abundances & Solar \citep{Anders89} \\
\hline
\end{tabular}
\end{table}

\subsection{Model fitting with \izi}
\label{sec:izi}

The classification process happens in two steps.
First, we compare the observations to the different grids and compute the evidence using the nested sampling algorithm.
Second, we use the evidence to compute the odds ratio and the probability for each region to belong to a specific class via Eq.~\ref{eq:odds} and Eq.~\ref{eq:oddsprob}.

To compare the regions to the models and compute the evidence, we modified the python version of \izi\ \citep{Blanc15} developed by \citet{Mingozzi20} \footnote{\url{https://github.com/francbelf/python_izi}} to allow for the computation of the evidence required for the model comparison process.

The original \izi\ code was designed to estimate the metallicity and ionisation parameter of \hii\ regions by doing a full likelihood calculation over the parameter space, given a photo-ionisation model grid and a set of observed extinction-corrected line fluxes and their associated errors. 
\citet{Mingozzi20} added the possibility to use a more efficient Markov Chain Monte Carlo (MCMC) algorithm instead of the full likelihood calculation and to fit the amount of dust extinction, modelled as a foreground screen with a fixed extinction law and parameterised by a reddening \ebv\ parameter following the original idea from \citet{Brinchmann04}. 
While MCMC is not ideal for computing evidences, \izi\ provides all the infrastructure we need to compare observations to models, including a self-consistent treatment of errors and upper limits.

To estimate the evidence, however, we changed the core fitting method from MCMC to nested sampling.
In particular, we use the implementation provided by the \verb!dynesty!\footnote{\url{https://dynesty.readthedocs.io}} python package \citep{Speagle20} which can simultaneously estimate both the evidence and the posterior PDF, allowing the software to be used for parameter estimation.
Another advantage of \verb!dynesty! is that it can automatically decide when the algorithm converges based on objective parameters, eliminating the uncertainties connected to the long time that an MCMC chain could take to reach convergence and to the arbitrary selection of a proper burn-in phase.
However, nested sampling must fully sample the prior volume, resulting in a loss of efficiency.
The change from MCMC to nested sampling required further modifications to the code, mainly related to how nested sampling (and \verb!dynesty! in particular) treats priors with respect to other commonly used MCMC-based algorithms.
Finally, we include the possibility of using model grids that depend on whatever couple of parameters, not only on the metallicity and ionisation parameter as in the original software.
This code version will not be released to the public yet because we are planning a complete refactoring of \izi, which will include these and other updates (e.g. the possibility to use n-dimensional grids).
The final version of the code will be released as a self-consistent pip package, together with an upcoming publication focused on the physical properties of the nebulae classified in this work.

\subsection{Comparison with the models}
\label{sec:comparison}

Our updated version of \izi\ compares the properties of each region to the prediction of the grids of models previously described and to prior information regarding the expected morphology and kinematics of different types of nebulae, to compute the evidence associated with each considered class.
In our case, we consider the principal lines used in the traditional diagnostic diagrams \citep[\Ha, \Hb, \oiii, \oi, \nii, \siia\ and \siib;][]{Baldwin81, Veilleux87}.
Both the observed and model fluxes are normalised to \Hb.
The observed line fluxes are corrected for Galactic extinction using the \citet{Cardelli89} extinction law, but not for internal extinction.
The internal reddening is self-consistently estimated by \izi\ when matching the model to the data, by using a standard \citet{Cardelli89} reddening law, as revised by \citet{ODonnel94}, with $\rm R_v = 3.1$.
As a consequence, for each family of models used during the classification, we estimate a separate value of the reddening (\ebv).
A comparison between these estimates and the one recovered via the Balmer decrement is reported in Sec.\,\ref{sec:EBV}.

We classify all the nebulae of the catalogue, without applying any cut related to the \sn\ ratio of any line.
We expect most of the regions to be detected with a \sn\ $>3$ only in a few emission lines (mostly \Ha), especially after the DIG correction.
Moreover, sources like PNe are expected to be mostly detected in \oiii, and a \sn\ cut would result in losing a significant fraction of these objects.
All lines with S/N $<3$ are considered as upper limits (with the associated error being considered as the upper limit), which \izi\ can treat appropriately during the likelihood and evidence calculations.

Line fluxes, however, are not the only quantities relevant for the classification of ionised nebulae.
In particular, PNe are expected to be compact objects, with diameters of the order of $1\,\si{pc}$ \citep[e.g.][]{Acker92, Bojicic21} and, therefore, unresolved in all PHANGS galaxies.
Similarly, SNR can reach sizes of the order of $100\,\si{pc}$ \citep{Asvarov14}.
They are larger than PNe but smaller than many \hii\ regions.
Another quantity that is relevant to the classification is the velocity dispersion of the emission lines.
\hii\ regions and PNe are usually characterised by low-velocity dispersion \citep[e.g.][]{Richer06, Hajian07, Schoenberner14, Dellabruna20, Dellabruna21, Spriggs20} and unresolved in medium-resolution data such as the one provided by MUSE.
On the other hand, SNRs are characterised by more complex kinematics, and they can show higher velocity dispersion \citep{Points19}.
We include the information on the size of the nebula and velocity dispersion of the gas as a prior during the calculation of the evidence.
In particular, we require PNe to be unresolved and with a velocity dispersion $< 25\,\si{km.s^{-1}}$ \citep{Richer06}, shock ionised regions to be smaller than $100\,\si{pc}$, and \hii\ regions to have a velocity dispersion $<24\,\si{km.s^{-1}}$ \citep{Dellabruna20, Dellabruna21}.
The prior is uniform below the limit and has an exponential decay above it.

Once all of the regions have been compared to the three grids of models and the evidence has been computed, Eq.\,\ref{eq:oddsprob} was used to recover the probability for each region to belong to one of the three classes of sources we are considering in this work.
The final classification of the nebulae is tied to these probabilities, but it is not simply determined by the class of nebulae with the highest probability.
Only those sources with a probability $\geq 90$\% in one of the three classes (\hii\ regions, PNe and Shocks) are considered `classified'.
Those sources where all probabilities are $\leq 90$\% are labelled as `ambiguous'.
This decision is driven by the fact that the odds ratio paradigm for model comparison is definitive (i.e.~produce reliable classifications) only when the results are overwhelmingly in favour of one class with respect to all the others.
Moreover, we expect the presence of regions where it was not possible to correctly deblend the contribution of nearby nebulae with different classification (e.g. \hii\ regions with embedded or nearby SNRs). 
If both nebulae have similar brightness, that is one of the nebulae is not overwhelmingly brighter with respect to the other, we expect them to fall into the ambiguous class.
Finally, there is a sub-sample of sources where the software did not converge in any of the three comparisons and, therefore, no probability has been computed.
These sources are labelled as `unclassified' and mainly consist of nebulae where no line was significantly detected.
Table\,\ref{tab:cat_example} summarises all the information included in the catalogue which is available through \verb!vizier! and the Canadian Astronomy Data Centre (CDAC)\footnote{\url{http://dx.doi.org/10.11570/23.0006}}.

\begin{table*}
\caption{Summary of the information included in the released catalogue of classified nebulae.}
 \label{tab:cat_example}
\centering
\begin{tabular}{ll}
\hline\hline
\multicolumn{2}{c}{Identification}\\
\hline
\texttt{index}& Unique identification number of the region\\
\texttt{gal\_name} & Name of the galaxy hosting the region\\
\texttt{region\_ID}& Identification number of the region within the galaxy\\
\texttt{DIST}& Distance of the galaxy in Mpc\tablefootmark{a}\\
\hline\hline
\multicolumn{2}{c}{Position and Morphology}\\
\hline
\texttt{ra}& Right ascension of the  \Ha\ luminosity-weighted centre of the region\\
\texttt{dec}& Declination of the \Ha\ luminosity-weighted centre of the region\\
\texttt{deproj\_dist}& Deprojected distance from the nucleus of the galaxy (in arcsec)\tablefootmark{a}\\
\texttt{deproj\_phi}& Deprojected position angle ($0$ -- $2\pi$)\tablefootmark{a}\\
\texttt{region\_area}& Area of the region in pixels\tablefootmark{b}\\
\texttt{region\_circ\_rad}& Circularised radius of the region in arcsec\\
\hline\hline
\multicolumn{2}{c}{Spectral properties}\\
\multicolumn{2}{c}{\small \texttt{lineid} should be replaced with the line ID reported in Table\,\ref{tab:lineid}}\\
\hline
\texttt{lineid\_FLUX}& Measured line flux in $10^{-20}\,\si{erg.cm^{-2}.s^{-1}}$\\
\texttt{lineid\_FLUX\_ERR}& Measured line flux error in $10^{-20}\,\si{erg.cm^{-2}.s^{-1}}$\\
\texttt{lineid\_SNR}& S/N of the line\\
\texttt{lineid\_FLUX\_DIG}& DIG corrected line flux in $10^{-20}\,\si{erg.cm^{-2}.s^{-1}}$\\
\texttt{lineid\_FLUX\_ERR\_DIG}& DIG corrected line flux error in $10^{-20}\,\si{erg.cm^{-2}.s^{-1}}$\\
\texttt{lineid\_SNR\_DIG}& S/N of the line after the DIG correction\\
\texttt{L\_HA}& \Ha\ luminosity in $\si{erg.s^{-1}}$\\
\texttt{L\_HA\_DIG}& DIG corrected \Ha\ luminosity  in $\si{erg.s^{-1}}$\\
\hline\hline
\multicolumn{2}{c}{Kinematics}\\
\multicolumn{2}{c}{\small The kinematics is reported only for the \Ha\ and \nii\ lines.}\\
\hline
\texttt{lineid\_SIGMA}& Velocity dispersion of the line in $\si{km.s^{-1}}$ \tablefootmark{c}\\
\texttt{lineid\_SIGMA\_ERR}& Error on the velocity dispersion of the line in $\si{km.s^{-1}}$\\
\texttt{lineid\_VEL}& Velocity of the line in $\si{km.s^{-1}}$\\
\texttt{lineid\_VEL\_ERR}& Error on the velocity of the line in $\si{km.s^{-1}}$\\
\hline\hline
\multicolumn{2}{c}{Classification}\\
\hline
\texttt{z\_hii}& Evidence of the \hii\ region models\\
\texttt{z\_pn}& Evidence of the PNe models\\
\texttt{z\_sh}& Evidence of the shock models\\
\texttt{p\_hii}& Probability of the \hii\ region models\\
\texttt{p\_pn}& Probability of the PNe models\\
\texttt{p\_sh}& Probability of the shock models\\
\texttt{c}& Final classification\\
\hline\hline
\multicolumn{2}{c}{Extinction}\\
\hline
\texttt{ebv}& Balmer decrement \ebv\ from measured fluxes\\
\texttt{ebv\_dig}& Balmer decrement \ebv\ from DIG corrected fluxes\\
\texttt{ebv\_sh}& \ebv\ estimated by IZI using shock models\\
\texttt{ebv\_hii}& \ebv\ estimated by IZI using \hii\ region models\\
\texttt{ebv\_pn}& \ebv\ estimated by IZI using PNe models\\
\hline
\end{tabular}
\tablefoot{
\tablefoottext{a}{Distances from \citet{Anand21},}
\tablefoottext{c}{pixel size: 0.2 arcsec$/$pix,}
\tablefoottext{b}{Corrected for instrumental broadening.}
}
\end{table*}

\section{Analysis}
\label{sec:analysis}

In this section, we analyse the results of the detection and classification processes.
We start by investigating how the distribution of the objects in the different classes changes as we perform different cuts to the sample to understand how different selections could bias the classification of a catalogue (Sec.~\ref{sec:reliability}).
Then, we apply to the same catalogue some of the most common traditional classification criteria (Sec.~\ref{sec:classical}), and we compare the results of the two different classification approaches to investigate similarities and differences (Sec.~\ref{sec:disc_trad}).

\subsection{Reliability of the classification}
\label{sec:reliability}

\begin{table}
\setlength{\tabcolsep}{5pt}
\caption{Distribution of the regions in the classes defined in Sec.\,\ref{sec:izi} as a function of the considered sample. Next to the absolute number, we report the fraction of nebulae in that class with respect to the total.}
\label{tab:subsamples}
\centering
\small
\begin{tabular}{lrrrr}
\hline\hline
Class & All & \Ha& [O~III] & BPT\\
\hline
\hii\ & 29986 (0.73) & 23939 (0.74) & 14313 (0.65) & 10993 (0.73)\\
PNe & 796 (0.02) & 535 (0.02) & 775 (0.04) & 32 (0.00)\\
Sh. & 6336 (0.15) & 4454 (0.14) & 4181 (0.19) & 2356 (0.16)\\
Amb. & 3722 (0.09) & 3222 (0.10) & 2556 (0.12) & 1546 (0.10)\\
Un. & 80 (0.00) & 60 (0.00) &71 (0.00) & 47 (0.00)\\
\hline
Total & 40920 & 32210 & 21896& 14977\\
\hline
\end{tabular}
\end{table}

\begin{figure*}[!htp]
\centering
\includegraphics[width=0.94\textwidth]{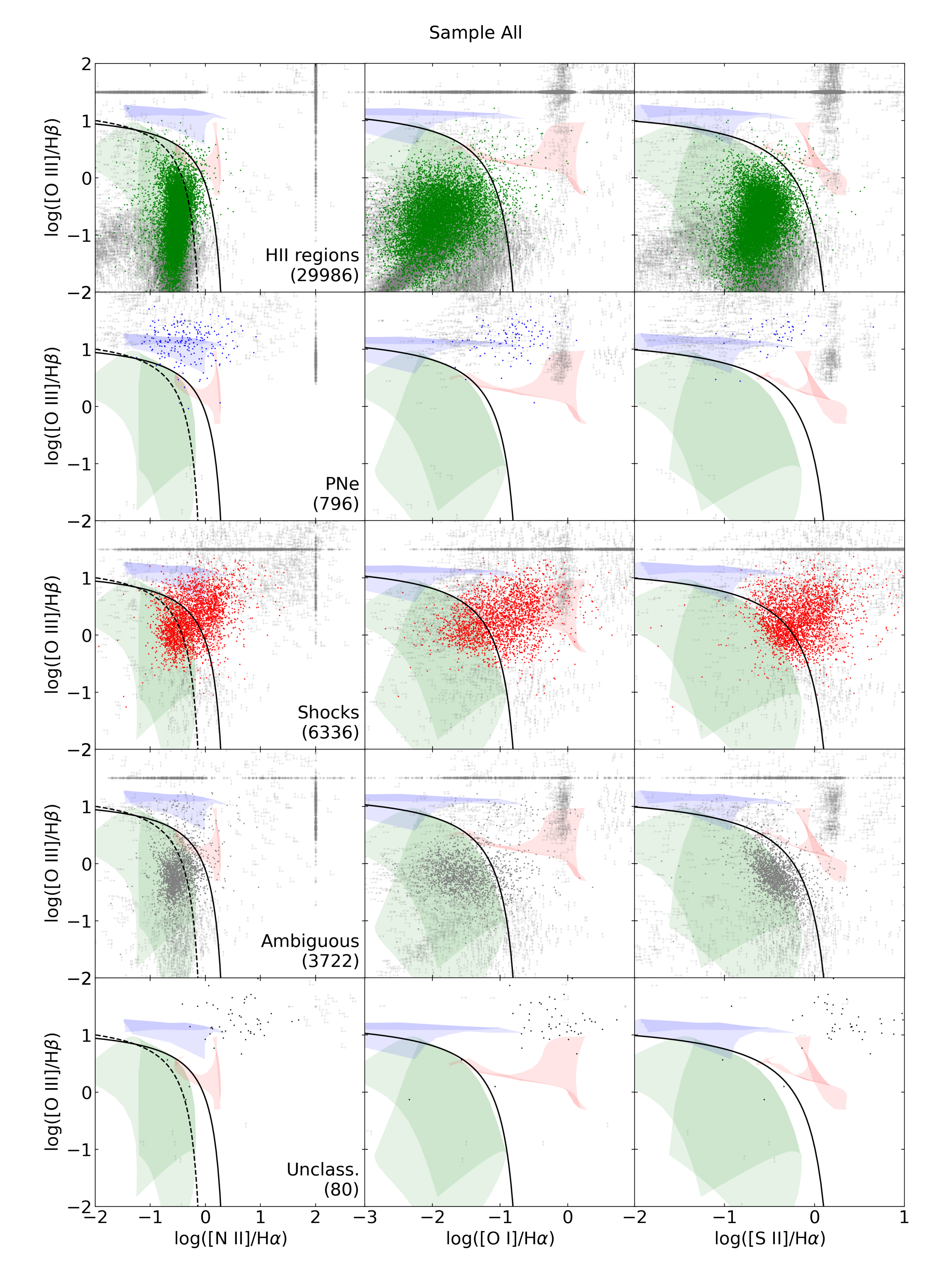}
\caption{
Diagnostic diagrams from \citet{Baldwin81} and \citet{Veilleux87} for the full sample. Lines and shaded areas are defined as in Fig.~\ref{fig:bpt_models}.
In each row only, from top to bottom, we show: \hii\ regions (green), PNe (blue), shocks (red), ambiguous (grey) and unclassified (black) objects. Light grey points in each plot represent upper or lower limits.}
\label{fig:bpt_full}
\end{figure*}

\begin{figure*}[!hp]
\centering
\includegraphics[width=0.94\textwidth]{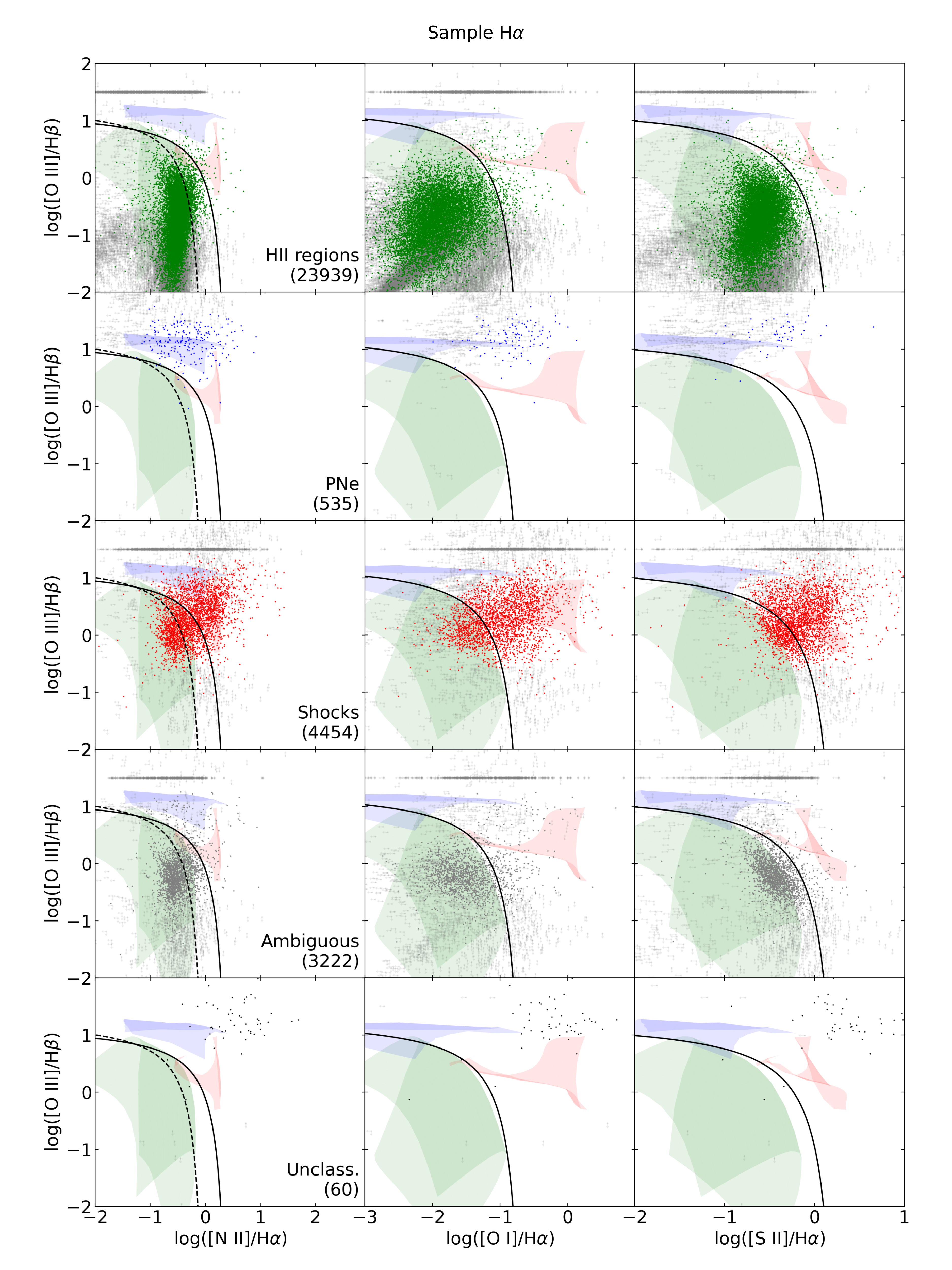}
\caption{Diagnostic diagrams from \citet{Baldwin81} and \citet{Veilleux87} for the \Ha\ sample. Colours and symbols are as in Fig.\,\ref{fig:bpt_full}.}
\label{fig:bpt_ha}
\end{figure*}

\begin{figure*}[!hp]
\centering
\includegraphics[width=0.94\textwidth]{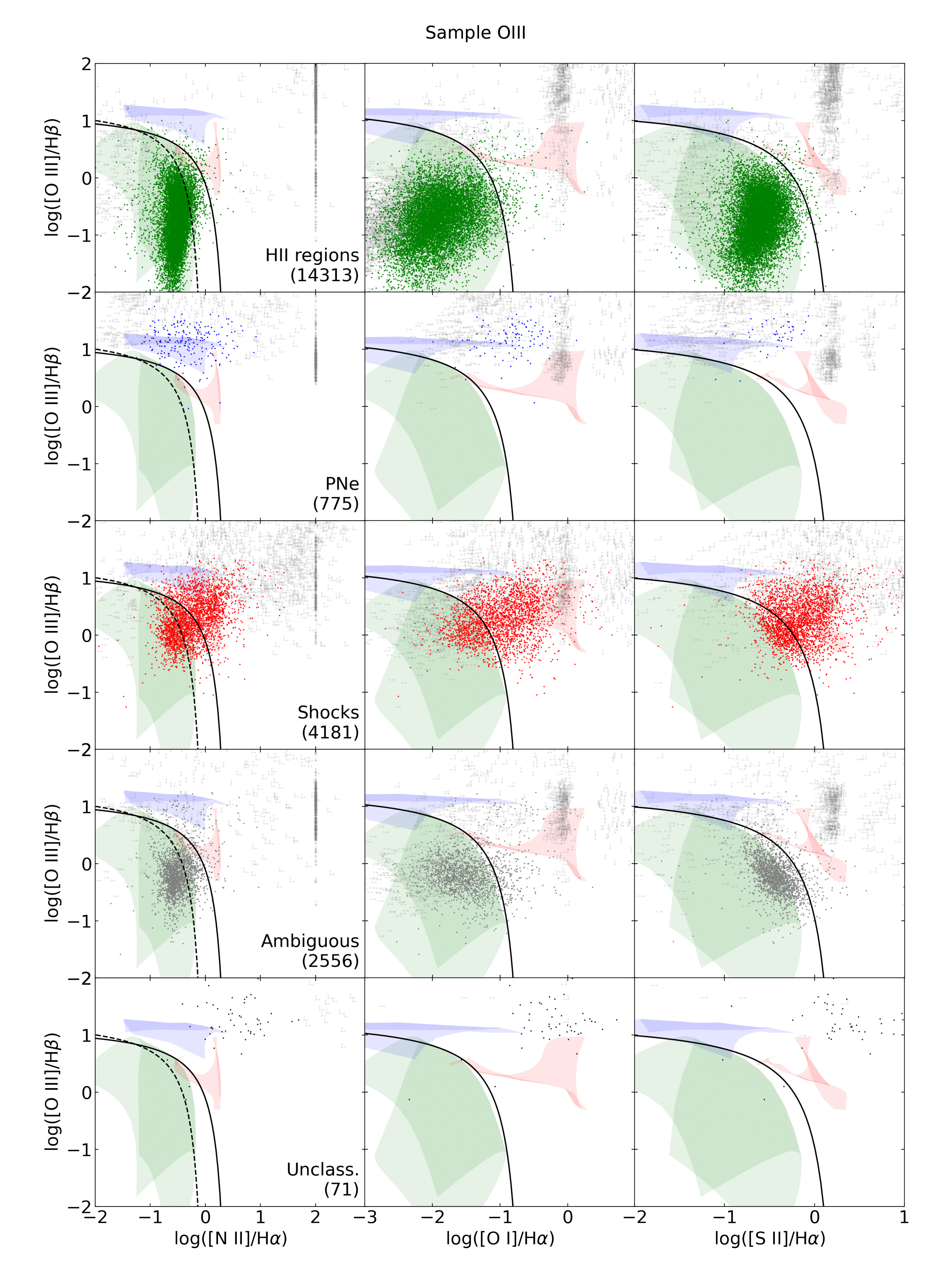}
\caption{Diagnostic diagrams from \citet{Baldwin81} and \citet{Veilleux87} for the OIII sample. Colours and symbols are as in Fig.\,\ref{fig:bpt_full}.}
\label{fig:bpt_o3}
\end{figure*}

\begin{figure*}[!htp]
\centering
\includegraphics[width=0.94\textwidth]{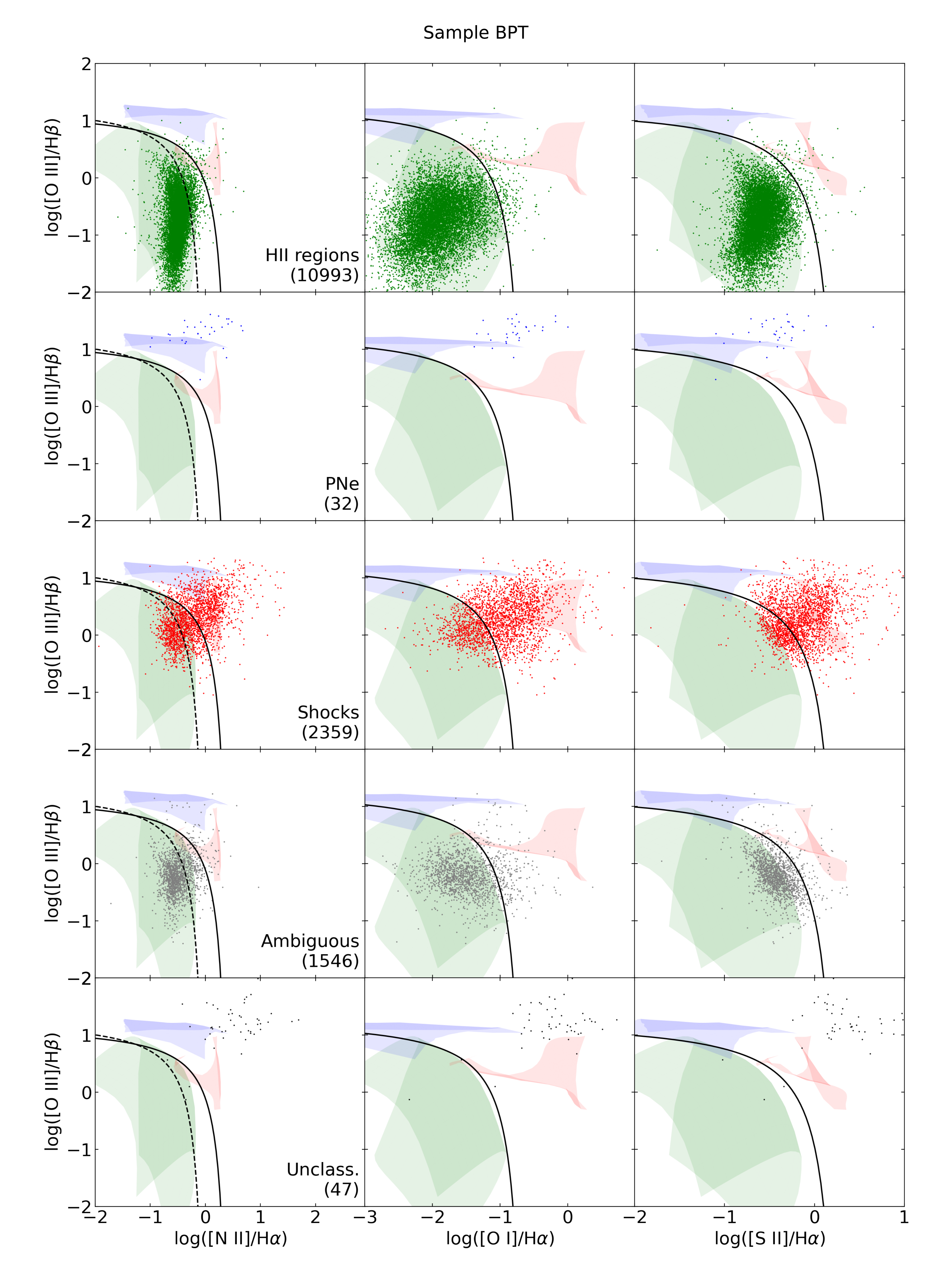}
\caption{Diagnostic diagrams from \citet{Baldwin81} and \citet{Veilleux87} for the BPT sample. Colours and symbols are as in Fig.\,\ref{fig:bpt_full}.}
\label{fig:bpt_bpt}
\end{figure*}

As introduced in Sec.~\ref{sec:comparison}, we run the classification algorithm indiscriminately on all the regions included in our catalogue, independently of the number of detected lines.
However, a lack of detection in certain lines can influence the final classification and how they distribute across the different classes.
To study how the number of detected lines affects the classification of the nebulae, we create three sub-samples by requiring $3\sigma$ detections for different subsets of emission lines. 
Therefore, in the following analysis, we consider four sub-samples of nebulae: a) full sample -- no requirements on line detection significance; b) the \textbf{\Ha} sample requires a $3\sigma$ detection of \Ha; c) the \textbf{OIII} sample requires a $3\sigma$ detection of the \oiii\ line; d) the \textbf{BPT} sample requires a $3\sigma$ detection of all the lines required to build BPT diagrams (\oiii, \Hb, \Ha, \nii, \siia, \siib, \oi).
Each sample's total number of regions is reported in Table~\ref{tab:subsamples}.

The BPT sample is the most restrictive one.
By design, it includes only the brightest and best-detected regions.
The \oiii\ line is relatively faint, so most of the regions in the OIII sub-sample are detected in several other lines ($91\%$ of the regions detected in \oiii\ are also detected in \Ha).
As a consequence, this cut is expected to be the second most restrictive one.
Finally, \Ha\ is the brightest line of the spectrum for the vast majority of ionised nebulae, so many objects in our catalogue are detected in \Ha\ even after removing the DIG contribution.
Figure~\ref{fig:bpt_full} shows the diagnostic diagrams of the full catalogue and compares the classification of each region with their expected classification according to their location in the diagrams.
Similar diagrams for the other sub-samples are reported in Figs.~\ref{fig:bpt_ha}--\ref{fig:bpt_bpt}.

The full catalogue contains 40920 regions, with the vast majority of them (29986, $73\%$ of the sample) classified as \hii\ regions.
As expected, these nebulae are mostly located in the diagnostic diagrams under (or close to) the \citet{Kewley01} lines (Fig.~\ref{fig:bpt_full}, top row).
However, there is a significant amount of points located in other regions of the diagrams ($25\%$).
The bulk of these `misplaced' \hii\ regions ($96\%$) actually corresponds to objects that lack detections in one or more emission lines, and therefore are plotted as limits in the BPT diagrams.
Comparing Fig.~\ref{fig:bpt_full} with Fig.~\ref{fig:bpt_ha}, we can see that the vast majority of these points ($\sim6000$ nebulae) disappear from the diagrams, meaning that their \Ha\ was not clearly detected.
Removal of these nebulae with no significant \Ha\ detection gives 23939 \hii\ regions (or $74\%$ of the \Ha\ sample).

The number of \hii\ regions further decreases if we require the detection of the \oiii\ line, a relatively strict requirement. 
In fact, the number of nebulae reduces to 14313 ($65\%$).
If we apply the final cut and look at the BPT sample, the number of \hii\ regions continues to decrease, but they still represent the vast majority of the nebulae (10993, $73\%$ of the BPT sample).
These are the better-detected sources, and they should have the more reliable classification.
In fact, $\sim98\%$ of them lie below the \citet{Kewley01} lines, where we expect star-forming regions to be.

The second most common class of nebulae are shock-ionised ones (6336 regions, $15\%$ of the sample).
The diagnostic diagrams (Fig.\,\ref{fig:bpt_full}, middle row) show that most of the regions classified as shocks overlap with the selected models, as expected.
There are some points sparsely distributed across the diagrams, but, similarly to what is observed with \hii\ regions, most of them are characterised by limits and, therefore, have a less reliable classification. 
When we require the detection of the \Ha\ line we recover only 4454 nebulae ($14\%$), while requiring the detection of the \oiii\ line we get 4181 nebulae ($19\%$).
Finally, in the BPT sub-sample we have a smaller number of shock ionised regions (2356, $16\%$).

Planetary nebulae is the class with the lowest number of associated regions (796, $2\%$). 
However, they seem to be relatively well clustered at the top of the diagnostic diagrams.
By requiring the detection of \Ha\, the number of nebulae slightly decreases (535, $2\%$), while, when we force the detection of the \oiii\ line, we practically recover the full sample (775, $4\%$).
Finally, if we further strengthen the constraints (requiring the detection of all the BPT lines), only a handful of nebulae are classified as PNe (32, $<1\%$). 
This confirms that most PNe are faint objects detected only in a few lines.
One of those lines, however, must be the \oiii, since this is typically the brightest line in PNe spectra due to the hard ionising radiation field emitted by their central stars.

The number of sources in the ambiguous class varies from 3722 ($9\%$) in the full catalogue to 1546 ($10\%$) in the BPT sample.
The ambiguity for most of these objects is between an \hii\ region or a shock classification, with most of them having higher probabilities of being \hii\ regions.
Only a minority of them show a different combination of non-zero probabilities.
In fact, the most prominent cloud of points is located close to the region of the diagnostic diagrams where the \hii\ region and shock models overlap (Fig.~\ref{fig:bpt_models} and Figs.\,\ref{fig:bpt_full}--\ref{fig:bpt_bpt}, fourth row).

Finally, we have a few unclassified objects in the catalogue (80).
This is a negligible amount of regions ($< 1$\%), and they are mostly located in the top right corner of the diagrams, in a region not covered by any grid of models.
The percentage of regions in this class grow slightly when moving from the full sample to the BPT sample (from $0.1$ of the full sample to $0.3\%$ of the BPT sample) but it always stays a negligible amount with respect to the total number of considered regions.
The reason why these regions are not classified is still not clear.
While there might be some exotic objects in this class (e.g. background galaxies or AGN), most of them look like ordinary nebulae in a visual inspection.

\subsection{Traditional classification criteria}
\label{sec:classical}


\begin{table}
\setlength{\tabcolsep}{5pt}
\caption{Distribution of nebulae in the different classes according to the traditional classification criteria. Next to the absolute number, we report the fraction of nebulae in that class with respect to the total.}
\label{tab:trad}
\centering
\small
\begin{tabular}{lrrrr}
\hline\hline
Class & All & \Ha& [OIII] & BPT\\
\hline
\hii\ & 20540 (0.50) & 20540 (0.64) & 12622 (0.58) & 10338 (0.69)\\
PNe & 647 (0.02) & 647 (0.02) & 456 (0.02) & 46 (0.00)\\
Sh. & 9328 (0.23) & 4123 (0.13) & 4110 (0.19) & 2211 (0.15)\\
Amb. & 8124 (0.20) & 4619 (0.14) & 2992 (0.14) & 1584 (0.11)\\
Un. & 2281 (0.06) & 2281 (0.07) & 1716 (0.08) & 798 (0.05)\\
\hline
Total & 40920 & 31110 & 21896& 14977\\
\hline
\end{tabular}
\end{table}

As introduced in Sec.\,\ref{sec:intro}, our classification algorithm is only the most recent one in a list of methods that have been developed in the literature.
The main novelty of the new algorithm is its use of models for a comparative approach to the classification of resolved nebulae, the assumption of a single ionisation mechanism for each nebula, the automatic inclusion of size and velocity dispersion in the procedure without any need for human interaction, and its probabilistic nature.
A comparison between the new classification and the classification that would have been obtained applying the traditional classification algorithms is fundamental to understanding the differences between the approaches and their reliability.

To traditionally classify \hii\ regions, we apply the criteria based on the BPT diagrams used by \citetalias{Santoro22} and \citet{Groves23} on a similar catalogue.
More explicitly, we classify as \hii\ region all the nebulae that are under the \citet{Kauffmann03} line in the \oiii $/$\Hb\ vs \nii $/$\Ha\ diagram and under the \citet{Kewley01} lines in the other two diagrams.
All three criteria must be satisfied simultaneously for a region to be classified as an \hii\ region.
For PNe we apply the method developed by \citet{Ciardullo02} and \citet{Herrmann08} based on the relative strength of the \oiii\ line with respect to \Ha.
All the nebulae classified as PNe must satisfy Eq.~\ref{eq:pne}, where $M_{[\mathrm{O~\textsc{iii}}]}$ is the \oiii\ absolute magnitude of the nebula computed starting from the apparent magnitude calibrated with Eq.~\ref{eq:mo3} from \citet{Jacoby89} and the distances reported in \citet{Anand21}.
\begin{equation}
\label{eq:pne}
\log{\frac{[\mathrm{O~\textsc{iii}}]5006}{\mathrm{H}\alpha+[\mathrm{N~\textsc{ii}}]6583}}>-0.37\times M_{[\mathrm{O~\textsc{iii}}]} -1.16,
\end{equation}
\begin{equation}
\label{eq:mo3}
    m_{[\mathrm{O\textsc{iii}}]} = -2.5\log_{10}\left(I_{[\mathrm{O\textsc{iii}}]}\right) -13.74.
\end{equation} 

We also require the nebulae to be unresolved, that is their apparent size should be smaller than the size of the PSF.
Lastly, we classify as shocks (or SNRs) all the regions with \sii $/\Ha> 0.4$, the most common criterion for identifying SNRs using narrow-band images \citep{Dodorico80}.
If a region shows one or more limits among its ratios, we classify the object if its ratios and limits are consistent with a specific class.

Also in this case, we include the unclassified or ambiguously classified classes. 
The first class includes all those objects that are not picked up by any criterion, while we consider ambiguously classified all those sources that are selected by more than one criterion.
We show in Table\,\ref{tab:trad} the distribution of the regions classified with the traditional methods, while Appendix~\ref{app:bpt_traditional} Figs.~\ref{fig:bpt_full_trad}--\ref{fig:bpt_bpt_trad} show the same diagnostic diagrams presented for our original classification method.

By comparing the BPT diagrams in Figs.~\ref{fig:bpt_full}--\ref{fig:bpt_bpt} with those in Figs.~\ref{fig:bpt_full_trad}--\ref{fig:bpt_bpt_trad} it is possible to see how the distribution of traditionally classified nebulae changes significantly with respect to what we obtain with the model-comparison-based classification algorithm.
In the traditional paradigm, the shock class collects most of the poorly detected regions. 
This happens because the traditional criteria do not correctly consider the non-detection of one or more lines needed for the classification.
Restricting the catalogue, the number of shocks decreases significantly, to reach a number comparable to what is classified by our algorithm.
Many points classified with the traditional criteria overlap with the models we used to represent shocks in the model-comparison-based classification. 
Still, the traditional criterion tends to classify as shock regions with significantly lower \oiii$/$\Ha\ than our automatic classifier and it extends into the region of the diagnostic diagrams typically occupied by \hii\ regions.
This results in a high fraction of ambiguously classified nebulae, which is a factor $\sim2$ times larger with respect to the results of the model-comparison-based algorithm (15--$20\%$ vs $10\%$ of the total sample of nebulae) and becomes comparable (although still 10--20$\%$ higher) only when applying very strict restrictions (i.e. in the [O III] or BPT sample).
Finally, the sharp boundary in the \sii$/$\Ha\ implies that the selection misses a large fraction of regions with \sii$/$\Ha\ below the cut, but that, according to our algorithm, are better modelled by shock models than \hii\ regions or PNe, as already shown by \citet{Kopsacheili20} and \citet{Long22}.
This highlights the need for a more refined classification method not based on simple, sharp criteria, but able to assign probabilities to ambiguous regions.

The traditional classification criteria also produce many more unclassified regions (a factor of 20--30 higher) with respect to what we find with the new algorithm.
This shows how there are regions of the parameter space of the properties of nebulae that are not covered by any criteria.
By looking at the distribution of the unclassified nebulae on the diagnostic diagrams, it is clear how most of them would be considered as \hii\ regions by using the \citet{Kewley01} line in the \nii$/$\Ha\ diagram instead of the \citet{Kauffmann03} one or by our new classification algorithm.

Most of the nebulae are classified as \hii\ regions also by the traditional criteria (20540, $50\%$), but they are only $\sim 70\%$ of the \hii\ regions classified by our algorithm in the full catalogue. 
However, we have already discussed in Sec.~\ref{sec:reliability} how the \Ha\ sample might provide a more reasonable comparison.
If we compare the results for the \Ha\ sample, the numbers are much more similar (20540, $64\%$, vs 23939, $81\%$), even though they are still significantly lower.
However, we must consider that many of the \hii\ regions classified by the model-comparison-based algorithm are classified as shocks, ambiguous and not classified at all in the traditional paradigm, as discussed above.

Finally, the traditional criteria identify a similar number of PNe with respect to our new algorithm (647, $2\%$).
Their distribution in the diagnostic diagrams matches relatively well the area covered by the models we are using for the classification, except for the \oi$/$\Ha\ diagram.
There, the nebulae show a significantly higher \oi$/$\Ha\ with respect to the models.
This problem is shared with the nebulae classified by the new algorithm.
This means that either we overestimate the brightness of the \oi\ line in our measurements, or that the models are not representative of the full population of PNe, which is possible, since we are not using the complete set of PNe models.
There is also a relatively large cloud of points with low \oiii$/$\Hb\ that overlap, at least partially, with the \hii\ regions models.
This could be a consequence of the fact that the traditional criterion does not exclude regions that are not PNe, but only select regions that are compatible with being PNe candidates \citep{Ciardullo02, Herrmann08}.

\subsection{\izi\ model comparison vs traditional criteria}
\label{sec:disc_trad}

\begin{figure*}
\centering
\includegraphics[width=0.45\textwidth]{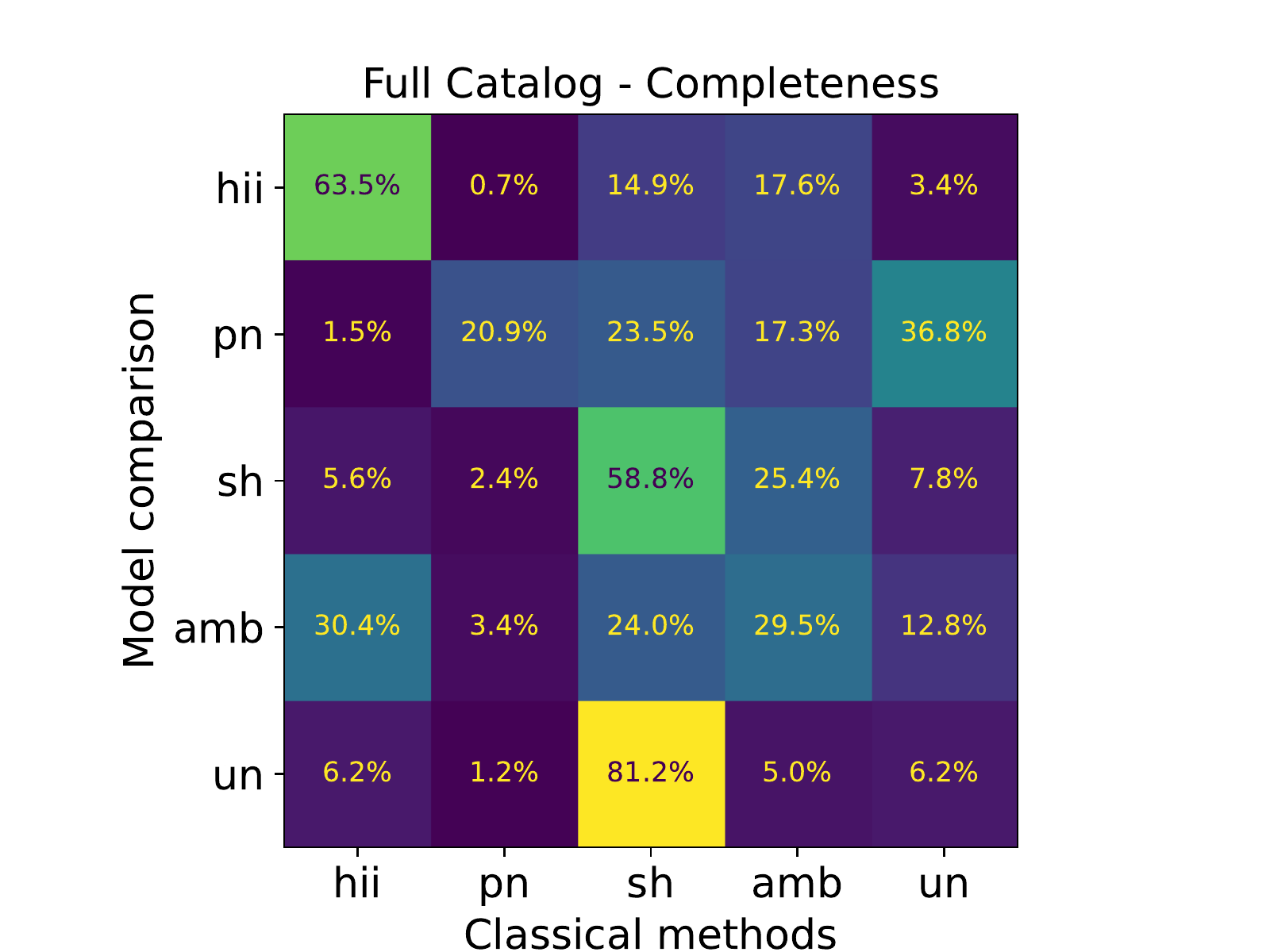} \quad
\includegraphics[width=0.45\textwidth]{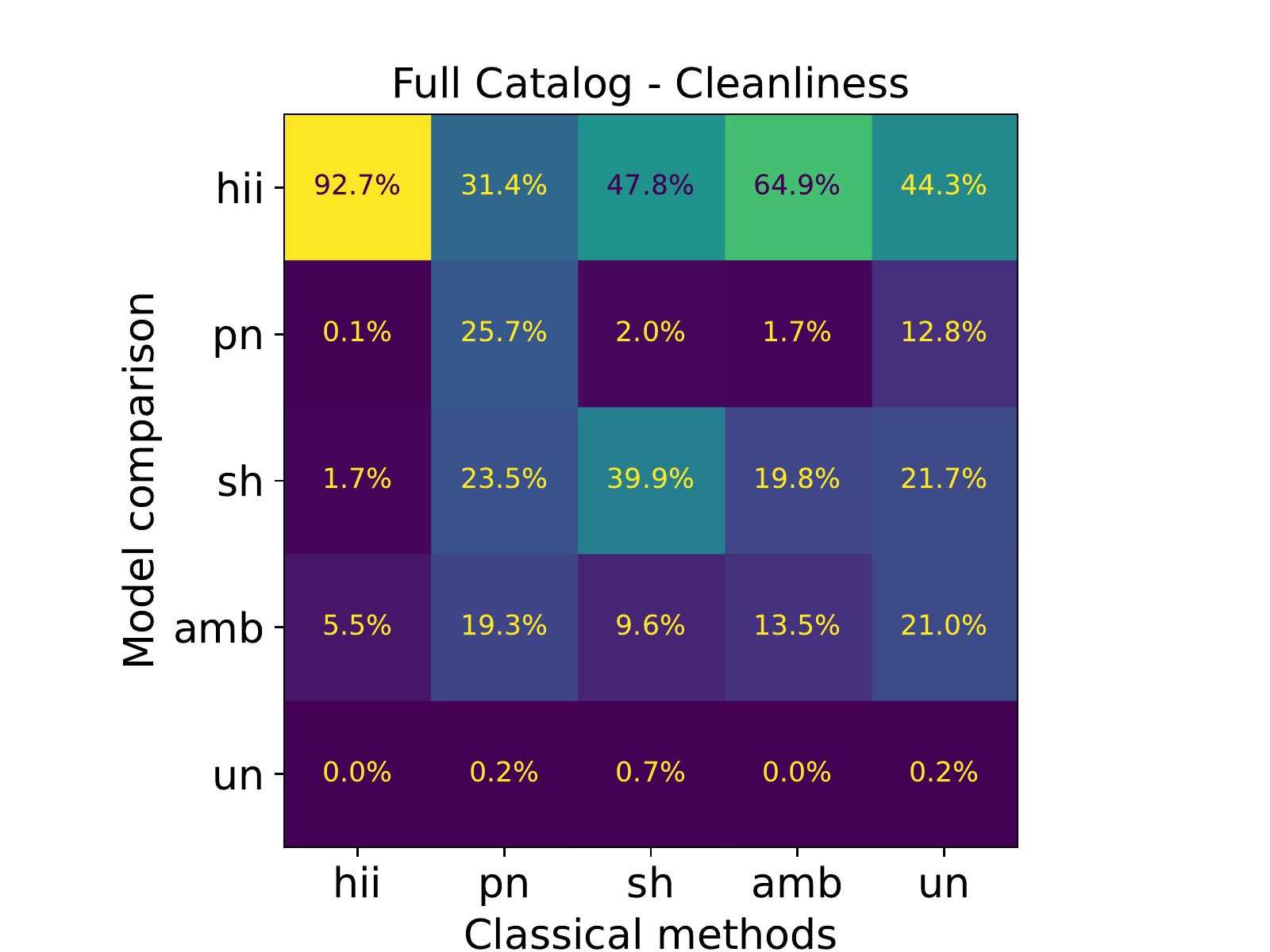}
\caption{Confusion matrices comparing the results of the model-comparison-based classification and traditional nebulae classification techniques for the full sample. The five classes included in the matrices are described in Sec.\,\ref{sec:reliability} and Sec.\,\ref{sec:classical} for the new and traditional classifier respectively. The left matrix reports the completeness of the traditionally classified sample in the diagonal, while the misclassification of each different class outside the diagonal. The right matrix, instead, shows how clean the traditionally classified sample is on the diagonal, and how much is contaminated by objects belonging to different classes outside of the diagonal.}
\label{fig:matrix_all}
\end{figure*}

\begin{figure*}
\centering
\includegraphics[width=0.45\textwidth]{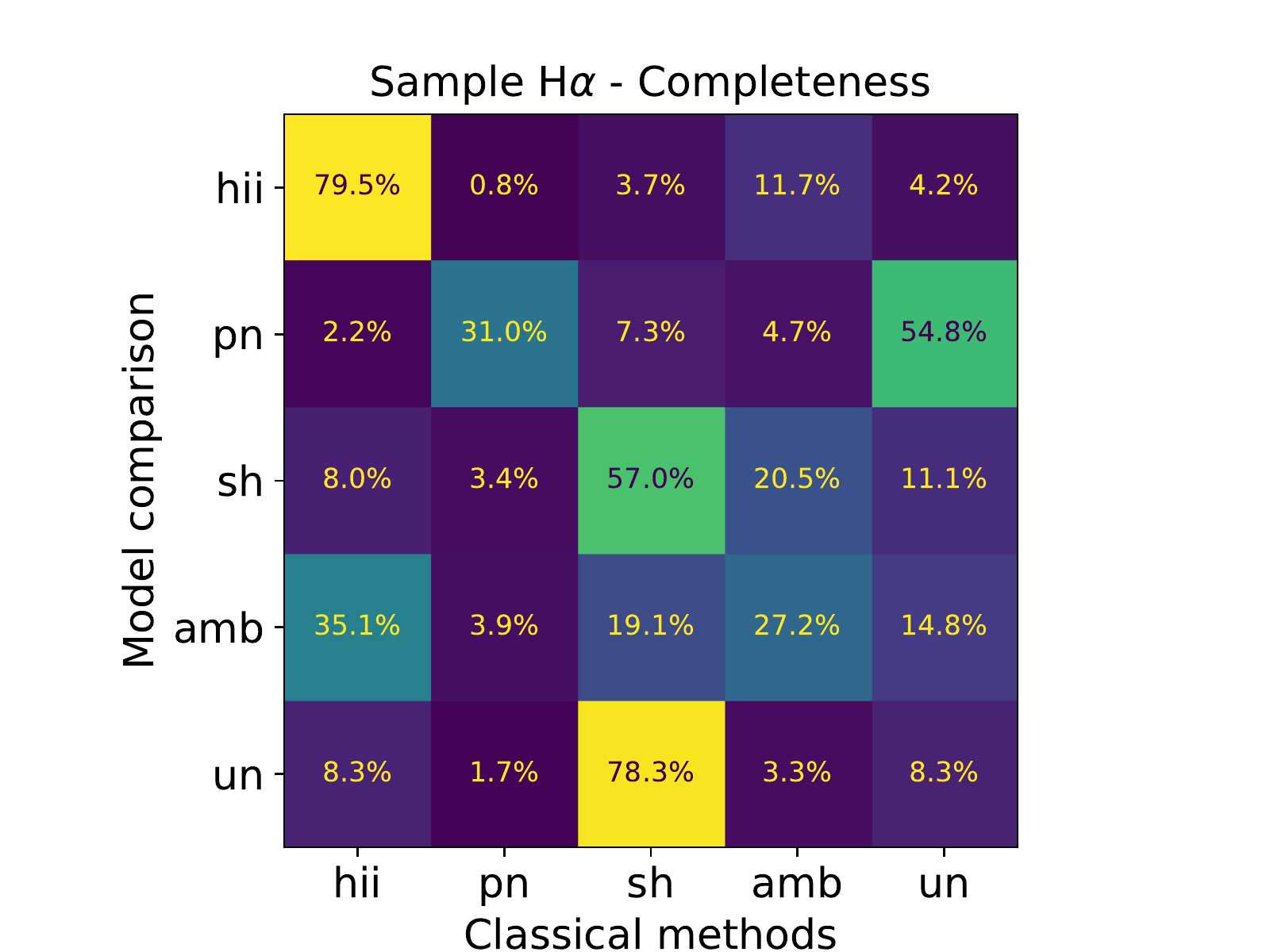} \quad
\includegraphics[width=0.45\textwidth]{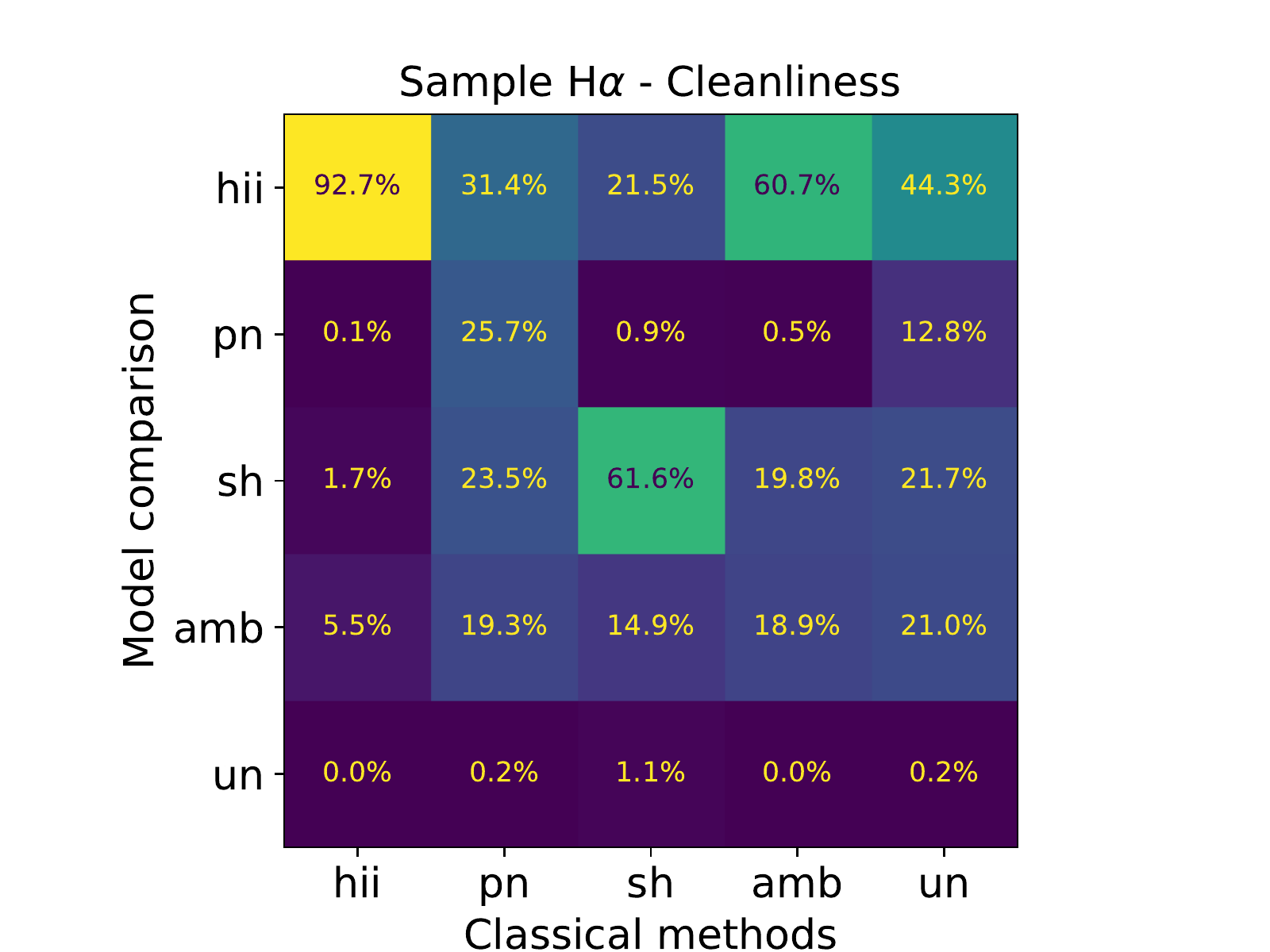}
\caption{Confusion matrices comparing the results of the new, model comparison-based classification and traditional nebulae classification techniques for the \Ha\ sample. Caption as in Fig.~\ref{fig:matrix_all}.}
\label{fig:matrix_ha}
\end{figure*}

\begin{figure*}
\centering
\includegraphics[width=0.45\textwidth]{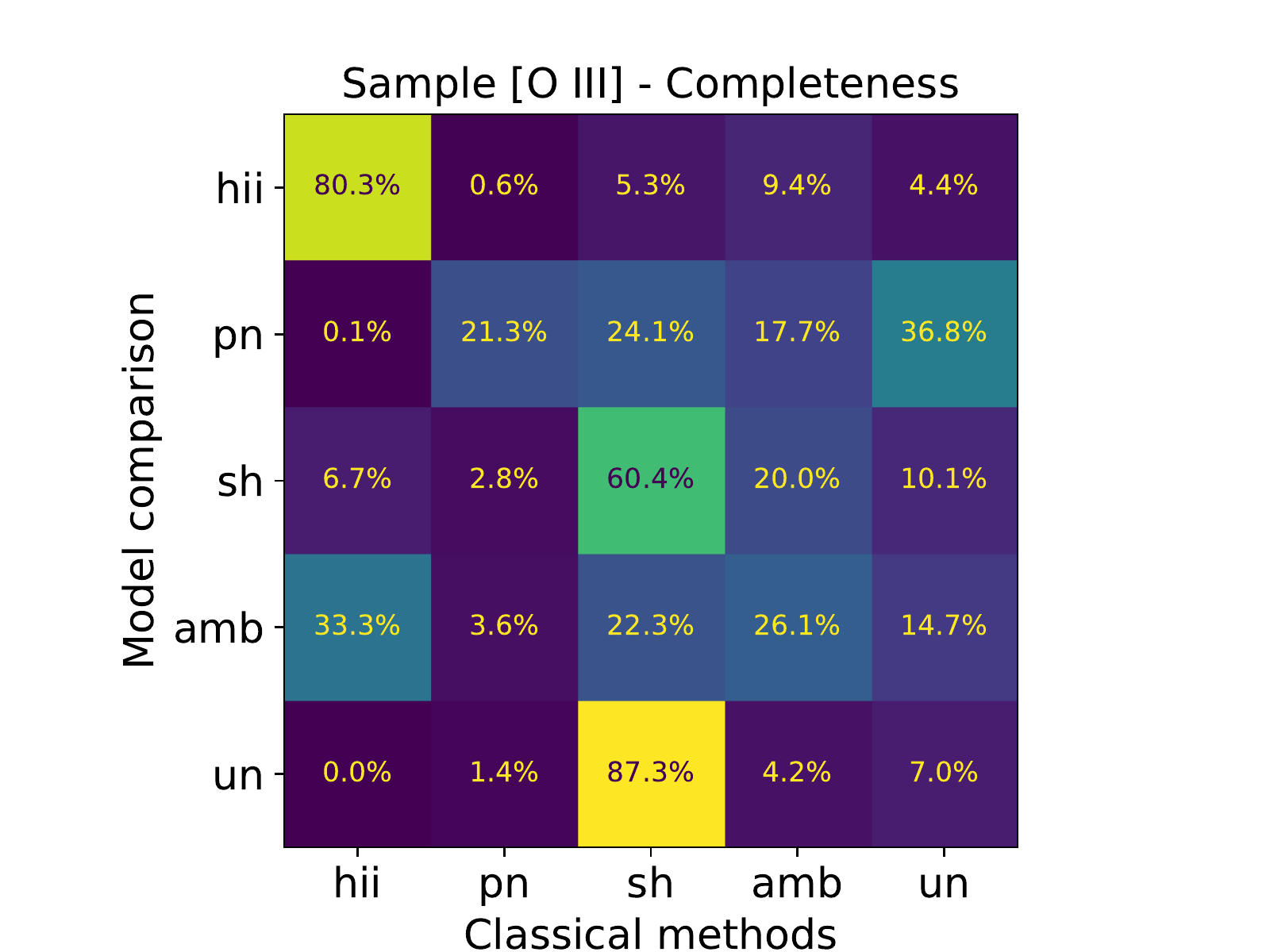} \quad
\includegraphics[width=0.45\textwidth]{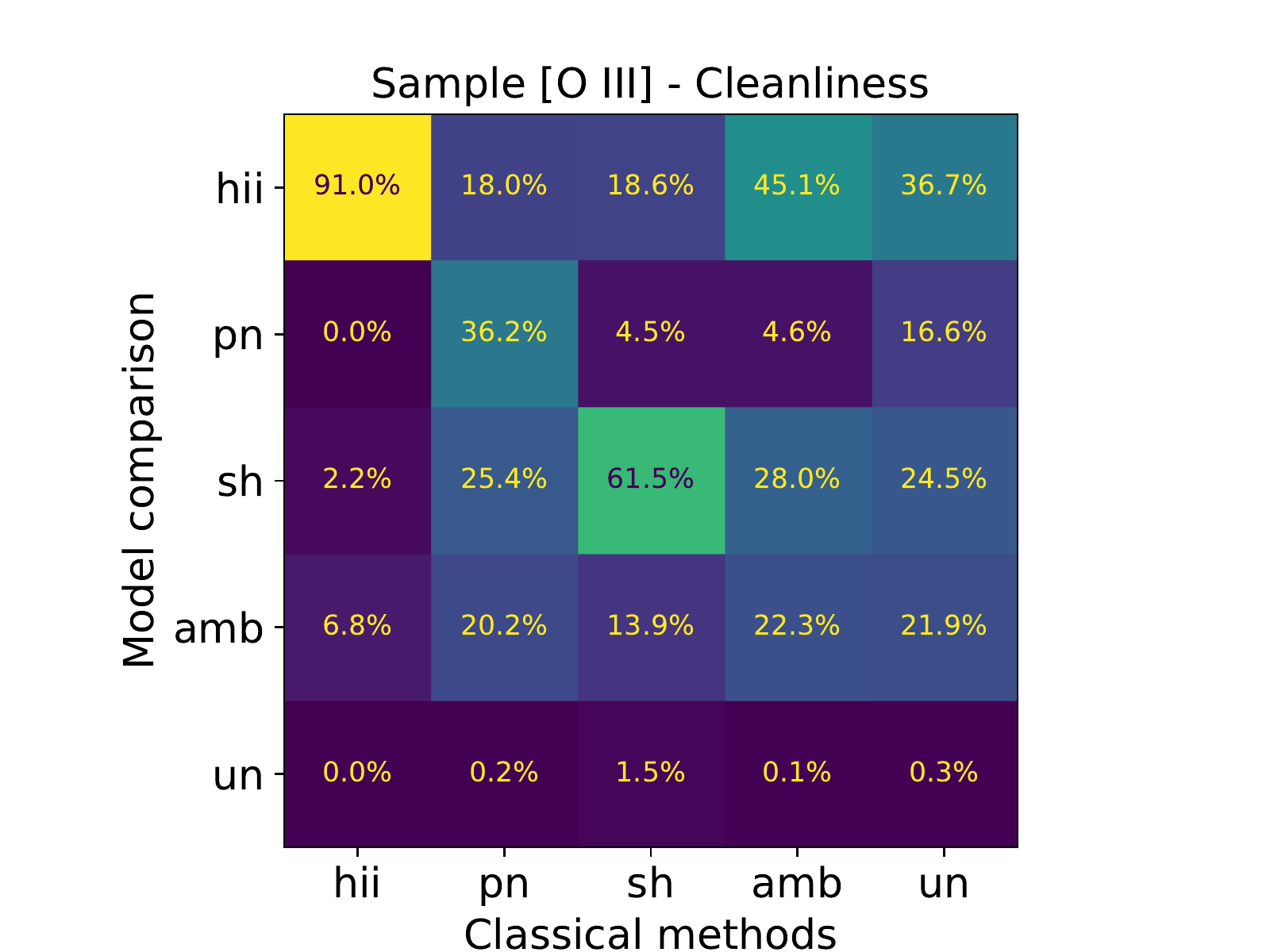}
\caption{Confusion matrices comparing the results of the new, model comparison-based classification and traditional nebulae classification techniques for the \Ha\ sample. Caption as in Fig.~\ref{fig:matrix_all}.}
\label{fig:matrix_o3}
\end{figure*}

\begin{figure*}
\centering
\includegraphics[width=0.45\textwidth]{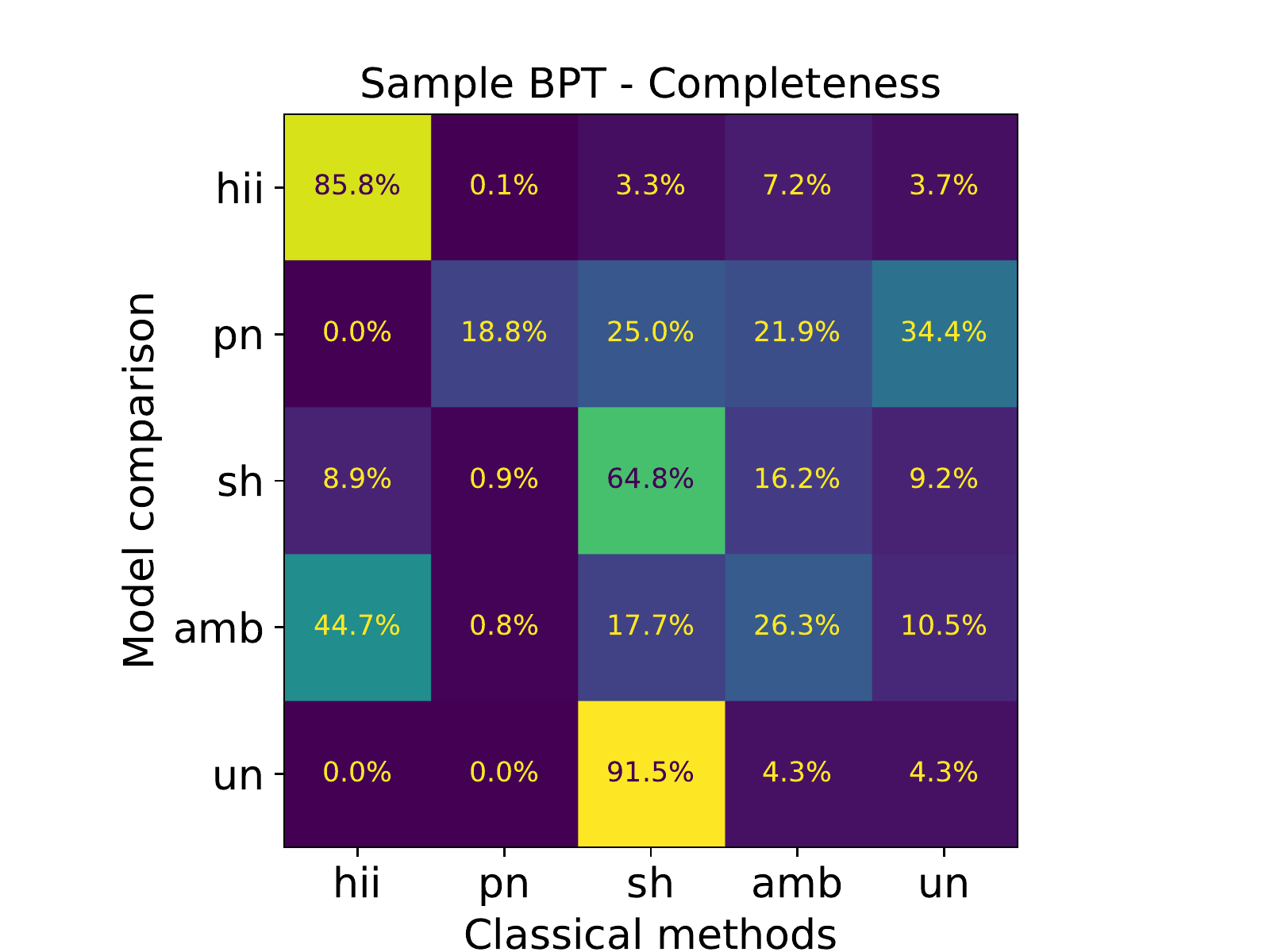} \quad
\includegraphics[width=0.45\textwidth]{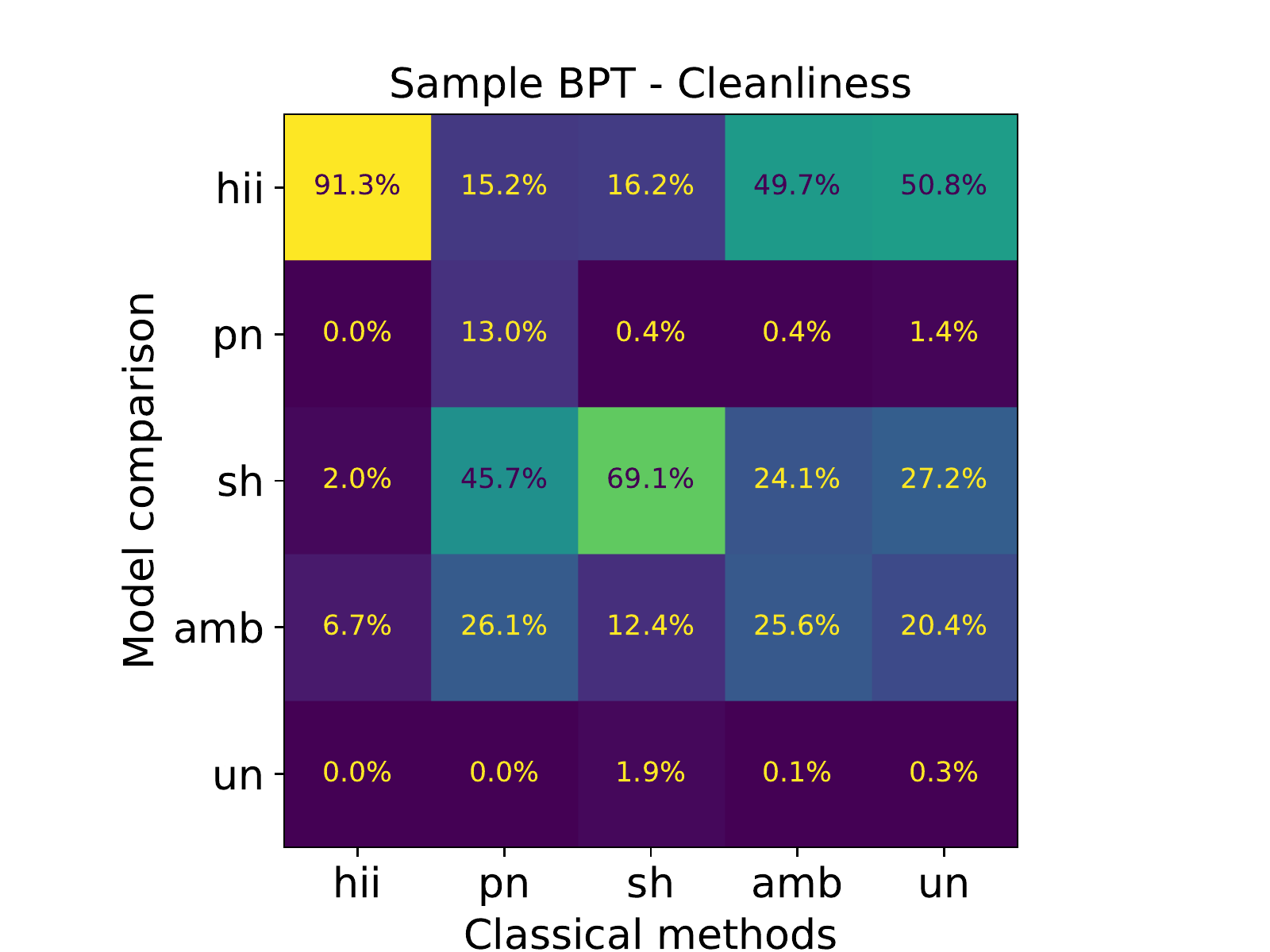}
\caption{Confusion matrices comparing the results of the new, model comparison-based classification and traditional nebulae classification techniques for the \Ha\ sample. Caption as in Fig.~\ref{fig:matrix_all}.}
\label{fig:matrix_bpt}
\end{figure*}

In Figs.~\ref{fig:matrix_all}--\ref{fig:matrix_bpt} we show a direct comparison between the results of our classification algorithm and the traditional criteria for each different sub-sample.
In each figure, the left matrix is normalised with respect to the \izi\ model-comparison classification, so it represents the fraction of the nebulae classified in a particular class by the model-comparison-based algorithm that falls on a different class in the traditional classification scheme.
Basically, these matrices show how the traditional classification criteria misclassify the nebulae, assuming that the \izi\ classification is correct. 
The right matrix, on the other hand, shows the opposite.
This can be interpreted as a way of evaluating how contaminated the samples of traditionally classified nebulae are, assuming, once again, that the \izi\ model-comparison-based classification is correct.
Looking at the left matrices, it is possible to see that the traditional criteria agree with the \izi\ classification for most \hii\ regions and shocks. 
In all the different samples, at least 60\% of \hii\ regions and $55\%$ of shock regions keep their classification using the traditional criteria.
In the BPT sample, the percentage increases to 86\% for \hii\ regions and 65\% for shocks.

The same is not true for PNe.
Typically, around 20--30\% of the PNe classified by the new algorithm keep their classification using the traditional criteria.
This test clearly shows that the two approaches are classifying different objects as PNe. 
The reason why this happens is still not clear.
The traditional criterion mostly relies on precise measurement of the \oiii\ and \Ha\ fluxes and their ratio.
It is possible to see from the literature on this topic that most of the effort when classifying PNe is dedicated to precisely measuring the line fluxes and their biases.
On the other hand, our new algorithm relies on comparing line ratios and other properties with models, which should allow us to perform the classification in less ideal conditions.

The unclassified and ambiguous classes produced by the two methods are not directly comparable, since they are defined differently.
However, it is interesting to see that the vast majority of the nebulae that our algorithm cannot classify are identified as shocks by the traditional classifier ($\sim80$--$90\%$).

The matrices on the right side of Figs.~\ref{fig:matrix_all}--\ref{fig:matrix_bpt} highlight other interesting effects.
While the sample of traditionally classified \hii\ regions seems to be quite clean, independently of which sub-sample we are considering, the same cannot be said of all the other classes.
Both PNe and shocks show significant contamination, with up to $50\%$ them being classified as \hii\ region by the model-comparison-based algorithm.
The situation improves significantly by moving from the full sample to the more restrictive sub-samples, especially for shocks.
If we consider only the BPT sub-sample, almost 70\% of the nebulae traditionally classified as shocks are considered shocks also by our algorithm.
In PNe the results are not as good, but we can still see some improvement.
Significantly, the best results for PNe are found using the OIII sample and not the BPT one, as expected since the number of PNe in the BPT sample is negligible.
Finally, the matrices also show that more than half of the ambiguous and unclassified regions found using traditional criteria are \hii\ regions according to the model-comparison-based algorithm.
\section{Discussion}
\label{sec:disc}

\subsection{Extinction estimates}
\label{sec:EBV}

\begin{figure*}
\centering
\includegraphics[width=0.9\textwidth]{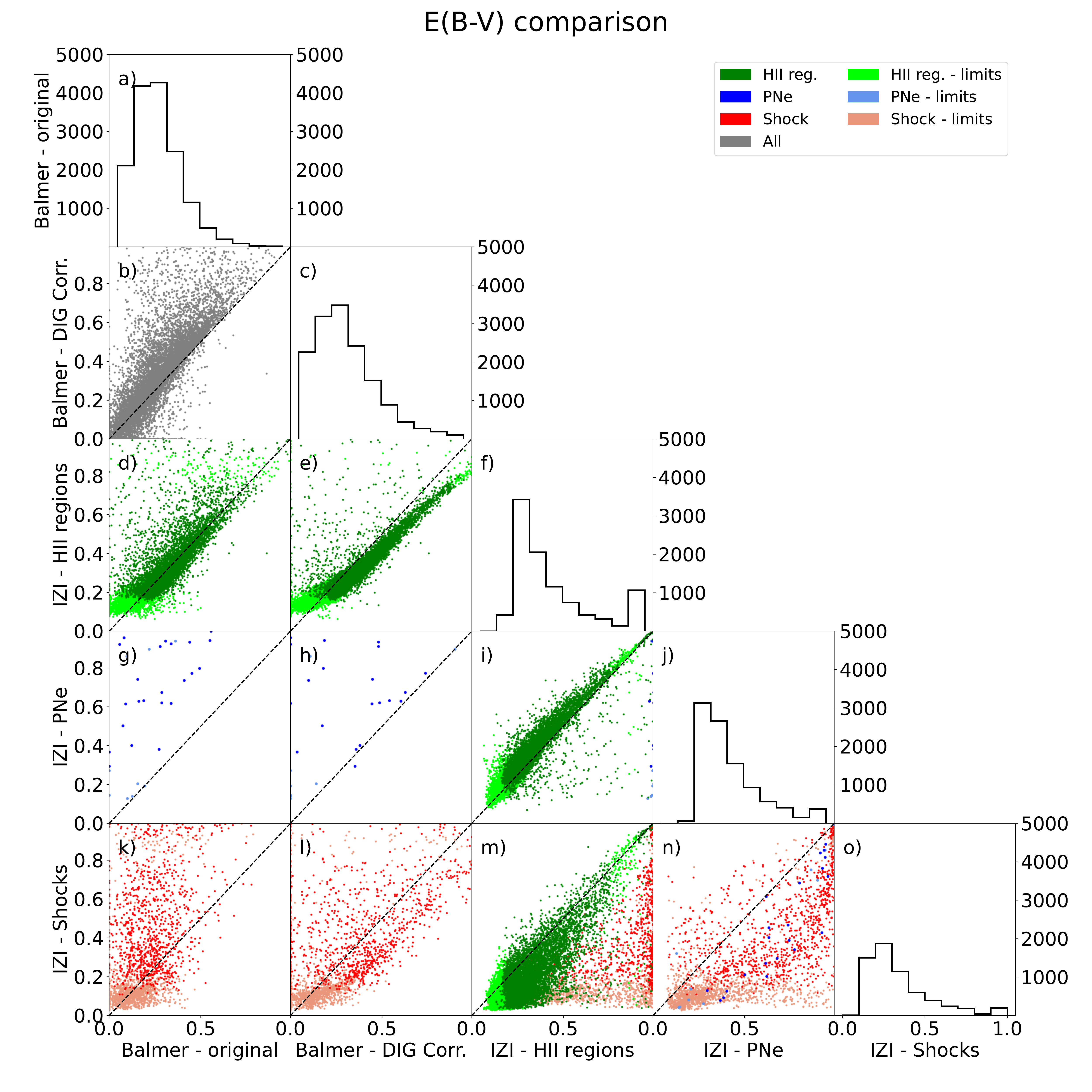}
\caption{Corner plot comparing the reddening value (\ebv) measured directly from the data (before and after the correction for DIG contribution) with the values obtained by IZI using the three different grids of models used in the classification algorithm (\hii\ regions, PNe and Shocks). Only the regions belonging to the BPT sample are considered in this plot (see Sec.\,\ref{sec:reliability}). In each plot, the coloured points represent the nebulae we could classify with our algorithm, following the usual convention. Only the points belonging to the grids considered by the specific panel are coloured. The plots on the matrix's diagonal show regions' distribution as a function of \ebv\ for each method.}
\label{fig:corner_ebv}
\end{figure*}

\begin{figure}
\centering
\includegraphics[width=0.45\textwidth]{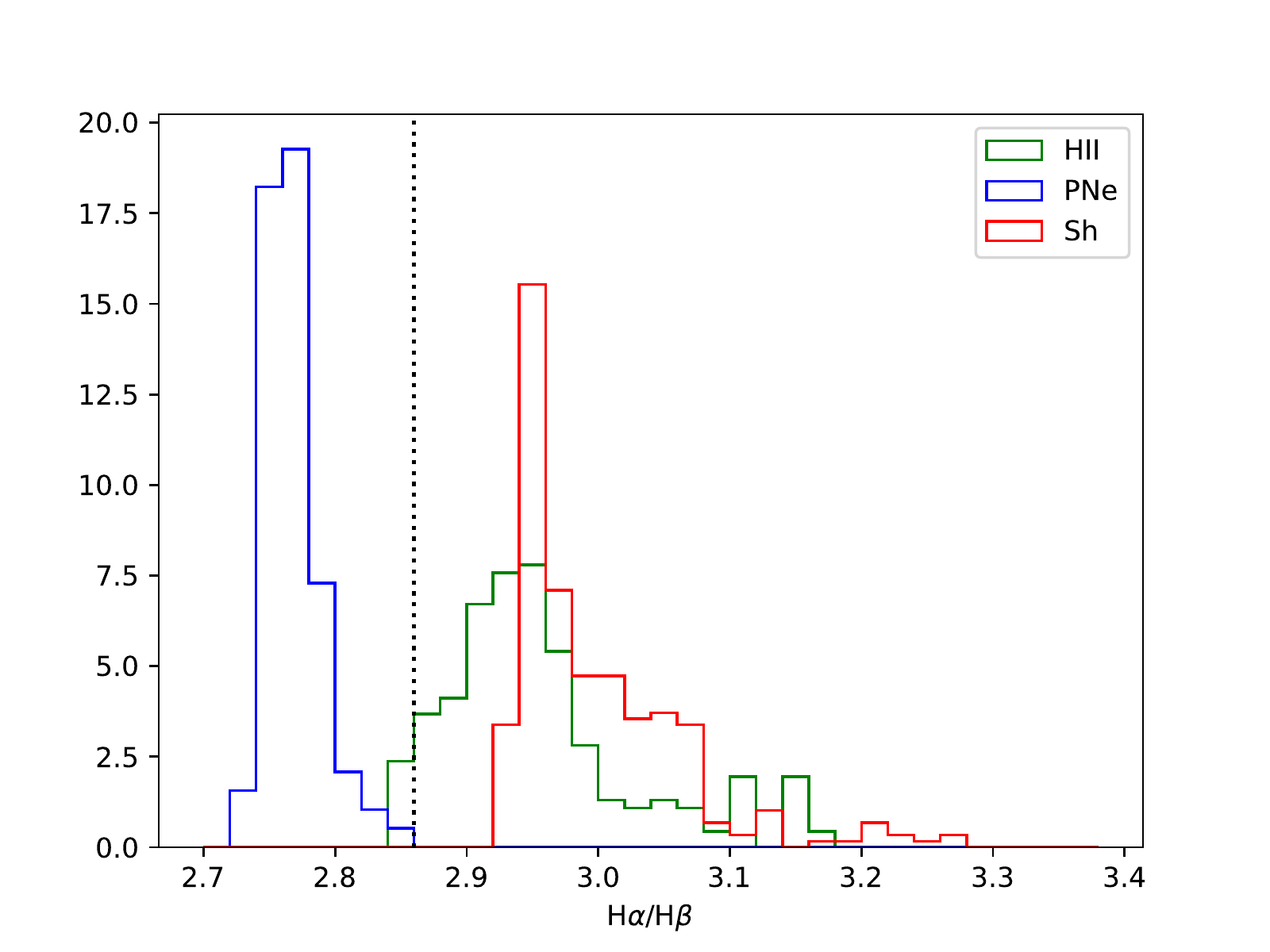}
\caption{Histogram showing the intrinsic \Ha$/$\Hb\ ratio for the different grids of models used in the classification procedure. Histograms are coloured following the usual convention for \hii\ regions, PNe and shocks. The vertical dashed line represents the theoretical value of the \Ha$/$\Hb\ ratio for Case B conditions and a temperature of 10$^4\,\si{K}$, typically assumed when recovering the extinction from the observed \Ha$/$\Hb\ ratio in \hii\ regions.}
\label{fig:ebv_hist}
\end{figure}

\begin{figure}
\centering
\includegraphics[width=0.45\textwidth]{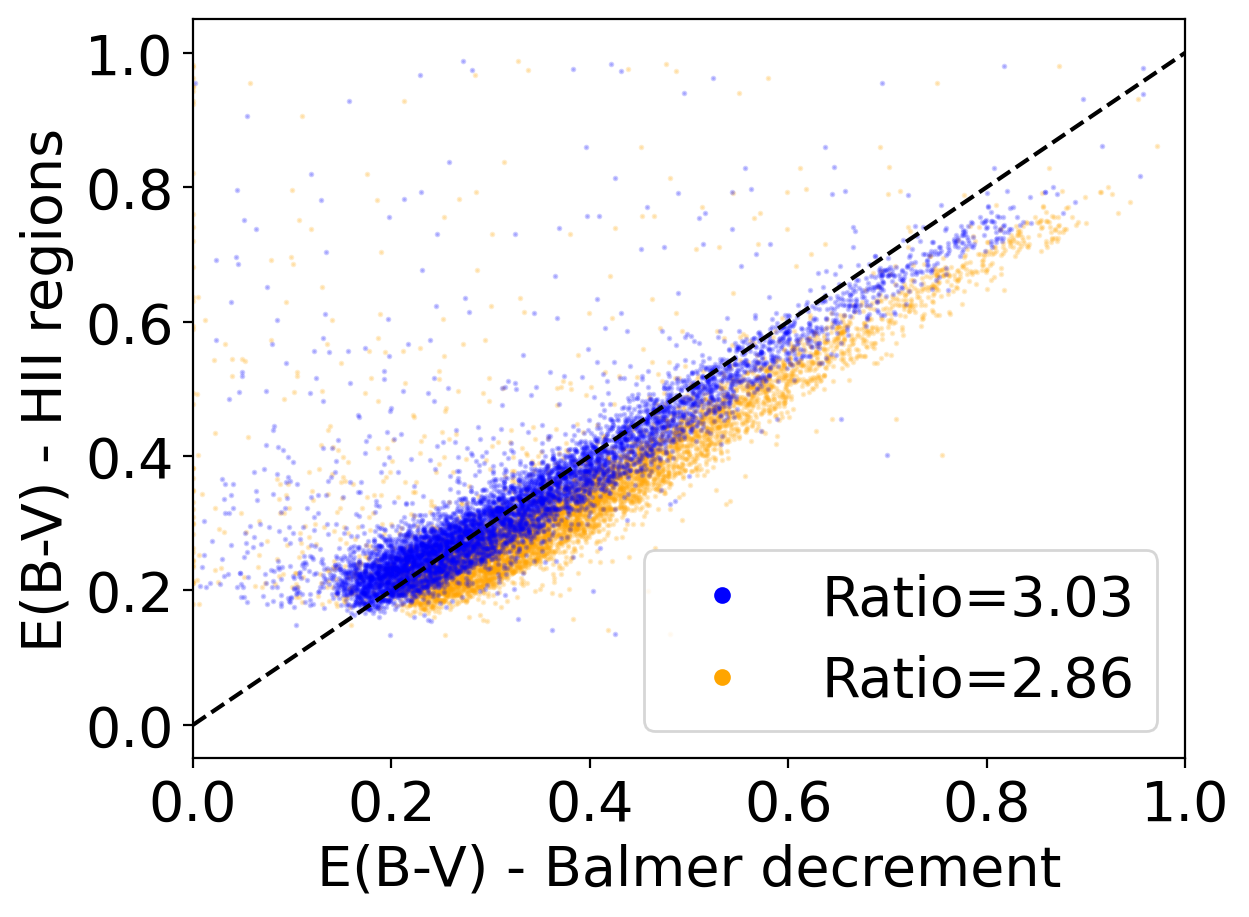}
\caption{Plot comparing the relation between the \ebv\ measured via the DIG corrected Balmer decrement and by \izi\ when using the traditional \Ha$/$\Hb\ $=2.86$ (orange points) and the new recommended value \Ha$/$\Hb\ $=3.03$ (blue).}
\label{fig:ebv_comp}
\end{figure}

Having a reliable estimate of the reddening in each region is essential when measuring quantities involving lines with wide wavelength separations (e.g. the ionisation parameter from the [S~\textsc{iii}]$\lambda9068/$\sii\ or \oiii$/$[O~\textsc{ii}]$\lambda3727$ ratios) or intrinsic extinction-corrected luminosities (e.g. star formation rate from \Ha).
The reddening is typically measured by comparing the observed \Ha$/$\Hb\ ratio with its theoretical value for a nebula with a temperature of $\SI{10000}{K}$, low-density limit and Case B conditions \citep[\Ha$/$\Hb\ $=2.86$,][]{Osterbrock06}.
We use this technique, as implemented in the {\tt pyneb} python package \citep{Luridiana15}, to estimate the colour excess (\ebv) in all nebulae where both \Ha\ and \Hb\ are detected at better than $3\sigma$, and we compute it both before and after the DIG correction.
The \ebv\ is a reddening tracer since it expresses how red an object is with respect to the expectation, and it is one of the free parameters considered by \izi\ when comparing the nebulae properties to the models.
\izi\ does not assume a fixed \Ha$/$\Hb\ to compute the \ebv\ but instead samples the parameter space to identify the reddening producing the best agreement between the observations and the models.
This is done for each grid of models, so, on top of the values obtained directly from the Balmer decrement ratio, we have three further measurements of the \ebv\ for each nebula, each one associated with the expectation value from the marginalised posterior probability distribution function in \ebv\ as determined by \izi\ for each class (\hii\ regions, PNe, and shocks).
For values of \ebv\ which are too close to the edge of the prior space (and in particular close to 0), \izi\ returns a limit.

In this section, we compare measurements of \ebv\ obtained with the different methods to understand differences and similarities.
We can make this comparison because all the different measurements are based on the same reddening law, that is the \citet{Cardelli89} law as revised by \citet{ODonnel94} with $\rm R(V) = 3.1$.
Figure~\ref{fig:corner_ebv} shows a corner plot comparing all the \ebv\ we measured.
We only show nebulae belonging to the BPT sub-sample because we are confident that their \Ha\ and \Hb\ are well detected, and that the \izi\ results should be the most reliable, since it can use the full set of lines when comparing the observations to the models.
In the figure, the regions for which \izi\ return an upper limit in \ebv, are shown in a lighter colour and are omitted from the following discussion.

We first analyse how the DIG correction influences the reddening correction.
Figure~\ref{fig:corner_ebv} (panel b) shows the comparison between the \ebv\ measured using the original and the DIG corrected Balmer decrement.
The two quantities show a well-defined relation, but there is a consistent scatter. 
In general, the \ebv\ measured after the DIG correction is higher than what was measured before.
This is consistent with a screen model for the dust distribution \citep[e.g.][]{Tomicic17}, which assumes that the DIG is distributed in a disk with a larger scale height than the dust and the other nebulae, mostly located in a thinner disk.
As a consequence, the DIG is less attenuated and results in the lower \ebv\ generally measured before the DIG correction.

Panels d, e, g, h, k, l, of Fig.~\ref{fig:corner_ebv} compare the \izi\ \ebv\ to the Balmer decrement \ebv\ measurements.
Here it is possible to see that the situation is more complex.
Panels d and e compare the \izi\ and Balmer decrement \ebv\ for \hii\ regions, which are the vast majority of the sample.
The two plots show two different behaviours.
When comparing \izi\ \ebv\ to the non-DIG corrected Balmer decrement \ebv\ (panel d), we obtain a relation that follows relatively well the 1-to-1 relation, albeit with a large scatter.
Moving to the Balmer decrement \ebv\ obtained using the DIG-corrected flux, we have the opposite situation.
The scatter is much less, but we are not so close to the 1-to-1 relation (it is important to remember that we run \izi\ using the DIG-corrected line fluxes).
In particular, the Balmer ratio systematically overestimates the extinction with respect to \izi, and it seems like the difference between the two relations increases when moving towards higher \ebv.
The most likely reason for this behaviour is that when measuring the extinction using the Balmer decrement, we assume a fixed theoretical \Ha$/\Hb$ ratio for all nebulae (2.86).
However, the underlying gas conditions assumed when using this value are not always satisfied.
In particular, the \Ha$/\Hb$ depends on the gas temperature, which can change significantly from cloud to cloud.
The lower the temperature, the higher the expected \Ha$/\Hb$ ratio \citep{Osterbrock06}.
For example, assuming a temperature of $\SI{5000}{K}$\footnote{All the computations involving the \Ha$/\Hb$ ratio and the gas temperature are performed via {\tt pyneb}} the theoretical \Ha$/$\Hb\ ratio is $3.04$ and results in a decrease of the \ebv\ of $\sim \SI{0.06}{mag}$.
Consequently, using a theoretical \Ha$/\Hb$ ratio that does not correspond to the actual conditions of the clouds results in a wrong extinction estimate.

The \hii\ regions models we use for the classification are not parameterised in terms of electron temperature. 
Their temperature profiles are set by the thermal equilibrium given by the cooling and heating processes incorporated in the CLOUDY calculations, which depend mainly on the chemical abundance, the electron density and the shape of the ionising spectrum. 
Therefore models with different metallicities and ionisation parameters will have different temperature profiles and different intrinsic \Ha$/\Hb$ ratios, as shown in Fig.~\ref{fig:ebv_hist}. 
Only a small fraction of the \hii\ region models have a \Ha$/\Hb$ ratio close to 2.86. 
The vast majority have higher values, indicating, on average, lower electron temperatures.

Selecting the correct theoretical value of the \Ha$/\Hb$ ratio is, however, not trivial.
In high \sn\ spectra, where it is possible to detect the auroral lines needed to measure the gas temperature, it is possible to apply an iterative process, to constrain the temperature (and the \Ha$/\Hb$ ratio) and the \ebv\ simultaneously.
However, such high-quality spectra are not always available and assuming a single theoretical value for all nebulae is often necessary.
By minimising the offset between the relation we observe in Fig.~\ref{fig:corner_ebv} (panel e) and the 1-to-1 relation, we can identify a new theoretical value that should result, on average, in better extinction estimates.
This is just a zero order correction, but it should still produce significant improvements.
A detailed analysis to understand the origin of the higher order trends and how they should be considered is outside the scope of this paper, and we leave it to a future work.
The new ratio resulting from the minimisation is \Ha$/\Hb = 3.03$ (Fig.~\ref{fig:ebv_comp}), which corresponds to a temperature of $\sim \SI{5200}{K}$, slightly lower than what is typically observed in nearby extragalactic \hii\ regions \citep[e.g.][]{Ho19, Berg20, Kreckel22}.
We recommend using this value instead of the classic \Ha$/\Hb = 2.86$ when applying the Balmer decrement to \hii\ regions in massive star-forming galaxies having nearly solar metallicity.
We also notice that this new approach works only if DIG-corrected fluxes are used to calculate the extinction.
If uncorrected fluxes are used, the dilution provided by the DIG contribution counteracts the effect of using the wrong theoretical ratio (Fig.~\ref{fig:corner_ebv}, panel d).
It is interesting how the observed agreement, and supposedly good performance of non-DIG corrected Balmer decrement extinction correction, seems to arise from a coincidence, where two different biases (underestimation of the extinction due to DIG contamination and overestimation of the extinction due to adopting the wrong intrinsic Balmer ratio) have similar magnitudes and act in opposite directions.

Figure~\ref{fig:corner_ebv} (panels k and l), indicate a similar, yet different, scenario for shock ionised regions.
Panel k shows that the \ebv\ measured by \izi\ using the shock models and that measured via the Balmer decrement using the original fluxes are not in good agreement.
\izi\ almost always finds much higher extinctions with respect to the Balmer decrement.
There seems to be a very steep relation between the two quantities, but the scatter is definitely high, and this relation's origin is unclear.
If we consider the DIG-corrected Balmer decrement, the picture changes.
There is a large scatter, and the smaller statistics prevent us from performing the same test we did for \hii\ regions, but it is possible to see a similar scenario.
The main difference is that the offset from the 1-to-1 relation is larger.
This is expected.
Shocks are a much more complex situation than photo-ionisation regarding the mechanism and the parameters that can influence the observed spectrum of a nebula, and Fig.~\ref{fig:ebv_hist} shows that the \Ha$/$\Hb\ of the models we are using are much higher than $2.86$.


Fig.~\ref{fig:corner_ebv} (panels g and h) shows the same plots again but for PNe.
Although, the number of PNe in this sub-sample is only 32, so we do not have sufficient statistics to perform any analysis.
We notice, however, that \izi\ recovers mostly high extinction for these objects, higher than what is found via the Balmer decrement, with the DIG-corrected Balmer decrement returning values that are more similar to the \izi\ ones. 
Planetary nebulae are known for having higher electron temperature and densities with respect to \hii\ regions, therefore \Ha$/$\Hb $=2.86$ is typically not a good approximation for these objects.
As it is also possible to see from Fig.~\ref{fig:ebv_hist} a lower reference ratio should be used.
Once again, in Galactic or nearby objects where it is possible to obtain high S$/$N spectra, the extinction has typically been measured assuming the appropriate \Ha$/$\Hb\ for the set of conditions found in the specific nebula \citep[e.g.][]{Meatheringham91, Meatheringham91b, Walsh16}.

Finally, Fig.~\ref{fig:corner_ebv} (panels i, m and n) compares the \ebv\ measured by \izi\ when applying two different models to the same nebulae to test how a misclassification would affect the measurement of the extinction.
For \hii\ regions misclassified as PNe, the final \ebv\ would be slightly overestimated.
In the opposite case, a PNe misclassified as an \hii\ region, the \ebv\ would be heavily underestimated.
This is in line with the distribution of the \Ha$/\Hb$ ratios of the models shown in Fig.~\ref{fig:ebv_hist}.
When comparing shocks and \hii\ regions (panel m), the \ebv\ of the shock ionised regions would be overestimated when misclassifying them as \hii\ regions
For most of the \hii\ regions, on the other hand, the impact of misclassification is much more limited, and would affect principally the nebulae with low extinction.
Concluding, panel n compares shocks and PNe.
While the statistics for PNe is still small, we can see that for most shocks, a misclassification would result in a larger measured \ebv.

\subsection{Physical properties of the nebulae}
\label{sec:physical_prop}

\begin{figure*}
\centering
\includegraphics[width=0.9\textwidth]{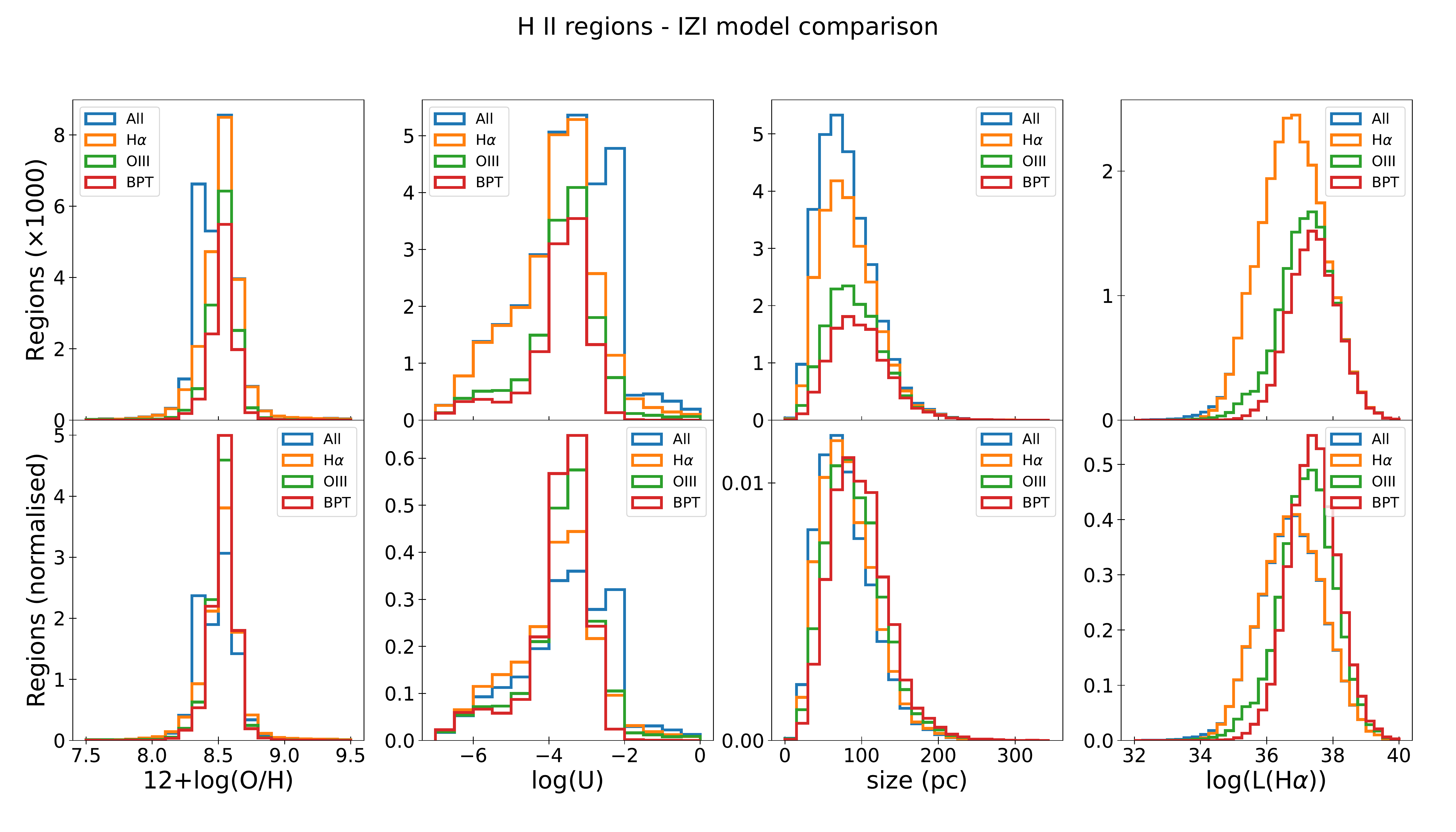}\\
\includegraphics[width=0.9\textwidth]{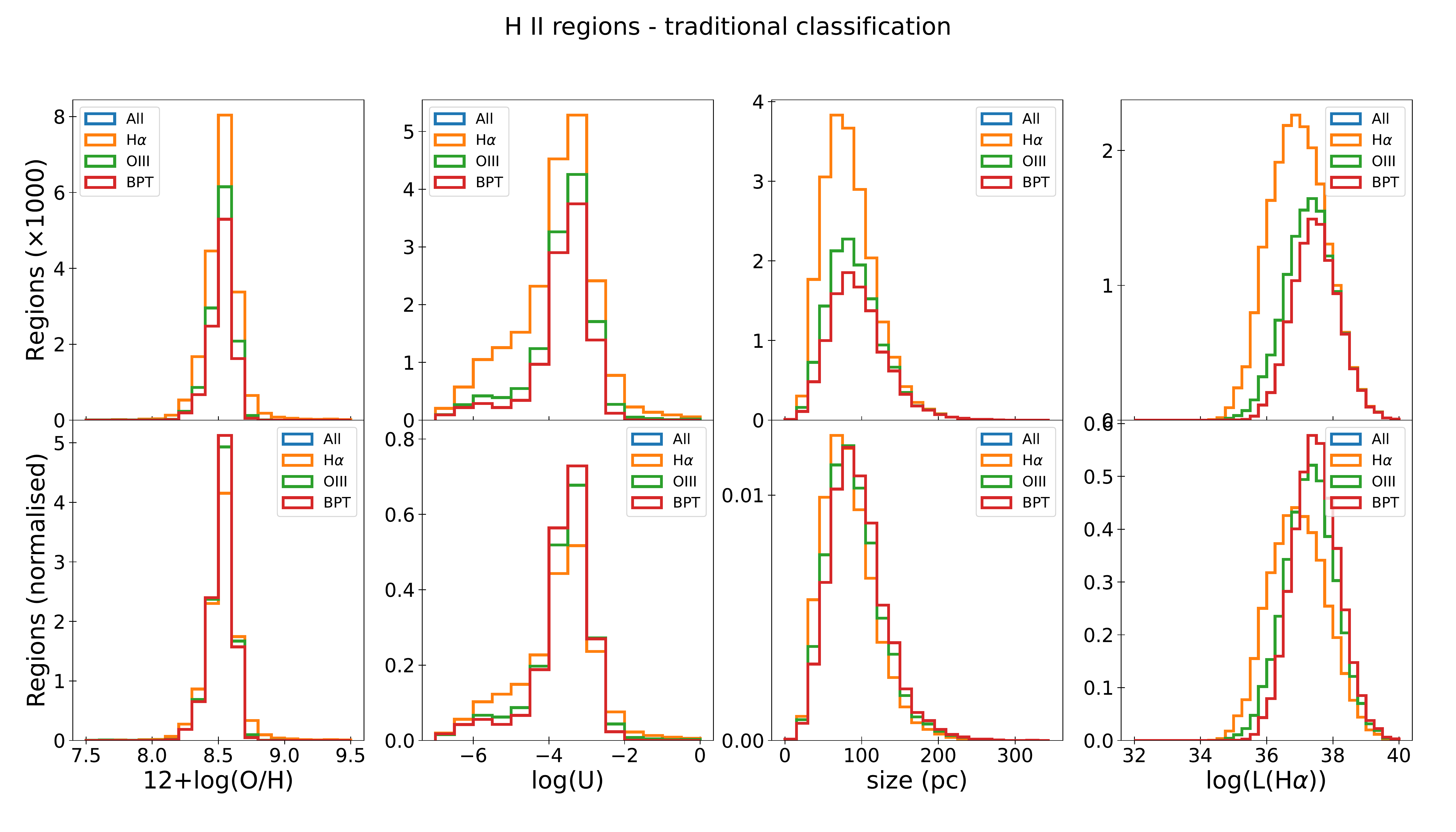}\\
\caption{Physical properties of \hii\ regions. \textbf{Top panel:} Histogram showing how the distribution of the main measured properties of \hii\ regions changes as a function of the considered sample. From the left to the right the following quantities are shown: S-calibration metallicities \citep{Pilyugin16}, ionisation parameter \citep{Diaz91}, circularised radius, \Ha\ luminosity. The top row shows the absolute distribution for each quantity, while the bottom row shows the normalised distribution. \textbf{Bottom panel:} same histograms as shown in the top panel, but for the \hii\ regions classified using the traditional classification criteria. Since there is no change in the number of \hii\ regions in the full catalogue and the \Ha\ sample, the two histograms overlap, and the blue line representing the full catalogue is not visible.}
\label{fig:hii_hist}
\end{figure*}

\begin{figure}
\centering
\includegraphics[width=0.45\textwidth]{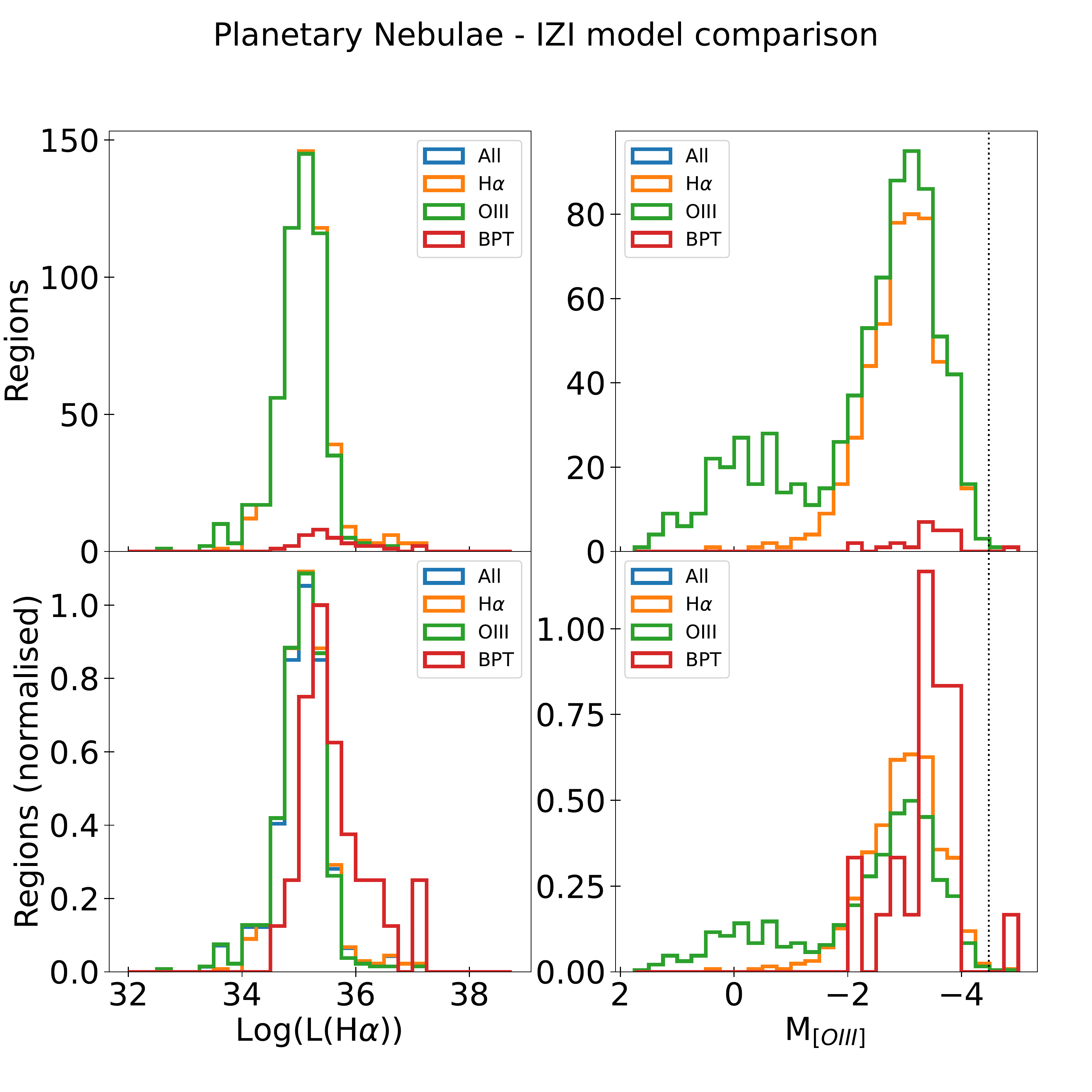}\\
\includegraphics[width=0.45\textwidth]{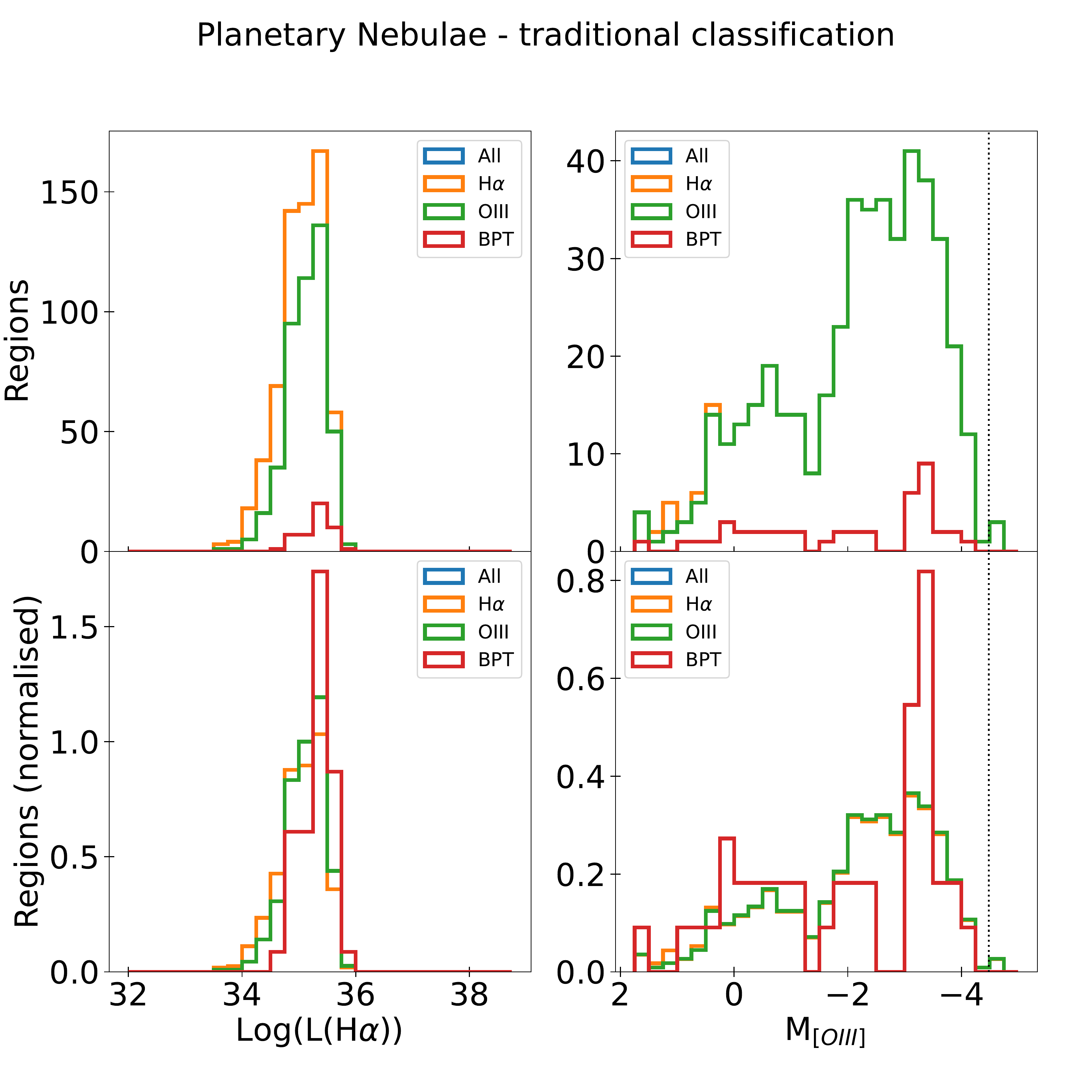}
\caption{Physical properties of PNe. \textbf{Top panel:} Histogram showing how the distribution of the main measured properties of PNe changes as a function of the considered sample. From the left to the right, the following quantities are shown: \Ha\ luminosity, \oiii\ absolute magnitude (adapted from Eq.~\ref{eq:mo3} using distances from Table~\ref{tab:sample}). The top row shows the absolute distribution for each quantity, while the bottom row shows the normalised distribution. \textbf{Bottom panel:} same histograms as shown in the top panel, but for the PNe classified using the traditional classification criteria.}
\label{fig:pn_hist}
\end{figure}

\begin{figure*}
\centering
\includegraphics[width=0.8\textwidth]{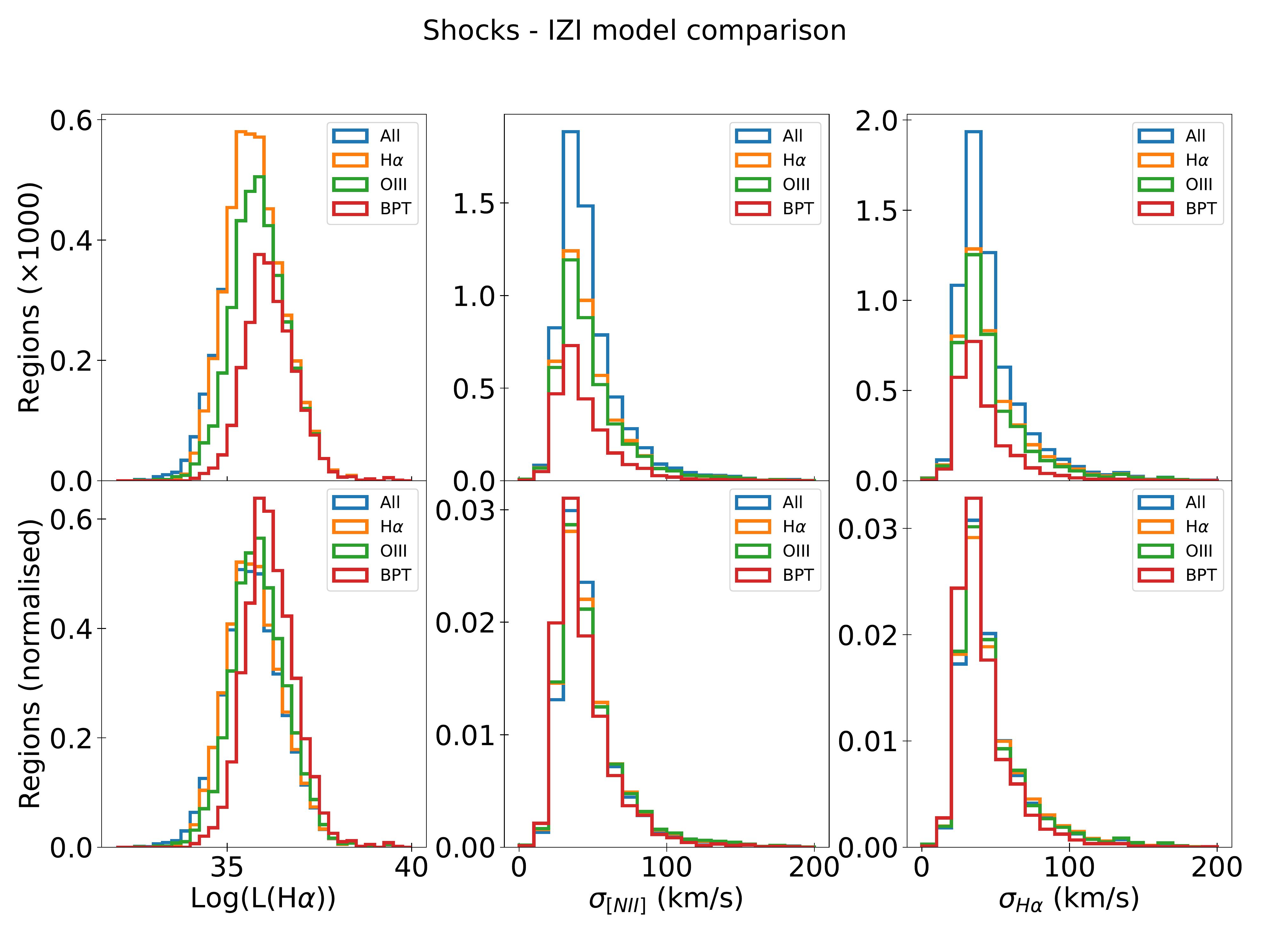}\\
\includegraphics[width=0.8\textwidth]{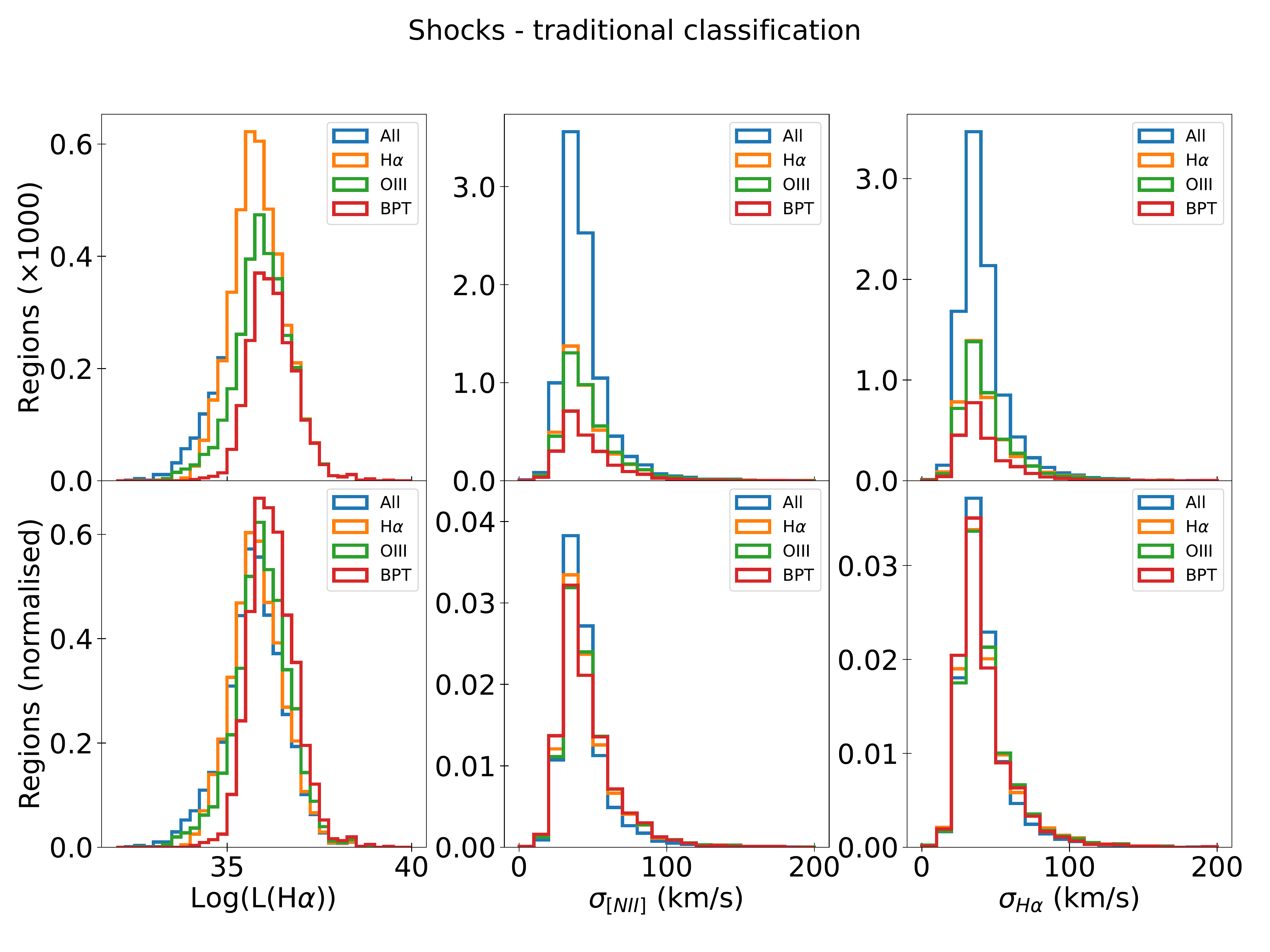}
\caption{Physical properties of shock-ionised regions. \textbf{Top panel:} Histogram showing how the distribution of the main measured properties of SNRs changes as a function of the considered sample. The following quantities are shown from the left to the right: \Ha\ luminosity, \nii\ velocity dispersion and \Ha\ velocity dispersion. The top row shows the absolute distribution for each quantity, while the bottom row shows the normalised distribution. \textbf{Bottom panel:} same histograms as shown in the top panel, but for the SNRs classified using the traditional classification criteria.}
\label{fig:sh_hist}
\end{figure*}

Thanks to the complete set of spectral information included in our nebulae catalogue, we can compute some regions' physical properties and study how their distribution changes depending on the selection criteria and the classification scheme.
This could help us identify biases introduced by different approaches to classifying and selecting sources when investigating ionised nebulae.
Some additional information could be recovered by comparing the values directly measured from the data with those obtained by \izi\ as we did for the extinction.
However, we decided not to perform this comparison since the set of models we are using for the classification is limited, and it has not been chosen to obtain precise and reliable estimates of the nebulae' physical parameters.
A similar analysis based on the advanced photo-ionisation and shock models will be the subject of a follow-up paper.

For \hii\ regions, we computed metallicities, following the S-calibrations by \citet{Pilyugin16}, the ionisation parameter from \citet[][$\log(u)$\footnote{This calibration requires both the [S~III]$\lambda$9068 and [S~III]$\lambda$9532, but only the former falls inside MUSE wavelength range. However, these two emission lines belong to the same doublet, and the following relation ties their flux: $I(9532)=2.47\times I(9068)$ \citep{Osterbrock06}.}]{Diaz91}, size (in pc) and \Ha\ luminosity.
For PNe, we show the \Ha\ luminosity and the \oiii\ absolute magnitude, the latter a critical quantity in estimating the distance of nearby galaxies via the PNe luminosity function (see Sec.~\ref{sec:pnelf} for a discussion on this specific quantity).
Finally, for the shock-ionised regions, we show their \Ha\ luminosity and the \Ha\ and \nii\ velocity dispersion, which represent the kinematics of the two groups of lines defined in Sec.~\ref{sec:post}.

All the quantities follow unimodal distributions, with a relatively narrow spread around a well-defined peak, for the \hii\ regions classified by both approaches (Fig.~\ref{fig:hii_hist}).
The only exception is the ionisation parameter, which shows a very sharp peak at high values.
The peak slowly disappears when moving to more restrictive samples suggesting that the feature is related to regions where the lines needed for the measurements are not well detected (i.e. the \siii\ line).
The figure also shows that when applying more restrictive cuts the distribution of metallicities changes, moving towards slightly higher metallicities.
Also, the shape of the size and \Ha\ luminosity distribution changes significantly.
As expected, more restrictive cuts mostly reject fainter and smaller regions, a sign that the most luminous regions are typically the largest ones.
As expected by the agreement between both classification paradigms when identifying \hii\ regions, the properties of these nebulae are distributed similarly in the two samples (top and bottom panels of Fig.~\ref{fig:hii_hist}).

Since the size of the PNe sample does not change much between the different sub-samples when applying the new, model-comparison-based classification algorithm, the \Ha\ luminosity and $M_{\rm [O~III]}$ distributions for these objects (Fig.~\ref{fig:pn_hist}, top panel) is relatively independent of the considered sub-sample.
The only exception is when we consider the BPT sub-sample.
It contains only a handful of objects, which are typically among the most luminous ones, as expected.
The situation changes significantly when considering the traditionally classified PNe (Fig.~\ref{fig:pn_hist}, bottom panels).
First of all, the \Ha\ distribution ends sharply at the bright end, as a consequence of the criterion used for the classification (Eq.~\ref{eq:pne}).
The PNe classified by the model-comparison-based algorithm seem to reach higher \Ha\ luminosities (up to $\sim10^{37}\,\si{erg.s^{-1}}$) and they do not show such a sharp cut.
Also the distribution of $M_{\rm [O~III]}$ changes significantly between the PNe identified by the two approaches.
The sample classified by the model-comparison-based algorithm shows a relatively symmetric and narrow peak between $-2$ and $-4$ mag.
The traditionally classified sample shows a similar peak, but much wider, and it decreases more slowly when moving towards fainter nebulae.
This histogram also shows the presence of a second peak of objects with relatively low \oiii\ emission.
It is likely dominated by contamination from \hii\ regions and therefore does not affect the knee of the luminosity function and the inferred distance so much.
All these differences highlight the discrepancies between the two classification approaches explored in this work.
Finally, in both cases, there are a few objects which surpass the theoretical limit in absolute magnitude defined by \citet{Ciardullo89} (dotted lines in Fig.~\ref{fig:pn_hist}).
Such findings are not uncommon, even though the nature of these objects is not clear \citep{Jacoby96, Scheuermann22}.

Lastly, we analyse the properties of the shock-ionised regions, which show unimodal distributions in all the considered quantities and samples (Fig.\,\ref{fig:sh_hist}, top and bottom panel).
Both velocity dispersion distributions have consistent shapes in all the considered sub-samples. 
Only the number of objects increases when relaxing the selection criteria.
The peak of the distribution is located at a relatively low velocity dispersion, but both \Ha\ and \nii\ show a prominent high-velocity tail.
The \Ha\ luminosity distribution of the regions does not show anything peculiar, but, as expected, when restricting the criteria we are removing nebulae from the faint end of the distribution.
Similarly to what happens with the \hii\ regions, there are no significant differences between the samples selected by the two classification approaches.

\subsection{Comparison with Santoro et al. (2022)}
\label{sec:compare_fs}

\begin{figure}
\centering
\includegraphics[width=0.45\textwidth]{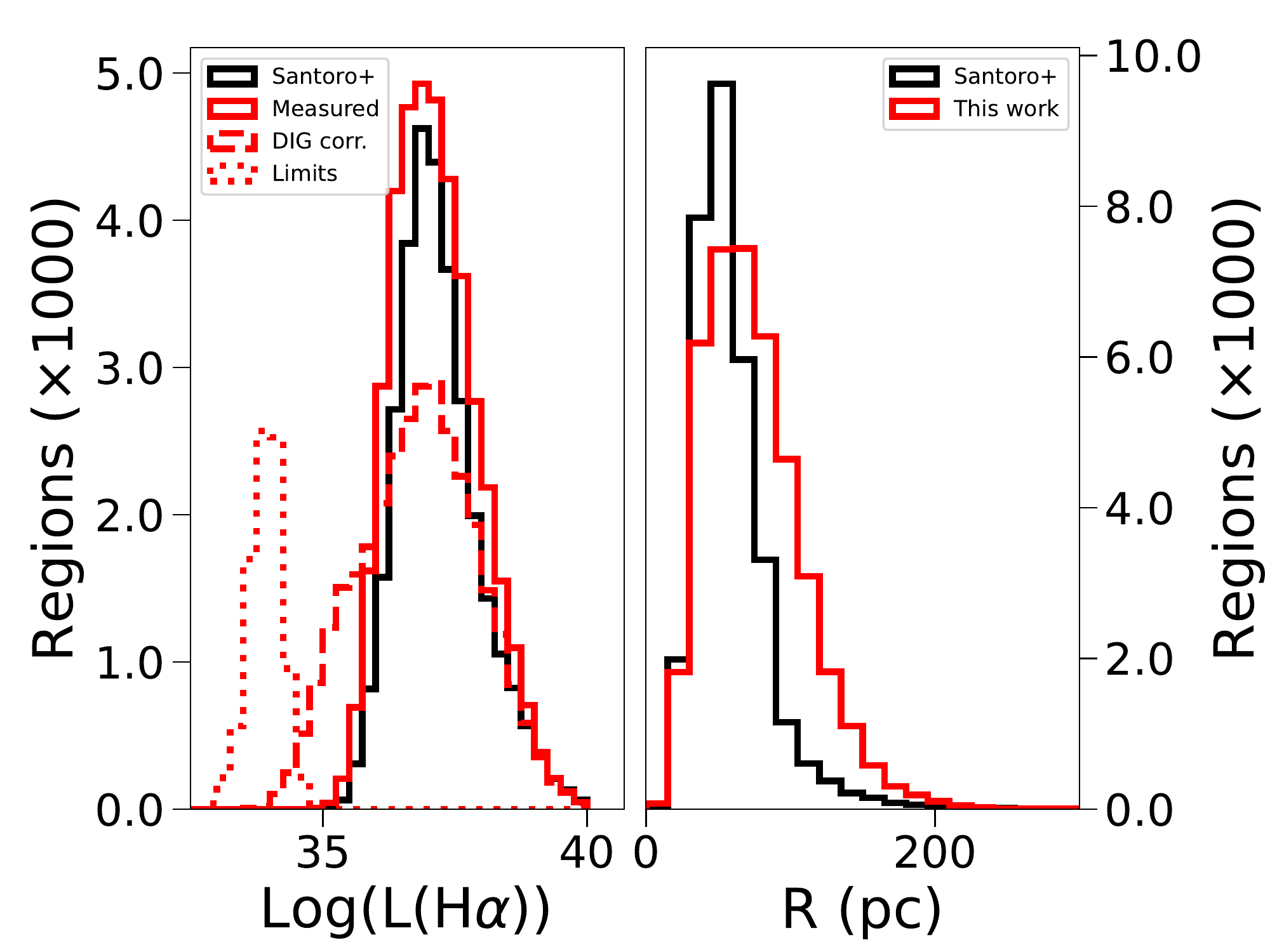}\\
\includegraphics[width=0.45\textwidth]{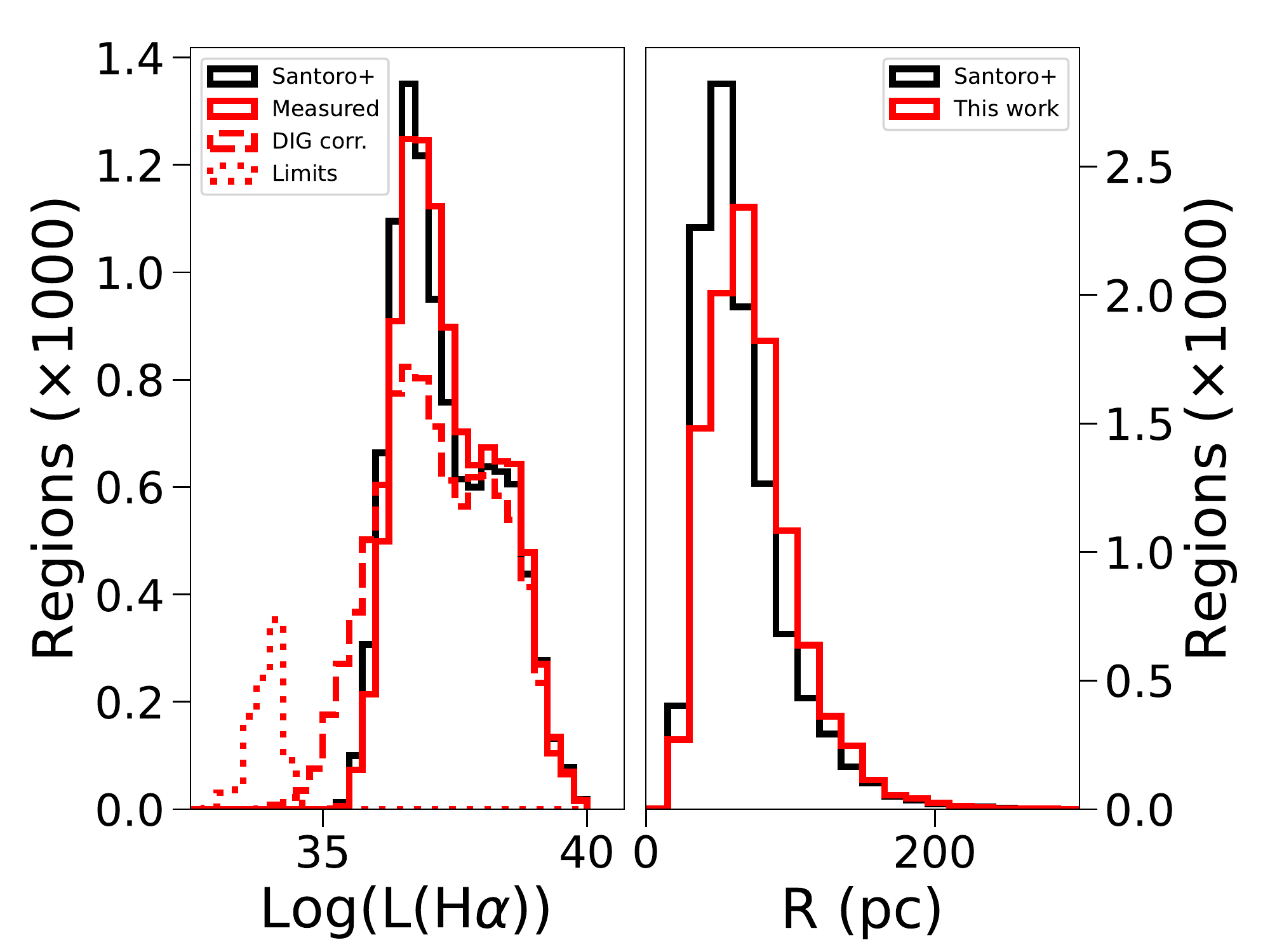}

\caption{Comparison with \citetalias{Santoro22} catalogue. \textbf{Top:} Histograms comparing the main properties of our catalogue to the ones of the nebulae catalogue published in \citetalias{Santoro22}. 
The left panel shows the \Ha\ luminosity distribution of the regions in the \citetalias{Santoro22} catalogue (black), and that of our regions before the correction for DIG contribution (red). The dashed histogram reports the \Ha\ luminosity distribution of our sample after the DIG correction, and the dotted one shows the upper limits for those regions where \Ha\ is not detected after the DIG subtraction.
On the right panel, we show the region size distribution in black for \citetalias{Santoro22} and in red for this work. For both catalogues, the reported size is the circularised radius of the regions.
\textbf{Bottom:} Similar histograms comparing the properties of the matched regions. Symbols and colours are the same as in the top panel.}
\label{fig:hist_fs_all}
\end{figure}

\begin{figure}
\centering
\includegraphics[width=0.45\textwidth]{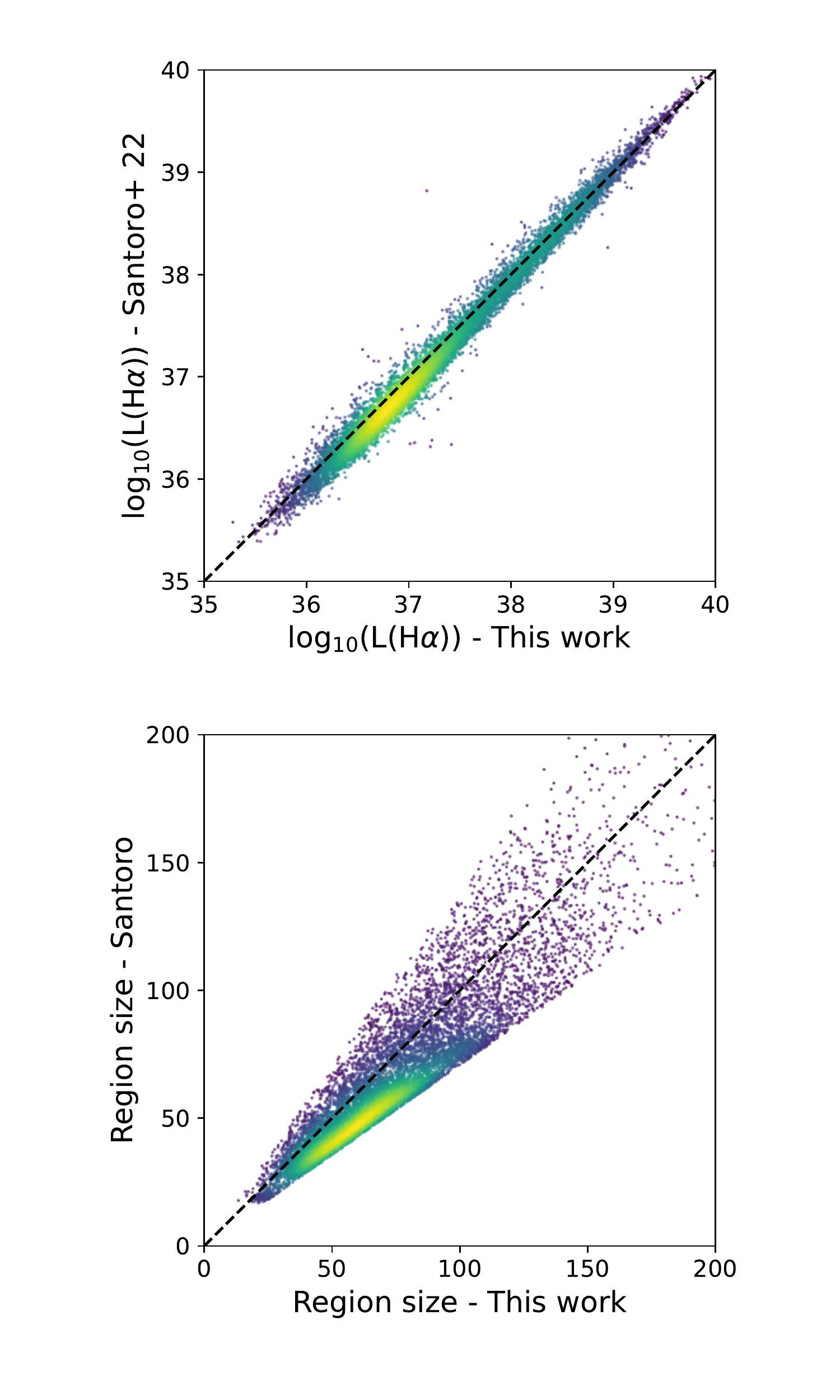}
\caption{Scatter plots showing a direct comparison between the properties of the regions that are present in both catalogues. The top plot shows a comparison between the \Ha\ luminosity, while the bottom one compares the circularised radii of the nebulae. The dashed lines in both plots represent a one-to-one relation. Colours visually represent the density of points in a specific plot area.}
\label{fig:scatter_fs_matched}
\end{figure}

\citetalias{Santoro22} recently used the PHANGS-MUSE data to compile a nebulae catalogue to study the \hii\ regions luminosity function (LF).
While based on the same data, \citetalias{Santoro22} aimed to produce a very clean sample of \hii\ regions, resulting in several differences between their catalogue and the one presented in this work.
Firstly, the \citetalias{Santoro22} catalogue is based only on the \Ha\ maps produced by the DAP, boosting the detection of \hii\ regions but reducing the possibility of detecting other classes of nebulae.
Secondly, the catalogue has been created using \textsc{HIIphot} \citep{Thilker00}, a common source detection algorithm specifically developed to detect \hii\ regions in two-dimensional images.
\textsc{HIIphot} detects nebulae by looking for morphologies typical of \hii\ regions in narrow-band images.
This contributes to the reduction of false detections (e.g. noise spikes), but it can also prevent the detection of regions whose morphology is not expected by the software.
Finally, \citetalias{Santoro22} cleans the catalogue by applying the \hii\ regions selection criteria based on the BPT diagrams we presented in Sec.~\ref{sec:classical} when investigating the LF properties. 
For a detailed view of how the catalogue was created and its general properties, we refer the reader to \citetalias{Santoro22}.
We compare the two catalogues in the following, highlighting similarities and differences.

The most noticeable difference between them is in the number of detected regions.
Our new catalogue contains 40920 nebulae, while \citetalias{Santoro22} detects 31497 (including those not classified as \hii\ regions).
This discrepancy can be attributed to the approaches used for nebulae detection.
We run our detection algorithm on a combination of different line maps (see Sec. \,\ref{sec:creation}).
This allows us to identify regions that are not detected in \Ha\ and favours the detection of faint sources that are easily missed by the \citetalias{Santoro22} approach.

To investigate how well the two compilations of nebulae overlap, we matched the two catalogues by comparing their associated segmentation maps.
For each pair of regions (one for each catalogue), we computed the fraction of overlapping pixels with respect to the area covered by each region. 
We consider a match only if both fractions are $0.5$ or larger.
We went for this approach since the diverse segmentation approaches prevent us to rely on the linear distance between the regions' centres to match them, as commonly done with point source catalogues.
We find matches for 10509 regions, roughly a third of the original \citetalias{Santoro22} catalogue and a fourth of our catalogue, which seems to suggest that the two catalogues detect entirely different sources and that just a small fraction of them is in common.
However, a comparison between the respective segmentation maps shows that, on average, $\sim86\%$ of the area covered by \citetalias{Santoro22} catalogue is also covered by our catalogue and that we recover a similar fraction of the total observed \Ha\ emission ($\sim 64\%$ for \citetalias{Santoro22} vs $\sim70\%$ for our catalogue).
On the other hand, only $\sim43\%$ of our segmentation maps is covered by \citetalias{Santoro22}'s ones.

These simple tests paint a scenario where both catalogues mainly contain the same nebulae (similar fraction of recovered \Ha), but the detection algorithms produce significantly different segmentation maps (Fig.~\ref{fig:segmentation}, panel c).
In particular, our catalogue contains more regions which are also generally larger, even though this results, on average, only in a marginal increase of recovered \Ha\ flux.
Figure\,\ref{fig:hist_fs_all} (top panel), where we compare the luminosity distribution (extinction corrected\footnote{We use the \ebv\ measured via the Balmer decrement with theoretical \Ha$/$\Hb $=3.03$}) and the size distribution of the regions in both catalogues clearly highlights these trends.
Considering the black and red (solid) histograms in the left panel of Fig.\,\ref{fig:hist_fs_all} (top), it is possible to see that the two distributions mostly match at the high-luminosity end but that moving towards fainter sources our catalogue detects significantly more regions in each luminosity bin.
However, the faint regions contribute less to the total amount of detected \Ha\ than the bright ones, resulting in similar \Ha\ flux detection fractions.
Finally, the right panel of Fig.\,\ref{fig:hist_fs_all} (top) shows how the size of the regions in our catalogue is typically larger than in \citetalias{Santoro22}.
All these properties make associating regions unequivocally quite challenging, explaining the low number of matches we find.
If we look only at the matching regions (Fig.\,\ref{fig:hist_fs_all}, bottom panel), it is possible to see that the \Ha\ luminosity distributions are much more consistent. 
In contrast, the size distribution still shows that our regions typically have slightly larger sizes, confirmed by the direct comparison presented in Fig.~\ref{fig:scatter_fs_matched}.
The figure shows that the nebulae from \citetalias{Santoro22} are $\sim 15\%$ less bright than the ones in our catalogue.
The size distribution is much more spread, with the regions in our catalogue showing, on average, circularised radii $\sim19\%$ larger than the corresponding nebula in \citetalias{Santoro22}.

Finally, Fig.\,\ref{fig:hist_fs_all} also shows how the luminosity distribution of our regions changes when we correct the \Ha\ luminosity for the DIG contribution (dashed and dotted red histograms).
While the top end of the distribution is not affected by the correction, generally, the shape of the distribution widens, and the peak moves towards fainter luminosities.
A significant number of regions ($\sim20\%$ for the full catalogue, $\sim10\%$ for the matched regions) are no longer detected in \Ha\ after the DIG correction.

\subsection{\hii\ regions luminosity function}
\label{sec:hii_lf}

The \hii\ regions LF is a fundamental tool for studying the properties of star-forming galaxies.
Typically, it is described by a simple power law with a slope of $\sim -2$ \citep{Elmegreen96}.
Variations of the steepness are often connected to both global properties of the host galaxy and local parameters that can influence the star formation process.
It is possible to build the \hii\ regions LF using a wide variety of tracers, from the UV emission \citep[e.g.][]{Cook16} to the radio or infrared emission \citep[e.g.][]{Mascoop21}.
However, the most commonly used tracer is their \Ha\ luminosity, since \Ha\ is a bright line, easy to observe with ground-based facilities.

While it is outside of the scope of this paper to discuss the \hii\ regions LF in the PHANGS-MUSE sample accurately, a topic already discussed in detail in \citetalias{Santoro22}, reproducing it with our catalogue is an interesting test to confirm if our approach is a viable option for these kinds of studies.
As a first step, to obtain results that are directly comparable to \citetalias{Santoro22} study, we recover a new sub-sample of nebulae from our catalogue following steps similar to what they describe in their work.
Firstly, we select only nebulae classified as \hii\ regions by the model-comparison-based algorithm.
Secondly, we select only nebulae with $3\sigma$ detections of both \Ha\ and \Hb\ after the DIG subtraction, and we correct our luminosities using the DIG corrected Balmer decrement \ebv.
To measure the \ebv, we assume a theoretical \Ha$/$\Hb\ ratio of $3.03$, following Sec.~\ref{sec:EBV}.
Finally, \citetalias{Santoro22} exclude the regions close to the centre of the galaxies where it is difficult to deblend the contribution of very bright and nearby regions.
Following their method, we use the environmental masks from \citet{Querejeta21} to reject these regions, which are only a minor fraction of the sample.
At the end of the selection process, we retain 22947 regions, distributed across the full sample of galaxies.
This means that $\sim 6000$ \hii\ regions have been rejected in total, the vast majority by requesting a 3$\sigma$ detection of both \Ha\ and \Hb.

\begin{figure*}
\centering
\includegraphics[width=0.9\textwidth]{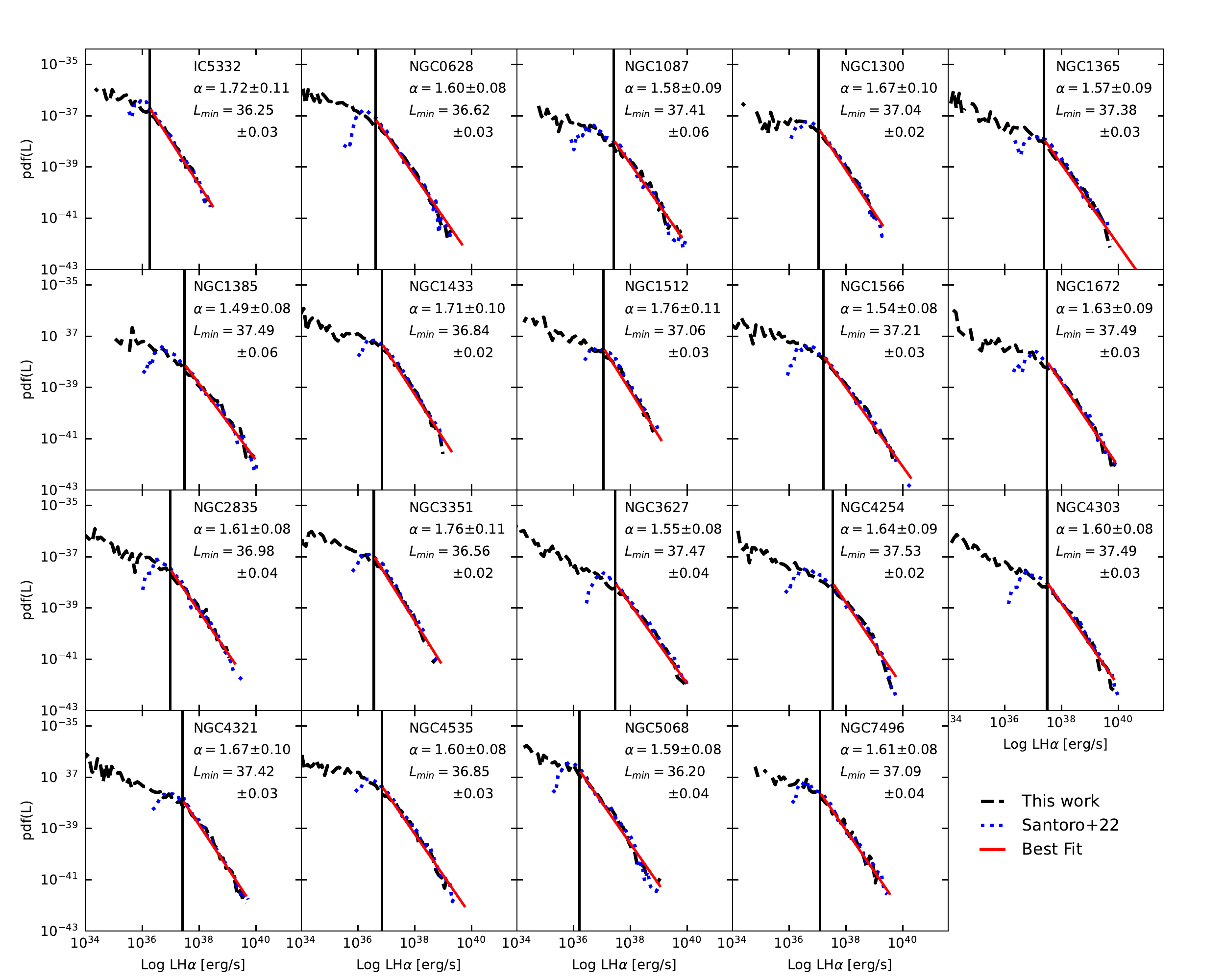}
\caption{\hii\ regions LF for each galaxy in the PHANGS-MUSE sample. Only objects satisfying the selection described in Sec.~\ref{sec:hii_lf} have been used to compute the LF. The black dashed lines represent the observed LF, the blue dotted lines represent the LF from \citetalias{Santoro22}, the red line is the fit obtained from our data, and finally, the black vertical line is the low luminosity cut (L$_{min}$). Only regions with L$>$L$_{min}$ are considered when fitting the LF.}
\label{fig:hii_lf}
\end{figure*}

We build and fit the LF following the procedure described by \citetalias{Santoro22} based on the work by \citet{Clauset09} and the \verb!powerlaw! python package \citep{Alstott14}, which uses a maximum-likelihood estimation method in combination with the Kolmogorov--Smirnov statistics to recover the slope $\alpha$ and the low luminosity cut of the LF (L$_{\mathrm{min}}$, the minimum luminosity to consider for the fit) simultaneously.
In the following, we parametrise the LF as $\propto \rm L^{-\alpha}$, so that $\alpha$ is always positive.
The results of this analysis are shown in Table~\ref{tab:hii_lf} (under the ``Custom'' columns) and Fig.~\ref{fig:hii_lf}.

First of all, the slopes of the LF are consistently between 1.50 and 1.80, with an average of 1.63 and a standard deviation of 0.07, in agreement with previous results in the literature \citep[e.g.][]{Kennicutt89, Elmegreen99} and with \citetalias{Santoro22} in particular.
Also considering each single galaxy, the agreement between the two works is remarkable.
Most of our slopes agree within $1\sigma$ with that measured in \citetalias{Santoro22}, with only a few being consistent only at a $3\sigma$ level.
These objects are those with the steepest slopes in \citetalias{Santoro22}, which can almost be considered outliers.
Also, the agreement between the actual LFs is excellent, as it is possible to see from Fig.~\ref{fig:hii_hist}.
The only difference is that our LF includes a larger number of regions at the low-luminosity end, which means that we are recovering a larger number of faint \hii\ regions.

We also tested how the properties of the LF change when not applying the previously mentioned cuts and when considering directly the \hii\ regions identified in the full catalogue and the other sub-samples described in Sec.\,\ref{sec:reliability}. 
Table~\ref{tab:hii_lf} shows that most of the changes to the slope and L$_{\mathrm{min}}$ are negligible when moving to the \Ha\ sample, and the largest ones are well within $1\sigma$.
This is expected, since most of the selection applied by \citetalias{Santoro22} cuts relies on the detection of the \Ha\ and \Hb\ and, therefore, rejects basically the same regions.
If we use the full sample of nebulae, the poorly detected regions affect the slope, which becomes flatter. 
When we move to the more restrictive samples (the OIII and the BPT samples), things start to change significantly.
Table~\ref{tab:hii_lf} shows how L$_{\mathrm{min}}$ and $\alpha$ moves towards higher luminosities after applying the more stringent cuts.
Therefore, the importance of the bright end of the luminosity function in the fit increases and also $\alpha$ moves towards higher values (steeper LF).

From this analysis, we can conclude that our approach to detecting nebulae, combined with the model-comparison-based classification algorithm, produces similar, if not better, results than more traditional approaches when investigating the \hii\ regions LF.
In particular, it seems that with our approach, we can sample the low luminosity end of the LF better, and measure more consistent slopes across the galaxies of the full PHANGS-MUSE sample with respect to what is found by \citetalias{Santoro22}.
We also see that not sampling well enough the low-luminosity end of the LF results in higher values of $\alpha$.

\begin{table*}
\caption{Values of the slope of the LF and L$_{\mathrm{min}}$ recovered by fitting a simple power-law to the data. The results recovered by \citetalias{Santoro22} are reported in the `Santoro' column as a reference.}
\label{tab:hii_lf}
\tiny
\setlength{\tabcolsep}{3.5pt}
\fontsize{7.5}{8}\selectfont
\centering
\begin{tabular}{lcccccc|cccccc}
\hline\hline
Galaxy& &\multicolumn{4}{c}{$\alpha$}& & &\multicolumn{4}{c}{L$_{\mathrm{min}}$} &\\
 & Santoro & Custom & Full& \Ha\ & OIII & BPT & Santoro & Custom & Full & \Ha\ & OIII & BPT\\
\hline
IC 5332 & 1.82$\pm$0.13 & 1.72$\pm$0.11 & 1.66$\pm$0.09& 1.71$\pm$0.10 & 1.82$\pm$0.13& 1.90$\pm$0.14 & 36.25$\pm$0.19 & 36.25$\pm$0.03& 36.12$\pm$0.04 & 36.23$\pm$0.03& 36.62$\pm$0.04 & 36.79$\pm$0.05\\
\ngc0628 & 1.71$\pm$0.10 & 1.60$\pm$0.08 & 1.50$\pm$0.08& 1.58$\pm$0.09 & 1.70$\pm$0.10& 1.76$\pm$0.11 & 36.63$\pm$0.08 & 36.62$\pm$0.03& 36.28$\pm$0.03 & 36.56$\pm$0.03& 37.08$\pm$0.03 & 37.29$\pm$0.03\\
\ngc1087 & 1.71$\pm$0.10 & 1.58$\pm$0.09 & 1.46$\pm$0.09& 1.57$\pm$0.09 & 1.64$\pm$0.09& 1.77$\pm$0.12 & 37.49$\pm$0.10 & 37.41$\pm$0.06& 37.00$\pm$0.06 & 37.38$\pm$0.05& 37.69$\pm$0.06 & 37.92$\pm$0.05\\
\ngc1300 & 1.77$\pm$0.12 & 1.67$\pm$0.10 & 1.64$\pm$0.09& 1.65$\pm$0.09 & 1.73$\pm$0.11& 1.78$\pm$0.12 & 37.08$\pm$0.25 & 37.04$\pm$0.02& 36.98$\pm$0.03 & 37.04$\pm$0.02& 37.38$\pm$0.03 & 37.52$\pm$0.03\\
\ngc1365 & 1.74$\pm$0.11 & 1.57$\pm$0.09 & 1.45$\pm$0.08& 1.52$\pm$0.08 & 1.57$\pm$0.09& 1.65$\pm$0.09 & 37.82$\pm$0.20 & 37.38$\pm$0.03& 36.99$\pm$0.04 & 37.36$\pm$0.03& 37.65$\pm$0.05 & 37.88$\pm$0.04\\
\ngc1385 & 1.52$\pm$0.08 & 1.49$\pm$0.08 & 1.38$\pm$0.08& 1.47$\pm$0.09 & 1.57$\pm$0.09& 1.64$\pm$0.09 & 37.29$\pm$0.17 & 37.49$\pm$0.06& 36.90$\pm$0.08 & 37.42$\pm$0.06& 37.82$\pm$0.06 & 38.05$\pm$0.07\\
\ngc1433 & 1.96$\pm$0.16 & 1.71$\pm$0.10 & 1.65$\pm$0.09& 1.69$\pm$0.10 & 1.75$\pm$0.11& 1.86$\pm$0.13 & 37.32$\pm$0.14 & 36.84$\pm$0.02& 36.72$\pm$0.02 & 36.82$\pm$0.02& 37.13$\pm$0.03 & 37.38$\pm$0.02\\
\ngc1512 & 2.04$\pm$0.17 & 1.76$\pm$0.11 & 1.61$\pm$0.08& 1.70$\pm$0.10 & 1.73$\pm$0.11& 1.80$\pm$0.12 & 37.41$\pm$0.10 & 37.06$\pm$0.03& 36.86$\pm$0.04 & 37.05$\pm$0.03& 37.25$\pm$0.04 & 37.48$\pm$0.04\\
\ngc1566 & 1.57$\pm$0.09 & 1.54$\pm$0.08 & 1.44$\pm$0.08& 1.53$\pm$0.08 & 1.58$\pm$0.09& 1.67$\pm$0.10 & 37.14$\pm$0.14 & 37.21$\pm$0.03& 36.74$\pm$0.04 & 37.15$\pm$0.02& 37.51$\pm$0.04 & 37.78$\pm$0.03\\
\ngc1672 & 1.67$\pm$0.10 & 1.63$\pm$0.09 & 1.46$\pm$0.08& 1.58$\pm$0.09 & 1.68$\pm$0.10& 1.70$\pm$0.10 & 37.55$\pm$0.12 & 37.49$\pm$0.03& 37.01$\pm$0.04 & 37.45$\pm$0.04& 37.78$\pm$0.03 & 37.93$\pm$0.03\\
\ngc2835 & 1.76$\pm$0.11 & 1.61$\pm$0.08 & 1.46$\pm$0.08& 1.59$\pm$0.08 & 1.76$\pm$0.11& 1.78$\pm$0.12 & 37.08$\pm$0.10 & 36.98$\pm$0.04& 36.45$\pm$0.06 & 36.94$\pm$0.04& 37.37$\pm$0.03 & 37.50$\pm$0.04\\
\ngc3351 & 1.98$\pm$0.16 & 1.76$\pm$0.11 & 1.64$\pm$0.09& 1.72$\pm$0.11 & 1.69$\pm$0.10& 1.76$\pm$0.11 & 36.84$\pm$0.13 & 36.56$\pm$0.02& 36.38$\pm$0.04 & 36.53$\pm$0.03& 36.73$\pm$0.05 & 37.03$\pm$0.04\\
\ngc3627 & 1.57$\pm$0.09 & 1.55$\pm$0.08 & 1.38$\pm$0.08& 1.53$\pm$0.08 & 1.59$\pm$0.09& 1.64$\pm$0.09 & 37.43$\pm$0.23 & 37.47$\pm$0.04& 36.74$\pm$0.06 & 37.39$\pm$0.05& 37.74$\pm$0.04 & 38.03$\pm$0.06\\
\ngc4254 & 1.61$\pm$0.08 & 1.64$\pm$0.09 & 1.47$\pm$0.09& 1.63$\pm$0.09 & 1.71$\pm$0.10& 1.78$\pm$0.12 & 37.36$\pm$0.12 & 37.53$\pm$0.02& 37.04$\pm$0.04 & 37.50$\pm$0.03& 37.73$\pm$0.03 & 37.89$\pm$0.03\\
\ngc4303 & 1.56$\pm$0.09 & 1.60$\pm$0.08 & 1.42$\pm$0.08& 1.57$\pm$0.09 & 1.69$\pm$0.10& 1.74$\pm$0.11 & 37.26$\pm$0.15 & 37.49$\pm$0.03& 36.88$\pm$0.04 & 37.43$\pm$0.03& 37.88$\pm$0.03 & 38.04$\pm$0.03\\
\ngc4321 & 1.95$\pm$0.15 & 1.67$\pm$0.10 & 1.49$\pm$0.08& 1.62$\pm$0.08 & 1.71$\pm$0.10& 1.79$\pm$0.12 & 37.95$\pm$0.27 & 37.42$\pm$0.03& 36.96$\pm$0.04 & 37.37$\pm$0.02& 37.72$\pm$0.04 & 37.90$\pm$0.04\\
\ngc4535 & 1.63$\pm$0.09 & 1.60$\pm$0.08 & 1.52$\pm$0.08& 1.58$\pm$0.09 & 1.69$\pm$0.10& 1.75$\pm$0.11 & 36.79$\pm$0.09 & 36.85$\pm$0.03& 36.58$\pm$0.04 & 36.79$\pm$0.03& 37.30$\pm$0.03 & 37.54$\pm$0.04\\
\ngc5068 & 1.58$\pm$0.09 & 1.59$\pm$0.08 & 1.52$\pm$0.08& 1.58$\pm$0.09 & 1.66$\pm$0.09& 1.75$\pm$0.11 & 35.92$\pm$0.06 & 36.20$\pm$0.04& 35.85$\pm$0.05 & 36.18$\pm$0.04& 36.57$\pm$0.04 & 36.81$\pm$0.05\\
\ngc7496 & 1.69$\pm$0.10 & 1.61$\pm$0.08 & 1.56$\pm$0.09& 1.60$\pm$0.08 & 1.69$\pm$0.10& 1.77$\pm$0.12 & 37.16$\pm$0.09 & 37.09$\pm$0.04& 36.91$\pm$0.04 & 37.08$\pm$0.04& 37.52$\pm$0.06 & 37.68$\pm$0.05\\
\hline
\end{tabular}
\end{table*}

\subsection{Comparison with Scheuermann et al. (2022)}
\label{sec:fabian}

\citet{Scheuermann22} performed a detailed search for PNe in the 19 PHANGS--MUSE galaxies to measure the galaxies' distance via the PNe luminosity function \citep[PNLF,][]{Jacoby89, Ciardullo02}.
They used traditional techniques (point-source detection, accurate aperture photometry) to identify and classify nebulae starting from \oiii\ maps at native resolution \citep[see][for a detailed description of the procedures]{Spriggs20, Roth21, Scheuermann22}.
The search resulted in a catalogue of 1049 sources, 899 of which have been confirmed as PNe while the remaining 150 have been classified as SNR.
The number is comparable to the number of nebulae we classify as PNe using our model-comparison-based algorithm in the full sample (796, Table \ref{tab:subsamples}).

To understand if there is a direct correspondence between the PNe identified by the two approaches, we match the two catalogues following the procedure developed in Sec.~\ref{sec:compare_fs}.
Since `segmentation maps' are not provided along with the \citet{Scheuermann22} catalogue, we start by building them by positioning at the coordinates of each nebula a circular mask representing the area used in the work to perform aperture photometry.
Then, we proceed as in Sec.~\ref{sec:compare_fs} but requiring less overlap to identify matches, since we expect our nebulae to be much more extended than the point-like sources in \citet{Scheuermann22} catalogue.
In particular, we consider a match if one of the nebulae in \citet{Scheuermann22} shares at least 50\% of its area with one of our nebulae and 10\% of our nebula is shared with \citet{Scheuermann22} nebula, or the overlap is higher than $90\%$ in any direction.

\begin{figure*}
\centering
\includegraphics[height=6cm]{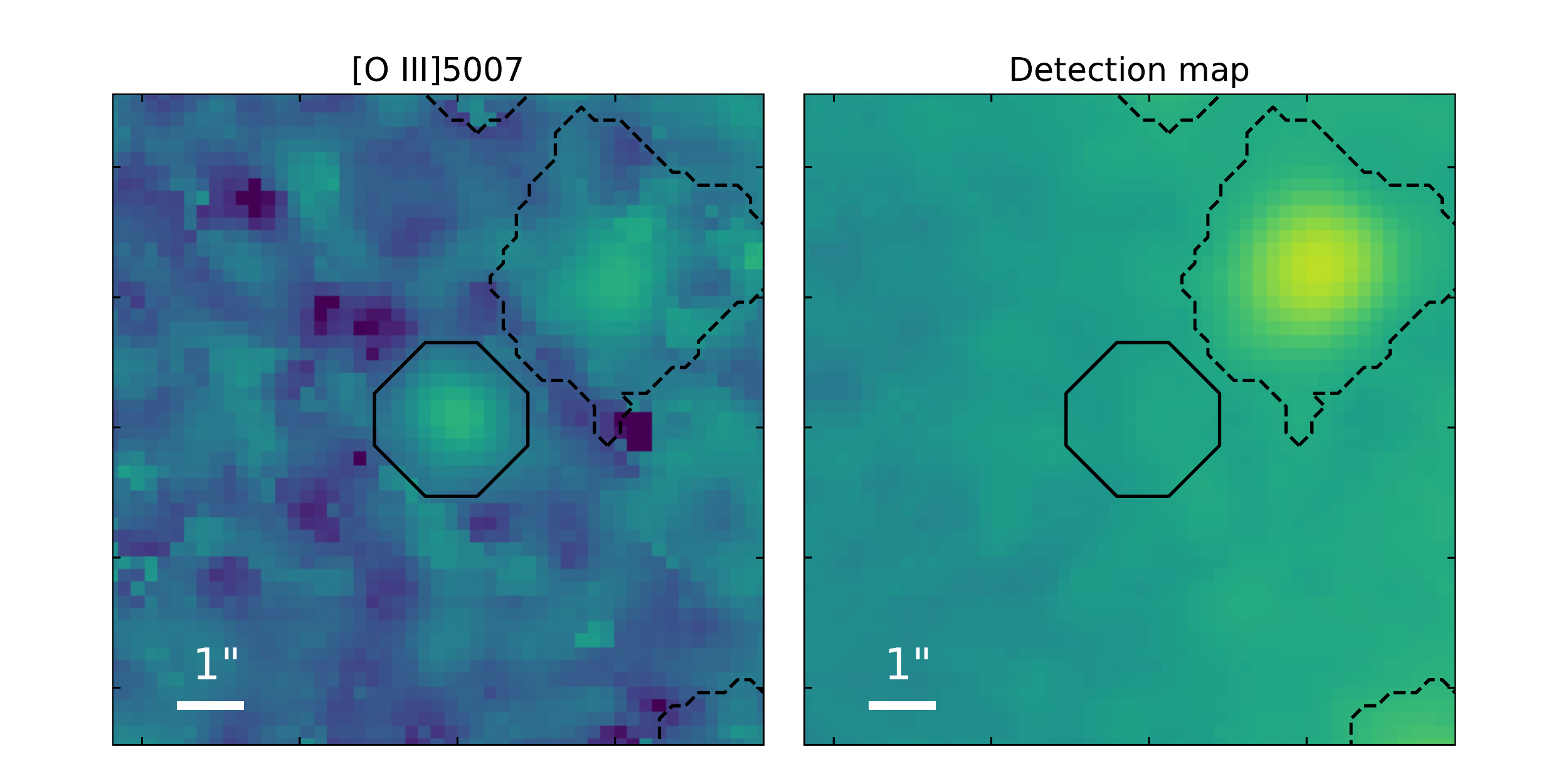}\unskip \vrule
\includegraphics[height=6cm]{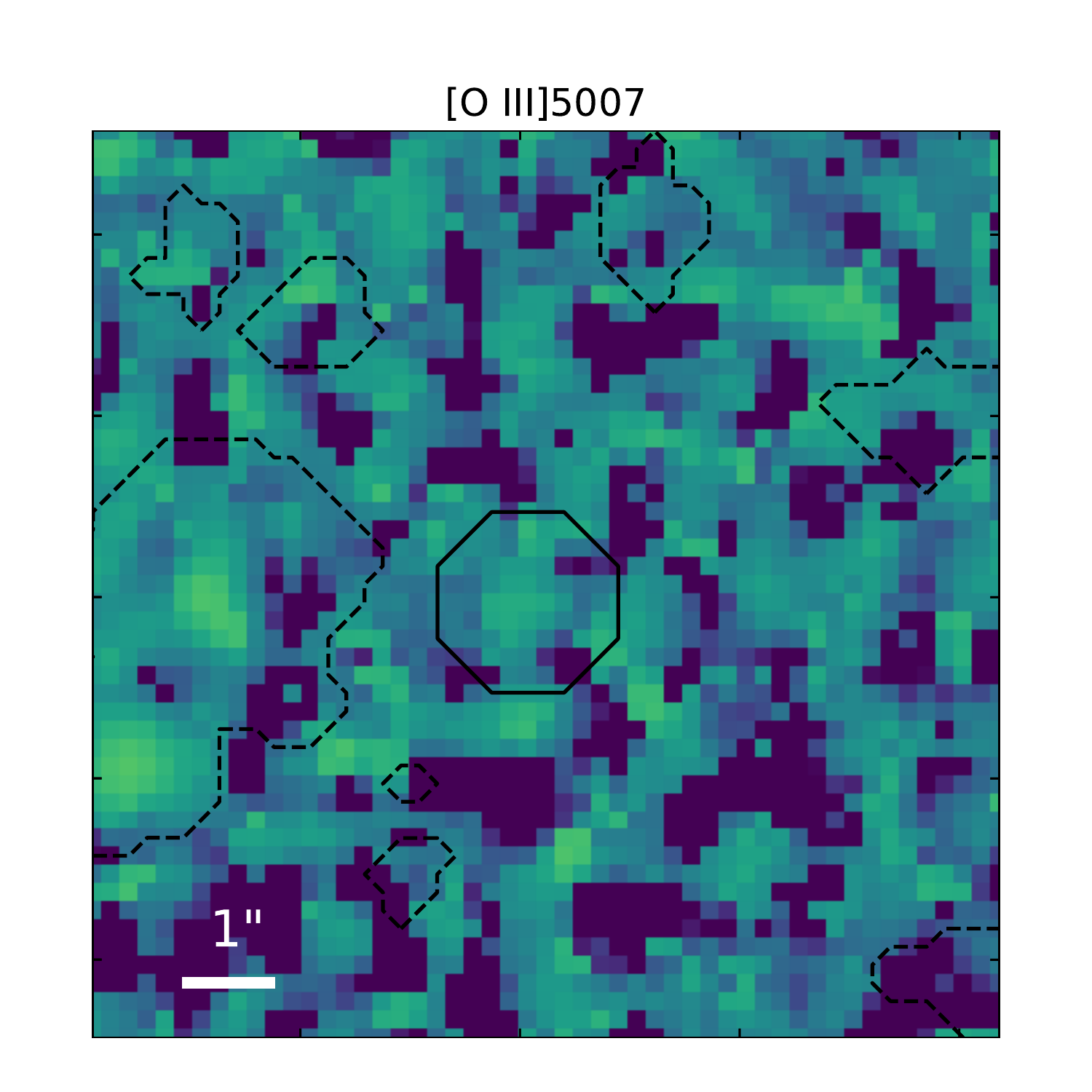}
\caption{Examples of nebulae in \citep{Scheuermann22} not included in our catalogue. \textbf{Left:} detail of \ngc1365 (region 10 in \citealt{Scheuermann22}) in the \oiii\ emission line map (left) and our detection map (right). In both plots, the solid black contours show the position of one of the nebulae detected in \citet{Scheuermann22} catalogue, while the dashed contours show the position of the nebulae in our catalogue. This figure shows how the nebula, clearly detected in the single line map, is not detectable in our detection map since it is blended with the relatively high background. $\SI{1}{\arcsec}$ in \ngc1365 corresponds to $\sim \SI{100}{pc}$. \textbf{Right:} Detail of the \oiii\ emission line map of IC~5332 (region 56 in \citealt{Scheuermann22}). The solid black contours show the position of one of the nebulae detected in \citet{Scheuermann22} catalogue, while the dashed contours show the position of the nebulae in our catalogue. $\SI{1}{\arcsec}$ in IC~5332 corresponds to $\sim \SI{45}{pc}$.}
\label{fig:pn_blended}
\end{figure*}

Matching the catalogue results in 567 (54.1\%) of the \citet{Scheuermann22} regions having a counterpart in our own catalogue, among which 232 have a ``perfect match'' (overlap $> 0.6$ on both directions).
The low yield seems to arise principally from the different approaches to nebulae detection.
\citet{Scheuermann22} performed a detailed search for this kind of object by directly examining the \oiii\ emission line map, significantly reducing crowding from other types of nebulae not particularly bright in \oiii.
On the other hand, we produced a detection map by combining multiple line maps (Sec.~\ref{sec:creation}).
This can blend the emission of nebulae in different lines, making the detection of faint PNe in crowded areas more challenging or even impossible (Fig.\,\ref{fig:pn_blended}).
On top of this, from a visual inspection, we estimate that $\sim 10\%$ of the regions detected by \citet{Scheuermann22} are not visible in the \oiii\ homogenised maps we are using in this work (Fig.\,\ref{fig:pn_blended}, right panel).
This could either result from the convolution, which spreads out the emission of faint nebulae, making them undetectable, or they are false detections.

We now compare how the nebulae included in both catalogues are classified by the two different approaches.
Table~\ref{tab:fabian} shows that half of the sample ($\sim47\%$) of \citet{Scheuermann22} PNe retains the same classification in our compilation.
Most of the remaining nebulae are either considered \hii\ regions ($21.5\%$) or shocks ($18.6\%$) by the model-comparison-based algorithm.
A lower fraction of objects fall in the ambiguous category ($12.5\%$), and only one object is unclassified.
The agreement between the model-comparison-based algorithm and the \citet{Scheuermann22} classification for PNe is much better than what we see between our model-comparison-based classification and our application of the traditional criterion for PNe. 
This suggests that the careful analysis performed by \citet{Scheuermann22} to ensure an accurate measurement of the \oiii\ fluxes and the point-like nature of the emission is essential for the traditional classification criterion to perform well.
However, the disagreement between the two approaches is still substantial.
The two methods do not agree for more than $50\%$ of the sample of PNe identified by \citet{Scheuermann22}.
On the other hand, $\sim73\%$ of \citet{Scheuermann22} SNRs are classified as shocks by our model-comparison-based algorithm.

In summary, on one side, the detection images we are using in this work prevent the detection of faint nebulae like PNe in crowded areas since their emission is blended with that of other regions bright in other lines.
On the other side, the two classification methods seem to not agree on identifying what is a PN, even though the agreement between the two paradigms is better than what is seen in Sec.~\ref{sec:classical}.
This discrepancy does not necessarily mean that our new classification criterion is failing, since the nebulae it classifies as PNe are located in the regions of the diagnostic diagrams where they are expected to be, and a visual inspection of the data reveals that these objects look as expected: unresolved \oiii\ bright nebulae.

\begin{table}
\caption{Comparison between \citet{Scheuermann22} PNe and SNR sample and our catalogue. Each column shows how the objects in each of the two \citet{Scheuermann22} samples are classified by our classification algorithm. Both the absolute number and the percentages (in brackets) are shown.}
\label{tab:fabian}
\centering
\begin{tabular}{lll}
\hline\hline
Class& PNe & SNR\\
\hline
\hii\ regions & 109 (21.6) & 4 (6.5)\\
PNe & 238 (47.1) & 6 (9.6)\\
Shocks & 94 (18.6) & 45 (72.6)\\
Ambiguous & 63 (12.5)& 4 (6.5)\\
Unclassified & 1 (0.2) & 3 (4.8)\\
\hline
Undetected & 394 & 88 \\
\hline
Total & 899 & 150\\
\hline
\end{tabular}
\end{table}

\subsection{Planetary nebulae luminosity function}
\label{sec:pnelf}

\begin{figure*}
\centering
\includegraphics[width=0.9\textwidth]{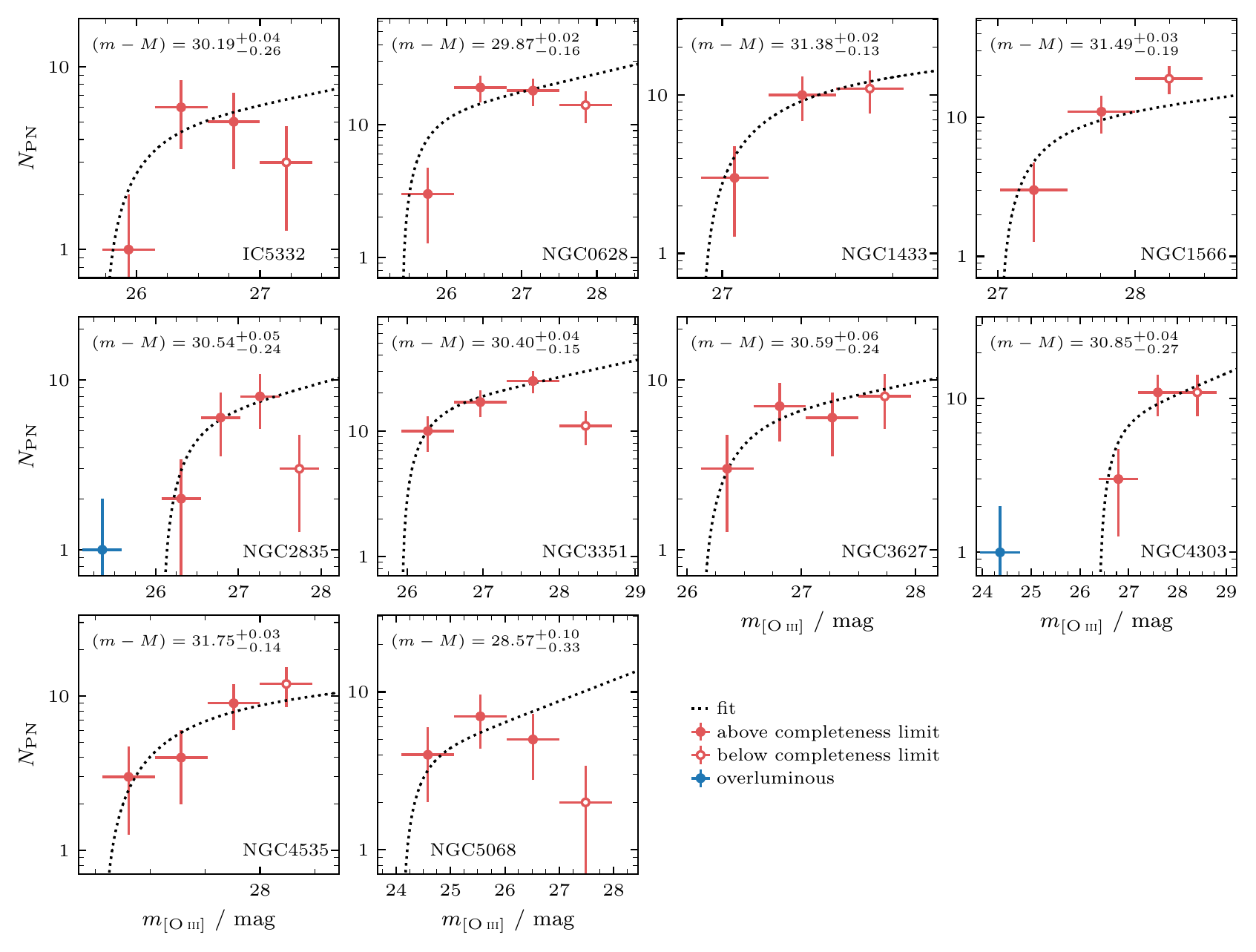}
\caption{PNLF for the galaxies where it was possible to perform the fit. The points with errors represent the observed PNLF, while the dotted line is the result of the fit. Only the points coloured in solid orange have been used in the fitting routine. The empty points are PNe below the completeness limit, while blue points represent discarded overluminous objects.}
\label{fig:pnelf}
\end{figure*}

Extragalactic PNe are often used for measuring the distance of their host via the PNLF \citep{Jacoby89, Ciardullo02}.
Although the sample of PNe recovered in this work is different from that compiled in \citet{Scheuermann22}, the nebulae we identify typically show the characteristics of PNe.
Therefore we can build the PNLF, estimate the distance of the galaxies in our sample and compare our results with the measurement from \citet{Scheuermann22} to evaluate if our approach is a viable option for these kinds of studies.

To reproduce the PNLF, we followed the procedure described in \citet{Scheuermann22} to enable a direct comparison between the results.
We use the Eq.~\ref{eq:mo3} \citep{Jacoby89} and the distance modulus definition (Eq.~\ref{eq:dmod}) to convert the nebulae DIG corrected, line flux ($I_{[\mathrm{O\textsc{iii}}]}$) in an absolute magnitude.
The fluxes are corrected for Milky Way extinction, but not for internal extinction.
\begin{equation}
\label{eq:dmod}
    \mu = m-M = 5\log_{10}(D) - 5.
\end{equation}
At this point, we fit the PNLF with the procedure described in \citet{Scheuermann22} and using the following functional form from \citet{Ciardullo89}:
\begin{equation}
    N(M_{[\mathrm{O\textsc{iii}}]}) \propto e^{0.307M_{[\mathrm{O\textsc{iii}}]}} \left(1-e^{3(M^*-M_{[\mathrm{O\textsc{iii}}]})}\right).
\end{equation}
$M^*$ is the zero-point of the luminosity function, and we assume it to be $-4.47$.

\begin{table}
\centering
\caption{Estimated distances of the galaxies where we could fit the PNLF compared with the results from \citet{Scheuermann22}. The columns report: the name of the galaxy, the number of PNe available for the fit, the distance modulus recovered with error, the distance of the galaxy in Mpc, the distance modulus recovered with error from \citet{Scheuermann22}, the distance of the galaxy in Mpc from \citet{Scheuermann22}.}
\label{tab:pndist}
{\small \renewcommand{\arraystretch}{1.4}
\begin{tabular}{lccccc}
\hline\hline
Galaxy & N & $\mu$ & D& $\mu_{\rm Sc22}$ & D$_{\rm Sc22}$\\
 & & mag & Mpc & mag & Mpc \\
\hline
IC 5332 & 12 & 30.19$^{+0.04}_{-0.26}$ & 10.94$^{+0.20}_{-1.33}$ & 29.73$^{+0.10}_{-0.20}$& 8.84$^{+0.39}_{-0.82}$\\
\ngc0628 & 40 & 29.87$^{+0.02}_{-0.16}$ & 9.43$^{+0.08}_{-0.69}$ & 29.89$^{+0.06}_{-0.09}$& 9.52$^{+0.26}_{-0.41}$\\
\ngc1433 & 13 & 31.38$^{+0.02}_{-0.13}$ & 18.84$^{+0.13}_{-1.16}$& 31.39$^{+0.04}_{-0.07}$& 18.94$^{+0.39}_{-0.56}$\\
\ngc1566 & 14 & 31.49$^{+0.03}_{-0.19}$ & 19.85$^{+0.26}_{-1.77}$& 31.13$^{+0.08}_{-0.17}$& 16.84$^{+0.60}_{-1.29}$\\
\ngc2835 & 16 & 30.54$^{+0.05}_{-0.24}$ & 12.81$^{+0.29}_{-1.44}$& 30.57$^{+0.08}_{-0.17}$& 13.03$^{+0.46}_{-1.04}$\\
\ngc3351 & 52 & 30.40$^{+0.04}_{-0.15}$ & 12.00$^{+0.25}_{-0.84}$& 30.36$^{+0.06}_{-0.08}$& 11.80$^{+0.31}_{-0.43}$\\
\ngc3627 & 16 & 30.59$^{+0.06}_{-0.24}$ & 13.12$^{+0.34}_{-1.45}$& 30.18$^{+0.08}_{-0.15}$& 10.88$^{+0.39}_{-0.77}$\\
\ngc4303 & 14 & 30.85$^{+0.04}_{-0.27}$ & 14.82$^{+0.25}_{-1.85}$& 30.65$^{+0.10}_{-0.26}$& 13.49$^{+0.64}_{-1.60}$\\
\ngc4535 & 16 & 31.75$^{+0.03}_{-0.14}$ & 22.41$^{+0.30}_{-1.47}$& 31.43$^{+0.06}_{-0.09}$& 19.29$^{+0.56}_{-0.82}$\\
\ngc5068 & 16 & 28.57$^{+0.10}_{-0.33}$ & 5.17$^{+0.24}_{-0.78}$ & 28.46$^{+0.11}_{-0.26}$& 4.93$^{+0.24}_{-0.59}$\\
\hline
\end{tabular}
}
\end{table}

The results are presented in Fig.~\ref{fig:pnelf} while in Table~\ref{tab:pndist} we directly compare our findings with the results from \citet{Scheuermann22}.
Only 10 out of 19 galaxies contain enough bright PNe to fit the PNLF.
In all cases, our measurements are within $3\sigma$ from what is reported by \citet{Scheuermann22}, with six galaxies being consistent within $1\sigma$.
We notice, however, that we generally measure larger distances.

We also compare our distances with the work from \citet{Anand21}, which provides a compilation of distances to PHANGS galaxies based on different methods (mostly tip of the red giant branch and Cepheids).
Also in this case we see a similar pattern, with most of our measurements being consistent within $3\sigma$ with their results.
The exception is \ngc4535, where the Cepheid-based distance reported in \citet{Anand21} is significantly lower also compared to \citet{Scheuermann22} results.

In summary, we can estimate a distance for just half of the galaxies in our sample, but our results generally agree with what is found in the literature and by \citet{Scheuermann22} in particular.
The problem arises from the low number of bright PNe detected, and, therefore, from their segmentation more than their classification, as discussed in Sec.~\ref{sec:fabian}.
While running a dedicated search for PNe is better, out approach is still useful to clean a sample of PNe from other sources.
Applying the segmentation algorithm only to the \oiii\ line map before applying the classification algorithm might produce better results.

\subsection{Shocks and supernova remnants}
\label{sec:shocks_dis}

\begin{figure}
\centering
\includegraphics[width=0.48\textwidth]{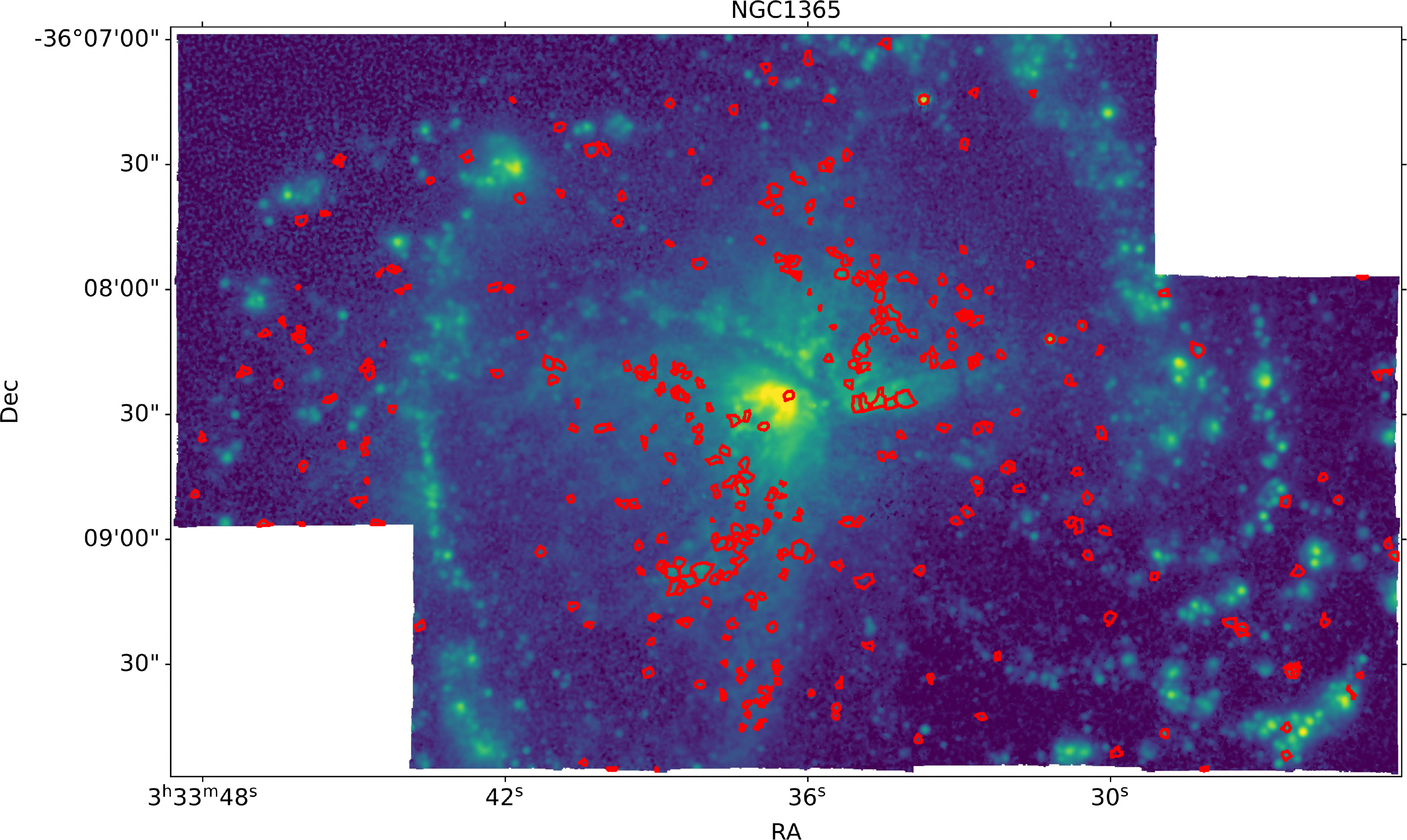}\\
\vskip0.3cm
\includegraphics[width=0.48\textwidth]{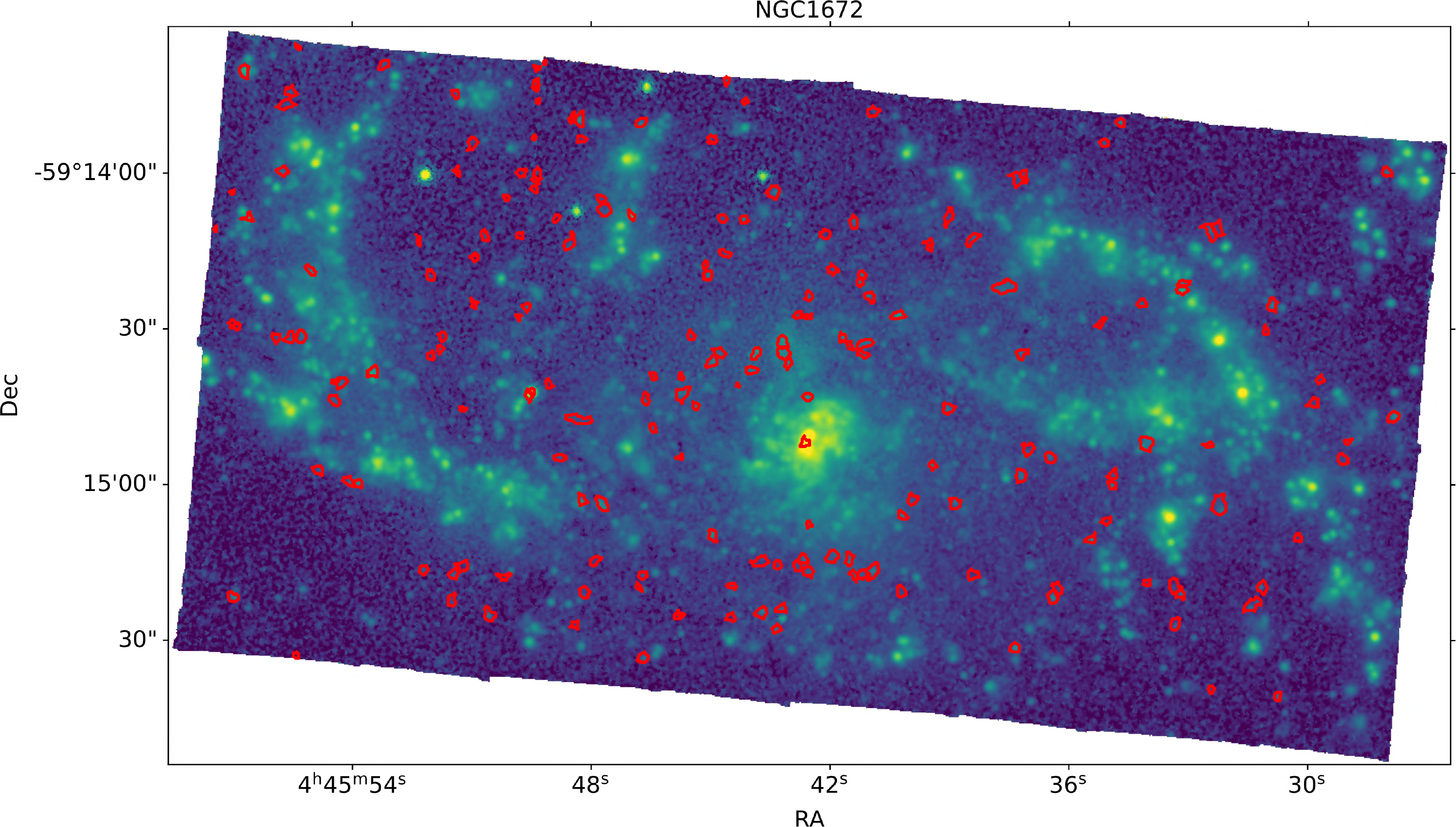}
\caption{Location of the regions classified as shocks in \ngc1365 (top) and \ngc1672 (bottom) on top of \oiii\ line maps.}
\label{fig:snr_1365}
\end{figure}

So far, we have identified all of the nebulae that can be described by the \citet{Allen08} models as shocks and not  as SNRs since we expect this category to be somewhat contaminated by the presence of AGN ionised regions.
However, this contamination can only occur if the galaxy hosts an AGN.
Four galaxies in our sample are known to host an AGN \citep[\ngc1365, \ngc1512, \ngc1566 and \ngc1672, e.g.][]{VeronCetty06,Belfiore22}, but only \ngc1365 is known to host an extended region of AGN-ionised gas \citep[also known as ionisation cones, e.g.][]{StorchiBergmann91,Venturi18}.
The other galaxies show some extended AGN- or LIER-like (low ionisation emission region) emission outside the nucleus, but the low equivalent width of \Ha\ seems to suggest that these are likely regions filled by DIG, which is ionised by the hard radiation produced by hot low-mass evolved stars \citep[HOLMES,][]{Belfiore22}.
Such behaviour is also observed in a few other galaxies with no known AGN in their nucleus (\ngc1300, \ngc1433, \ngc3351, \ngc3627).
The AGN-SNR degeneracy problem can be significant in \ngc1365 and in the very central regions of the other three known AGN, but it should not produce significant contamination in all the other objects.
In fact, the top panel of Fig.~\ref{fig:snr_1365} shows how there is an over-density of bright shock-ionised nebulae along the ionisation cones of the galaxy, which are easily visible in the \oiii\ line map shown in the figure, while no ionisation cones nor over-densities of shocks nebulae are observed in another AGN, for example \ngc1672.
We can conclude that while there is contamination by a certain number of AGN-ionised regions, it is limited to \ngc1365 and the nucleus of a few other galaxies.

Unfortunately, SNRs (and shocks) are the least well-studied class of sources in the extragalactic environment.
There are only a handful of SNRs catalogues in galaxies outside the Local Group, and none of them covers any of the galaxies included in the PHANGS-MUSE survey.
As a consequence, we cannot perform the same analysis we did for PNe and \hii\ regions to understand how well our segmentation and classification algorithms are performing.
However, we can compare the number of SNRs we find in our galaxies with what is available in the literature.
We consider five recent catalogues, each one focused on a different, nearby, main sequence galaxy: M31 \citep{Lee14}, M33 \citep{Lee14b}, M83 \citep{Long22}, \ngc3344 \citep{Moumen19} and \ngc4030 \citep{Cidfernandes21}.
The first two used narrow-band images to identify SNR candidates using the \sii$/$\Ha\ ratio together with some additional criteria based on their expected shell-like morphology and on the presence or absence of blue stars at the candidate position \citep[e.g.][]{Lee14, Lee14b}.
\citet{Long22} uses emission-line maps from the MUSE mosaic produced by \citet{Dellabruna22} and other narrow-band images from Magellan and HST to identify SNR candidates, that are then classified using a dedicated analysis of their spectra acquired by MUSE or other instruments (e.g. GMOS).
\citet{Moumen19} used SITELLE, the imaging Fourier Transform spectrograph of the Canada-France-Hawaaii Telescope (CFHT), to obtain emission-line maps of the galaxy in multiple emission lines.
They selected the SNRs candidates using a combination of criteria: \sii$/$\Ha\ $>0.4$, high S/N for the two \sii\ lines, size of the region $<120~\si{pc}$ and finally they require concentrated regions based on their pseudo-Voight profile.
Finally, \citet{Cidfernandes21} used a principal component analisys on a sample of nebulae recovered via the analysis of MUSE data, combined with the \sii$/$\Ha/ vs. \nii$/$\Ha/ diagram from \citet{Kopsacheili20}, to identify a sample of 59 SNRs.
Following the described criteria, \citet{Lee14} reports 156 SNR in M31, \citet{Lee14b} 199 SNR in M33, \citet{Long22} 366 in M83, \citet{Moumen19} 129 in \ngc3344\footnote{We are considering the total number of SNR candidates identified by \citet{Moumen19}. However, they further distinguish between ``confirmed'' SNRs (42), ``probable'' SNRs (45) and ``less likely'' SNRs (42).} and \citet{Cidfernandes21} 59 in \ngc4030.
In Table~\ref{tab:snr_table} we show the number of SNR for each galaxy in our sample.
It is possible to see that we find, on average, 271 SNR per galaxy, with a maximum of 496 in \ngc1566 and a minimum of 78 in \ngc7496.
This is perfectly in line with what is reported by the literature, especially considering the different observation techniques, depth of the data, different surveyed area and detection methods. 
These numbers are also comparable to the 294 known Galactic SNRs \citep{Green19}, most of which have been identified through their radio emission.
If we consider only the BPT sub-sample, which should contain only the most reliable detections, we detect $\sim100$ SNRs per galaxy on average, still on the same order of magnitude as what is found in the literature.
We can also compare the distribution of the \Ha\ luminosity for our sample (Fig.~\ref{fig:sh_hist}) of SNRs with that reported by \citet{Lee14b} for their SNR candidates in M31 and M33.
Both distributions peak around 10$^{36}\,\si{erg.s^{-1}}$ and they have a similar shape, even though it is clear that we have much better statistics.
Also, our data are significantly deeper, since we have a significant amount of nebulae below 10$^{35}\,\si{erg.s^{-1}}$ even when we apply the more restrictive cuts, while they barely report any below this limit.
We also have a certain number of brighter nebulae, which can be connected to multiple aspects.
On one hand, the sample size definitely increases the possibility of finding brighter objects.
On the other hand, the different methods used to define the size of the nebulae, especially considering that they are observing much closer galaxies, with higher spatial resolutions, and the background subtraction, can influence the measurement of the \Ha\ luminosity.

Finally, we must note that, in a few objects, the spatial distribution of the shock classified regions is peculiar.
The atlas presented in Appendix~\ref{sec:atlas}, shows that in some galaxies (e.g. \ngc3627) these nebulae fill the interarm region while in other objects (e.g. \ngc1433, \ngc3351) there is a peculiar overdensity of shock classified regions just outside of the galaxies' bulge.
These seems regions permeated by the DIG emission, which shows line ratios typical of LI(N)ERS \citep{Belfiore22} and similar to what is observed in SNRs and shocks.
It is possible, therefore, that these circumnuclear or interarm regions are not SNRs, but particularly bright DIG clouds and that they can constitute a significant source of contamination for the shock sample.
We also have to consider that other mechanisms can produce shocks in the ISM (e.g. bars: \citealt{Athanassoula92}, nuclear spirals: \citealt{Kim17}), so it is possible that some of the nebulae are shock ionised by these kinds of processes.
For example, Fig.~\ref{fig:NGC1300_map} clearly shows some spiral structures outside of the circumnuclear ring where the vast majority of the nebulae are classified as shocks.

While it is clear that this analysis cannot give us a picture as complete as we have for the other types of nebulae, it still suggests that we are recovering a sample of SNRs which is reasonably comparable with what is found by other works in the literature, both in size of the sample (for each galaxy) and \Ha\ luminosity distribution, even though there might be some source of contamination (bright DIG clouds in particular).
A detailed analysis of our shock or SNRs sample will definitely be interesting since the literature on these objects is relatively scarce, but it is outside the scope of this work.
An ongoing work will construct tailored SNR catalogues using classical methods (Li et al. in prep) and will be well suited for comparison with our catalogue of shock-dominated objects.

\begin{table}[]
    \centering
    \caption{SNRs included in our sample per each galaxy. \ngc1365 is not shown here, since we expect its sample to be heavily contaminated.}
    \label{tab:snr_table}
    \begin{tabular}{lcccc}
    \hline\hline
    Galaxy & Full &\Ha & OIII& BPT\\
    \hline
    IC 5332 & 184 &142 &139& 101 \\
    \ngc0628 & 344 &197 &194 &111 \\
    \ngc1087 & 131 &98 &97 &73 \\
    \ngc1300 & 117 &98 &88 &45 \\
    \ngc1385 & 136 &97 &104 &77 \\
    \ngc1433 & 349 &276 &254 &119 \\
    \ngc1512 & 121 &89 &93 &33 \\
    \ngc1566 & 496 &302 &324 &156 \\
    \ngc1672 & 210 &128 &123 &74 \\
    \ngc2835 & 286 &214 &204 &157 \\
    \ngc3351 & 205 &149 &140 &67 \\
    \ngc3627 & 349 &206 &197 &83 \\
    \ngc4254 & 361 &207 &226 &115 \\
    \ngc4303 & 424 &213 &196 &99 \\
    \ngc4321 & 289 &134 &142 &56 \\
    \ngc4535 & 250 &160 &162 &73 \\
    \ngc5068 & 464 &383 &367 &304 \\
    \ngc7496 & 78&55 &55 &38 \\
    \hline
    \end{tabular}
\end{table}

\section{Conclusions}
\label{sec:conc}

This work presents a new catalogue of ionised nebulae in nearby galaxies, containing spectral (line fluxes, kinematics) and spatial (position, shape) information for over $40,000$ nebulae distributed across the 19 galaxies of the PHANGS-MUSE sample.
The nebulae have been classified using a new approach based on comparing their observed properties with grids of models representing the three main classes of ionised nebulae found in galaxies: \hii\ regions, PNe and shock-ionised nebulae.
This makes our catalogue the largest compilation of nebulae with a reliable classification and a uniform set of information (independent from their classification) available in the literature to date.

The algorithm exploits the principle of the odds ratio to assign to each nebula the probability it belongs to each one of the three considered classes.
Thanks to its comparative and probabilistic nature, the algorithm overcomes some of the main limitations characterising the criteria traditionally used in the literature to classify nebulae, such as their binarity, the sharp limits used for the classification and the limited amount of information that they rely on to perform the classification.
Given the large volume of data on ionised nebulae that will be produced in the near future by instruments like MUSE and surveys like the Local Volume Mapper survey, this algorithm is an important step toward more complex and robust automatic approaches to the classification of any type of nebulae.

The analysis of the catalogue and the comparison between the results of the model-comparison-based classification with what can be obtained from the traditional criteria reveals that:
\begin{itemize}
    \item The DIG emission can significantly alter the observed line ratios of the nebulae, particularly in the faintest ones where the DIG emission is dominant. A DIG correction is therefore essential to classify the nebulae correctly.
    \item The model-comparison-based classification algorithm works remarkably well for \hii\ regions, as shown by the analysis of the \hii\ regions LF in Sec.~\ref{sec:hii_lf}. Also, the agreement between the traditional criteria and the new approach is good for this class of objects, but the model-comparison-based algorithm can identify and classify \hii\ regions in areas of the BPT diagrams that are typically ignored by the traditional criteria or in competition with other classes of nebulae.
    \item A comparison between the \ebv\ measured via the Balmer decrement and that estimated by \izi\ during the evidence estimation for \hii\ regions revealed that the extinction is systematically overestimated by $\sim 0.05\,\si{mag}$ when applying the Balmer decrement technique to DIG corrected fluxes. The origin of this discrepancy arises from the theoretical \Ha$/$\Hb\ ratio assumed during the computation. The typical value for standard conditions (low-density limit, $T_e \sim 10000\,\si{K}$, Case B) \Ha$/$\Hb$=2.86$ is not representative of the true \Ha$/$\Hb\ ratio according to the model grid we are considering. A theoretical ratio of $3.03$ (Case B conditions, low-density limit, $T_e \sim 5200\,\si{K}$) seems to minimise the offset between the \ebv\ measured by \izi\ and that recovered via the Balmer decrement.
    \item We also observe that the agreement, and good performance of non-DIG corrected Balmer decrement extinction correction, seems to arise from a coincidence. The underestimation of the extinction due to DIG contamination and its overestimation due to adopting the wrong intrinsic Balmer ratio, produces biases of similar magnitudes but act in opposite directions, cancelling each other.
    \item The sample of PNe we recover seems to be incomplete when compared with other works in the literature studying the same galaxies \citep{Scheuermann22}. However, the issue seems to be mostly related to limitations of the source detection method, while the classification of the detected objects appears robust. 
    \item For 10 out of 19 galaxies, we were able to fit the PNLF, and the results are comparable with the literature, and \citet{Scheuermann22} in particular.
    \item We identify a rather large population of nebulae classified as shocks (10--15\% of the sample). Despite for a few contaminants, like AGN-ionised nebulae in a few galaxies of the sample bright DIG clouds, and dynamically shocked regions, we expect this to be the largest compilation of extragalactic SNRs available in the literature to date. This catalogue will open the possibility for a systematic and detailed study of this interesting, but still relatively poorly understood population of objects.
\end{itemize}

Concluding, we showed in this work that our model-comparison-based algorithm is an invaluable tool, that can significantly simplify and objectify the classification and characterisation of large samples of ionised nebulae. 
Moreover, with its number and variety of nebulae, our catalogue opens the doors to a wide variety of further studies, but also to the possibility to develop even more automatised, machine learning powered, identification and classification algorithms.

\begin{acknowledgements}
The authors of the paper want to thank the anonymous referee for their helpful suggestions.
Based on observations collected at the European Southern Observatory under ESO programmes 1100.B-0651, 095.C-0473, and 094.C- 0623 (PHANGS-MUSE; PI: Schinnerer), as well as 094.B-0321 (MAGNUM; PI: Marconi), 099.B-0242, 0100.B-0116, 098.B-0551 (MAD; PI: Carollo) and 097.B-0640 (TIMER; PI: Gadotti).
Science-level MUSE mosaicked datacubes and high-level analysis products are provided via the ESO archive phase 3 interface\footnote{\url{https://archive.eso.org/scienceportal/home?data_collection=PHANGS}}. 
A full description of the first PHANGS data release is presented in \cite{Emsellem22}.
This work was carried out as part of the PHANGS collaboration.
E.C. and G.A.B. acknowledge support from ANID Basal projects ACE210002 and FB210003.
K.K., F.S., and O.E. gratefully acknowledge funding from the German Research Foundation (DFG) in the form of an Emmy Noether Research Group (grant number KR4598/2-1, PI Kreckel). 
H.A.P. acknowledges support by the National Science and Technology Council of Taiwan under grant 110-2112-M-032-020-MY3.
S.C.O.G. acknowledges funding from the European Research Council via the ERC Synergy Grant ``ECOGAL -- Understanding our Galactic ecosystem: From the disk of the Milky Way to the formation sites of stars and planets'' (project ID 855130) 
and from the Heidelberg Cluster of Excellence (EXC 2181 - 390900948) ``STRUCTURES: A unifying approach to emergent phenomena in the physical world, mathematics, and complex data'', funded by the German Excellence Strategy. 
K.K., E.J.W. and S.C.O.G. acknowledge support from the Deutsche Forschungsgemeinschaft (DFG) via the Collaborative Research Center (SFB 881 -- 138713538) ``The Milky Way System'' (subprojects A1, B1, B2 and B8, P1 and P2). 
E.S. acknowledges funding from the European Research Council (ERC) under the European Union’s Horizon 2020 research and innovation programme (grant agreement No. 694343).
F.B. acknowledges funding from the European Research Council (ERC) under the European Union’s Horizon 2020 research and innovation programme (grant agreement No.726384/Empire)
The Starlink software \citep{Currie14} is currently supported by the East Asian Observatory.
This research made use of Astropy, a community-developed core Python package for Astronomy \citep{Astropy13, Astropy18}, Matplotlib \citep{Hunter07}, NumPy \citep{Harris20}, SciPy \citep{Virtanen20}, Scikit-learn \citep{Scikit11}.
\end{acknowledgements}

\bibliographystyle{aa} 
\bibliography{bibliography} 

\begin{appendix}
\onecolumn  
\section{Diagnostic diagrams with traditional classification}
\label{app:bpt_traditional}

\begin{figure*}[!ht]
\centering
\includegraphics[width=0.9\textwidth]{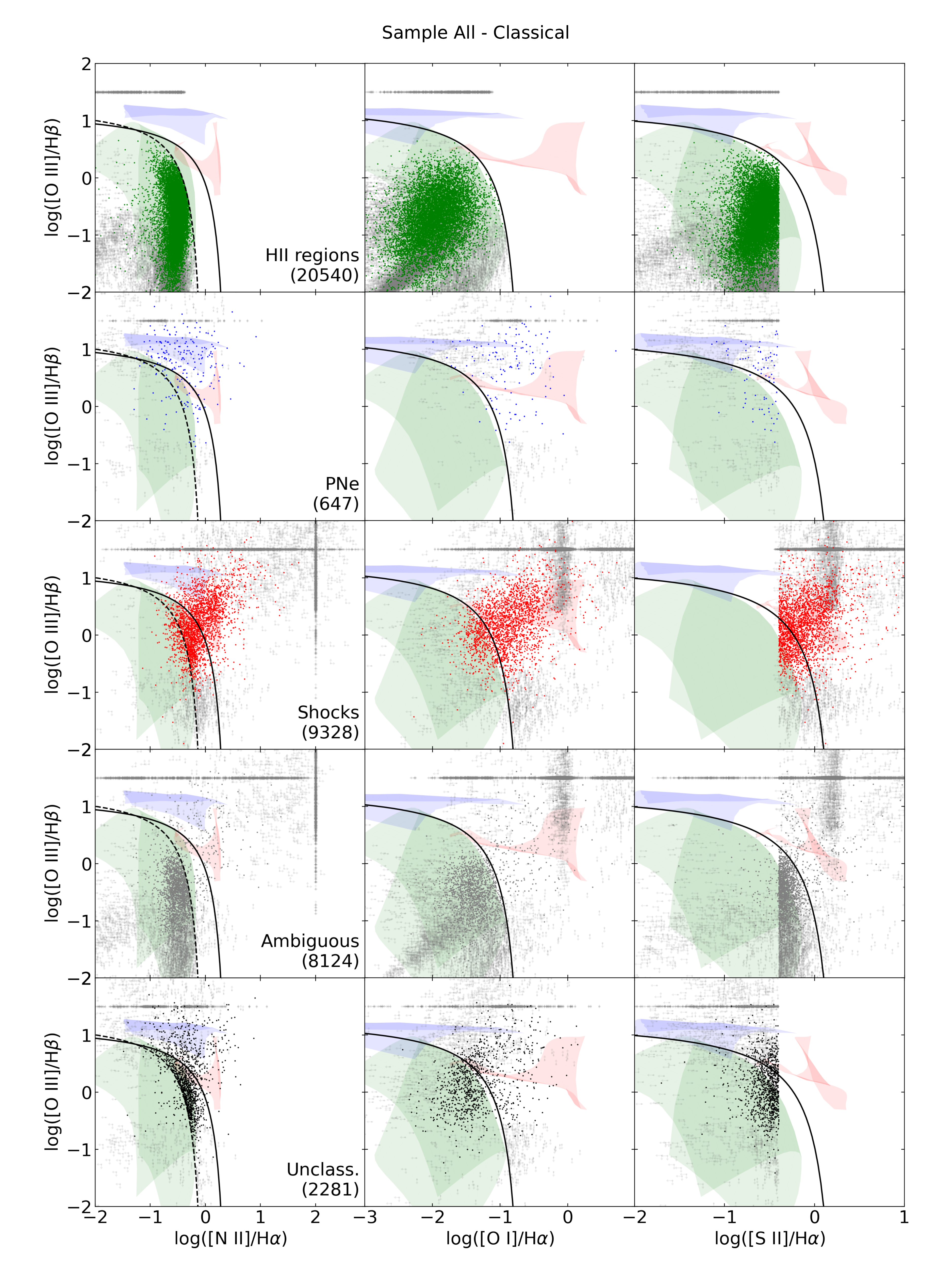}
\caption{Diagnostic diagrams from \citet{Baldwin81} and \citet{Veilleux87} for the full sample. Lines and shaded areas are defined as in Fig.~\ref{fig:bpt_models}. Here, the regions are classified according to the traditional classification criteria.
In each row only, from top to bottom, we show: \hii\ regions (green), PNe (blue), shocks (red), ambiguous (grey) and unclassified (black) objects.
}
\label{fig:bpt_full_trad}
\end{figure*}

\begin{figure*}[!htp]
\centering
\includegraphics[width=0.9\textwidth]{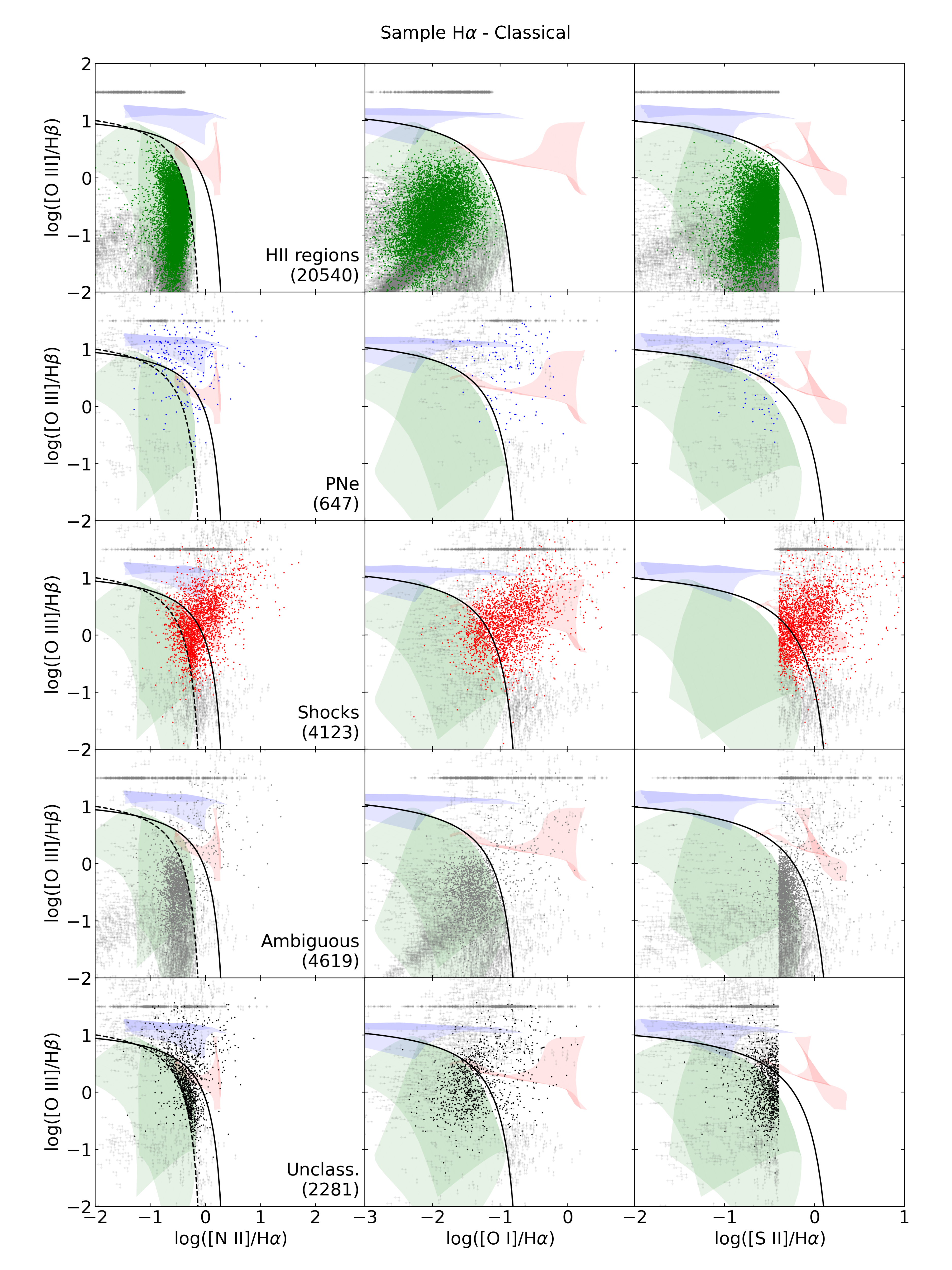}
\caption{Diagnostic diagrams from \citet{Baldwin81} and \citet{Veilleux87} for the \Ha\ sample. Colours and symbols are as in Fig.\,\ref{fig:bpt_full_trad}.}
\label{fig:bpt_ha_trad}
\end{figure*}

\begin{figure*}[!htp]
\centering
\includegraphics[width=0.9\textwidth]{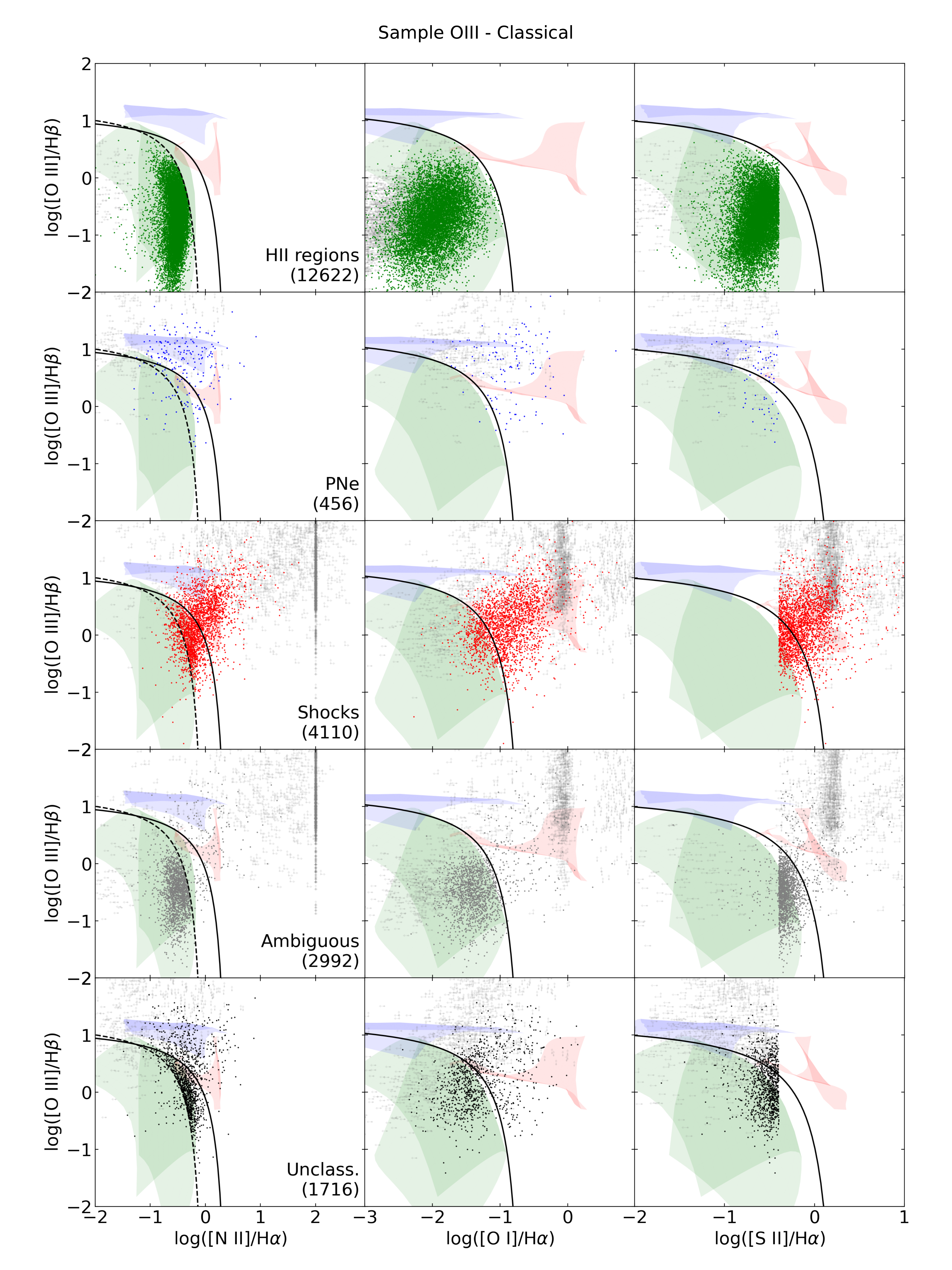}
\caption{Diagnostic diagrams from \citet{Baldwin81} and \citet{Veilleux87} for the OIII sample. Colours and symbols are as in Fig.\,\ref{fig:bpt_full_trad}.}
\label{fig:bpt_o3_trad}
\end{figure*}

\begin{figure*}[!htp]
\centering
\includegraphics[width=0.9\textwidth]{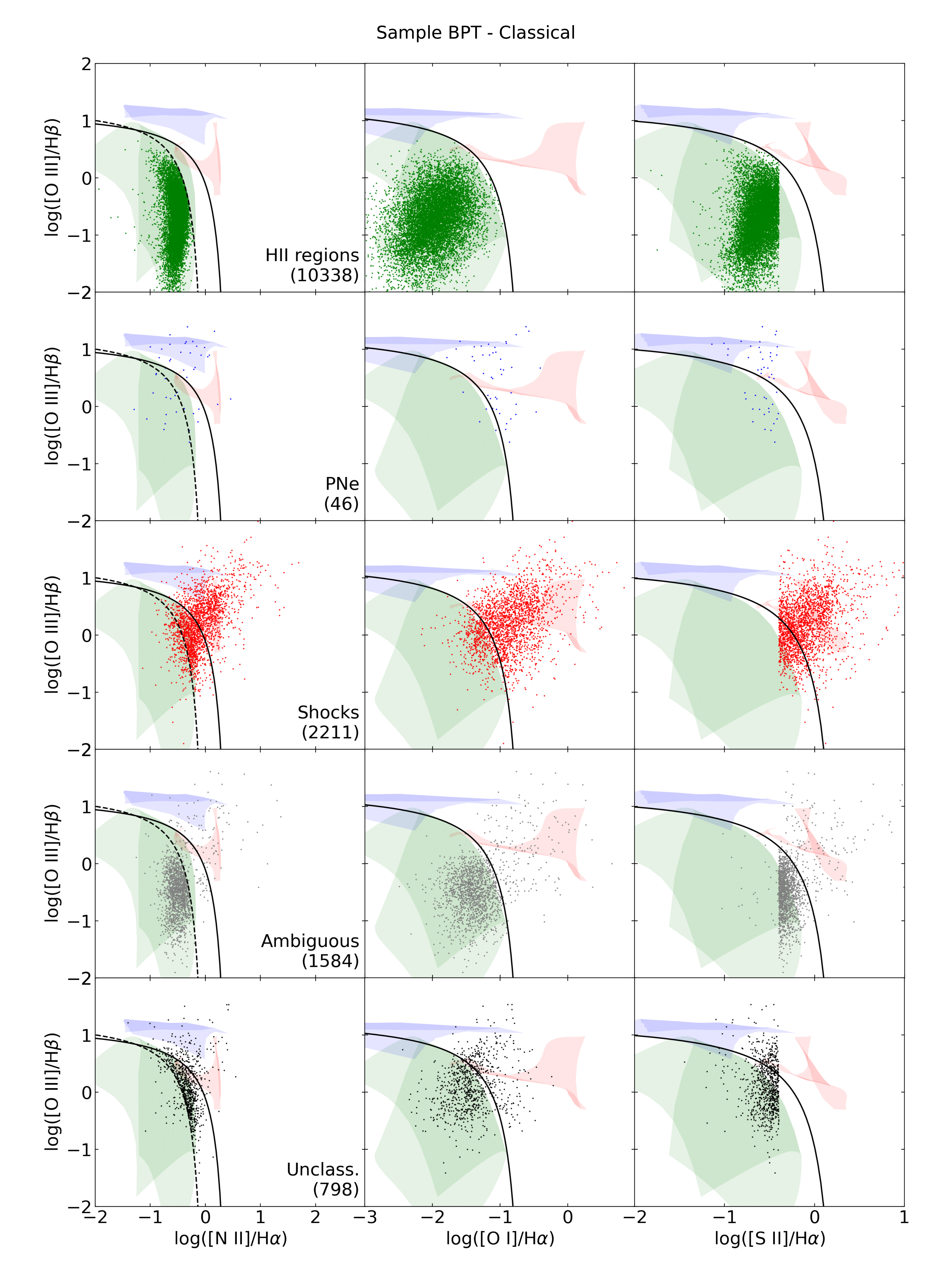}
\caption{Diagnostic diagrams from \citet{Baldwin81} and \citet{Veilleux87} for the BPT sample. Colours and symbols are as in Fig.\,\ref{fig:bpt_full_trad}.}
\label{fig:bpt_bpt_trad}
\end{figure*}
\FloatBarrier

\section{Atlas of nebulae}
\label{sec:atlas}

\begin{figure*}
\centering
\includegraphics[width=0.94\textwidth]{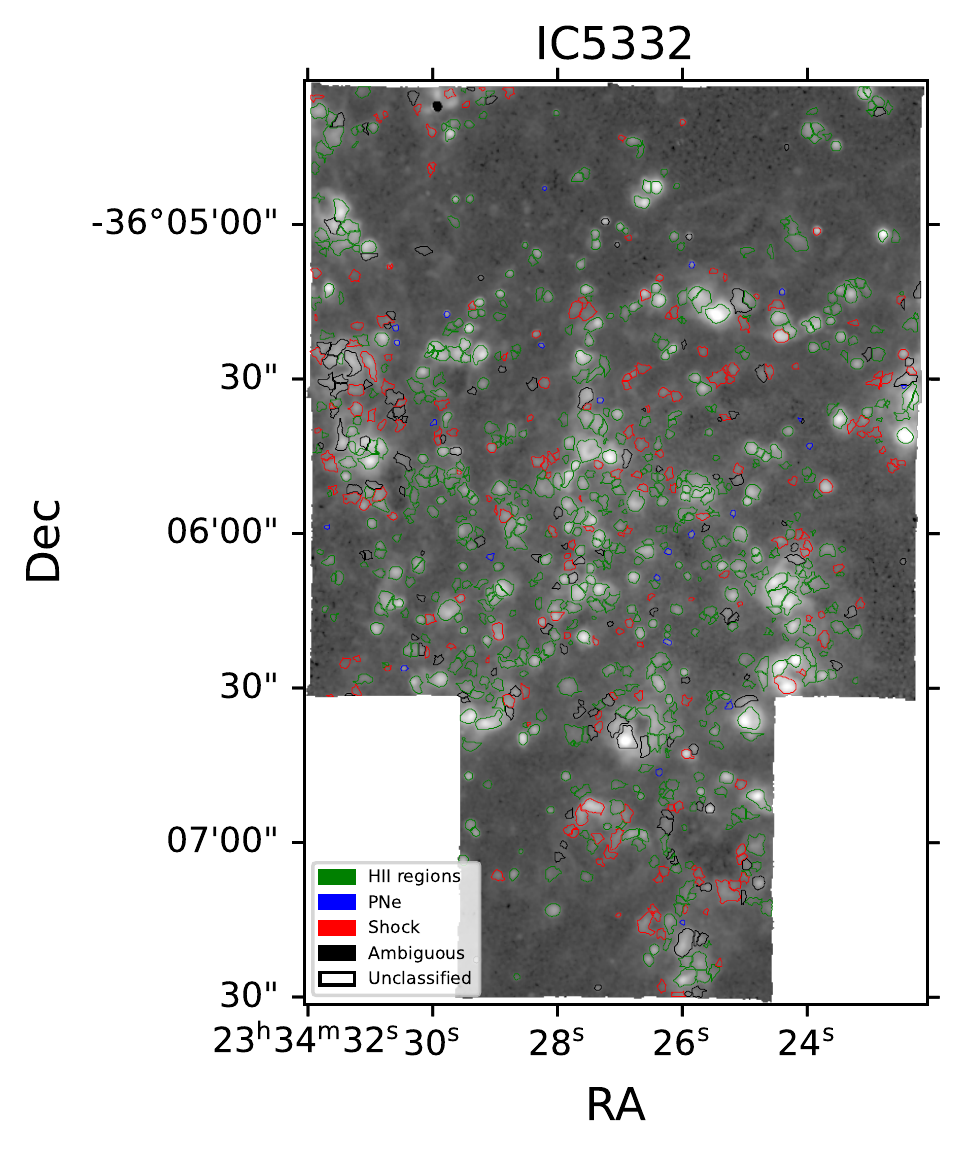}
\caption{Boundaries of the nebulae detected in IC 5332 superimposed on the detection map for the galaxy. The colour of the contour represents the classification of the nebula according to our model-comparison-based algorithm.}
\label{fig:IC5332_map}
\end{figure*}

\begin{figure*}
\centering
\includegraphics[width=0.94\textwidth]{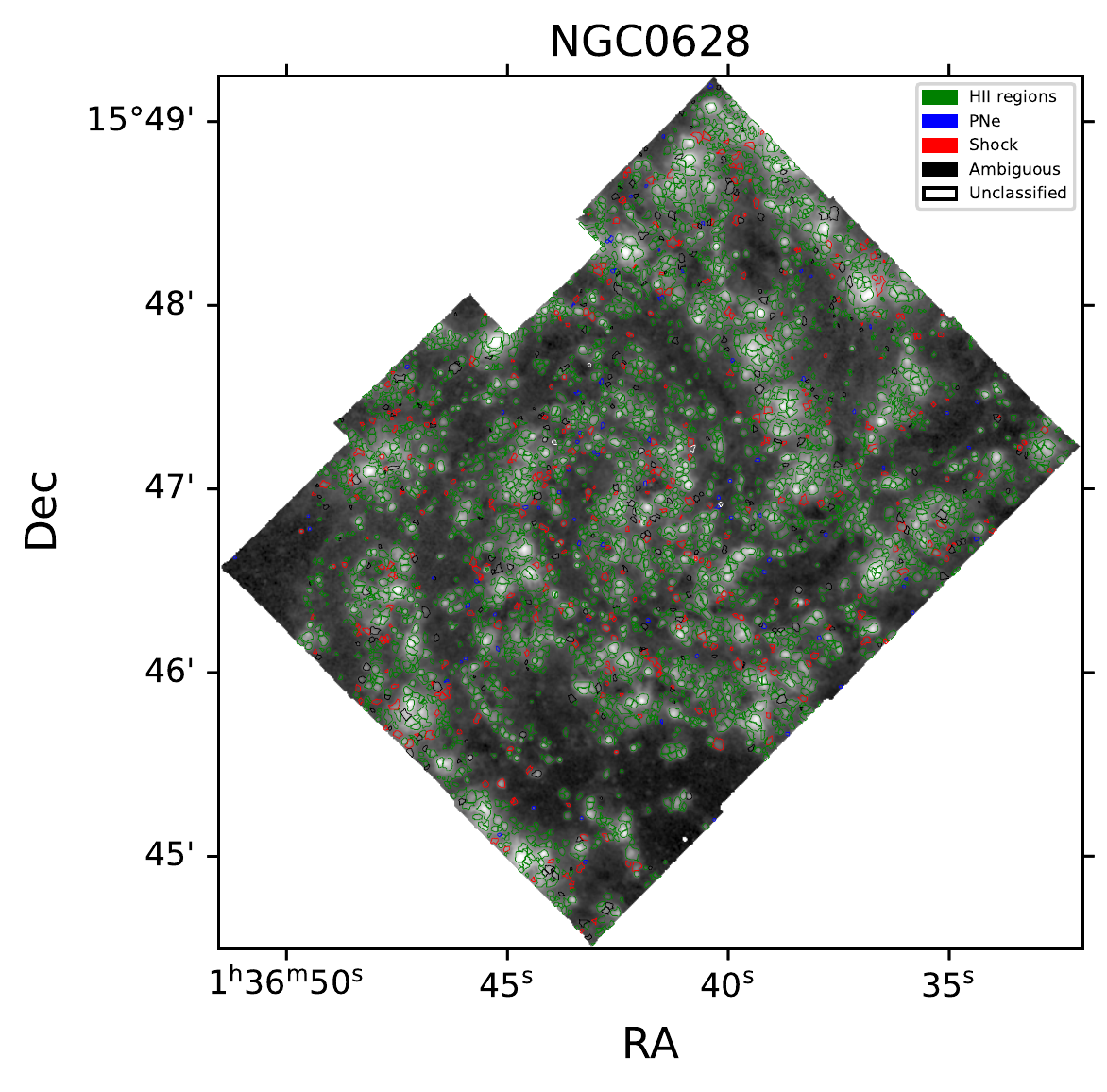}
\caption{Boundaries of the nebulae detected in \ngc0628 superimposed on the detection map for the galaxy. The colour of the contour represents the classification of the nebula according to our model-comparison-based algorithm.}
\label{fig:NGC0628_map}
\end{figure*}

\begin{figure*}
\centering
\includegraphics[width=0.94\textwidth]{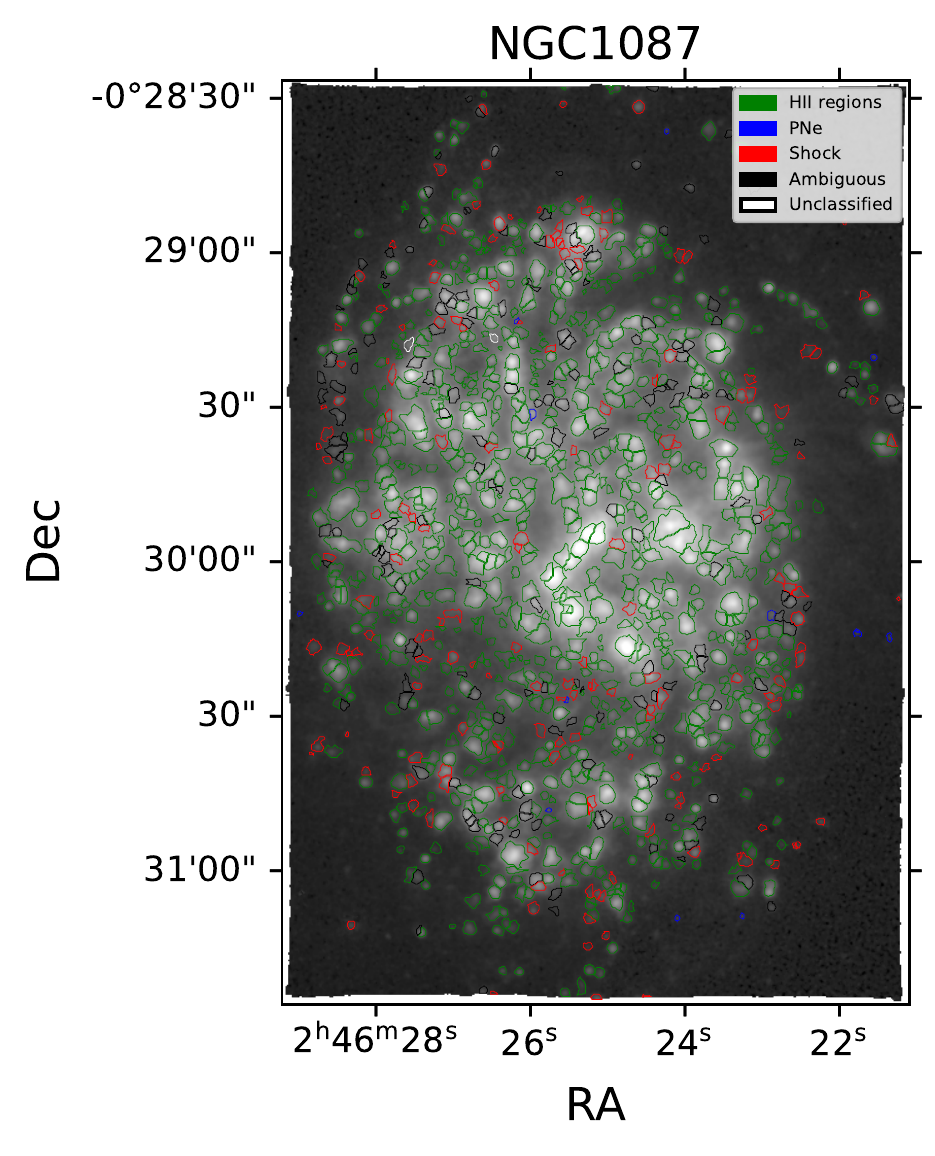}
\caption{Boundaries of the nebulae detected in \ngc1087 superimposed on the detection map for the galaxy. The colour of the contour represents the classification of the nebula according to our model-comparison-based algorithm.}
\label{fig:NGC1087_map}
\end{figure*}

\begin{figure*}
\centering
\includegraphics[width=0.94\textwidth]{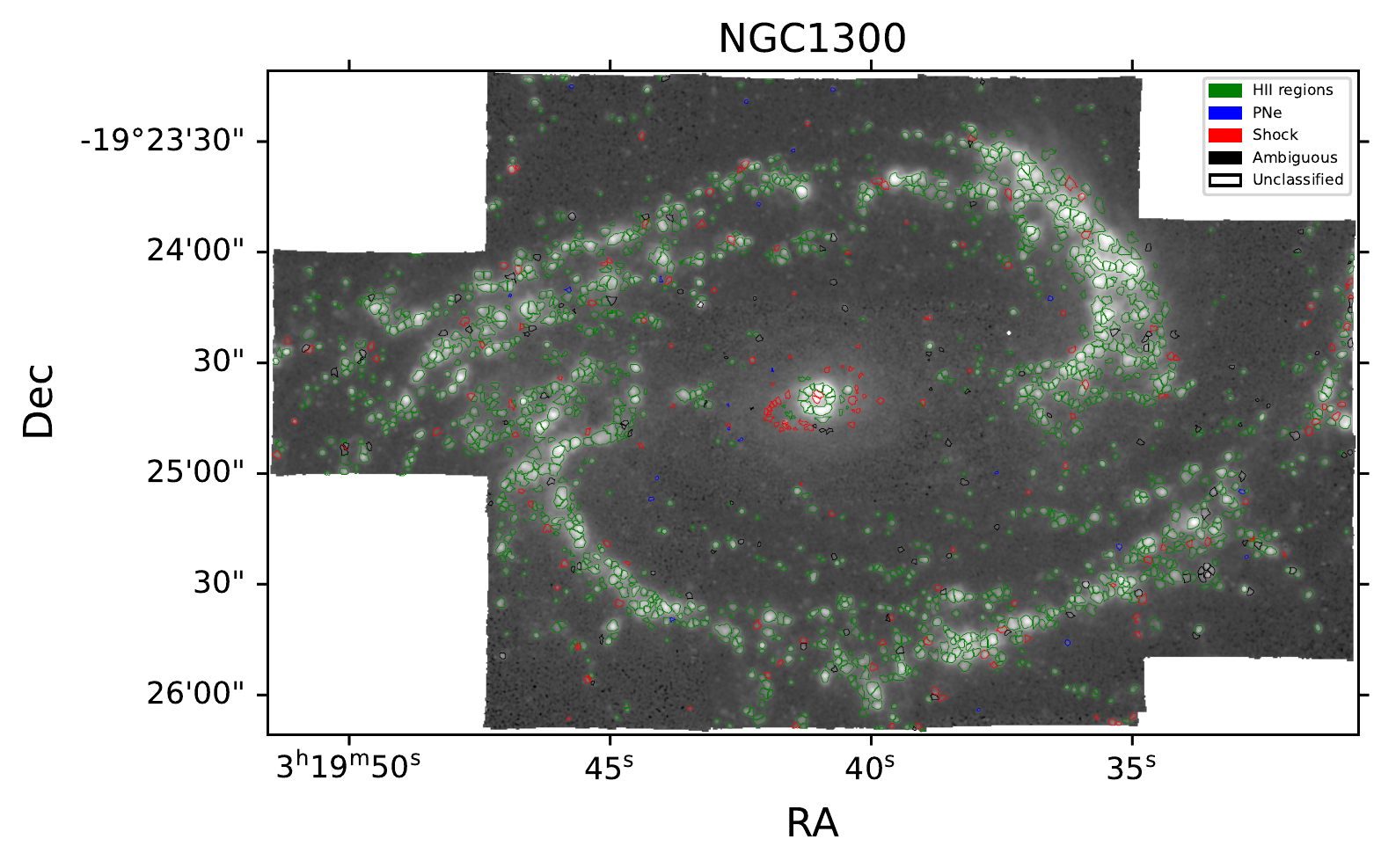}
\caption{Boundaries of the nebulae detected in \ngc1300 superimposed on the detection map for the galaxy. The colour of the contour represents the classification of the nebula according to our model-comparison-based algorithm.}
\label{fig:NGC1300_map}
\end{figure*}

\begin{figure*}
\centering
\includegraphics[width=0.94\textwidth]{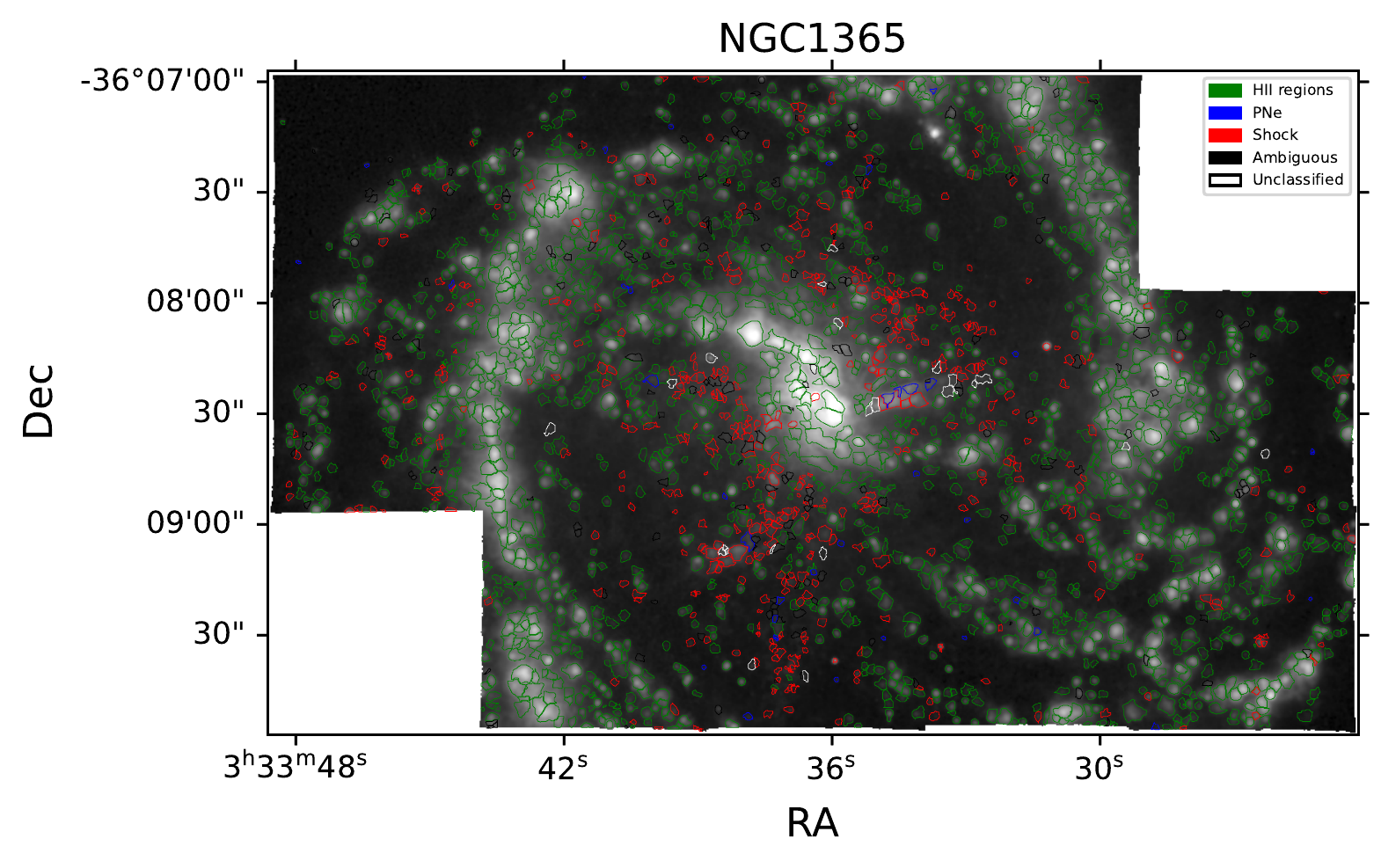}
\caption{Boundaries of the nebulae detected in \ngc1365 superimposed on the detection map for the galaxy. The colour of the contour represents the classification of the nebula according to our model-comparison-based algorithm.}
\label{fig:NGC1365_map}
\end{figure*}

\begin{figure*}
\centering
\includegraphics[width=0.94\textwidth]{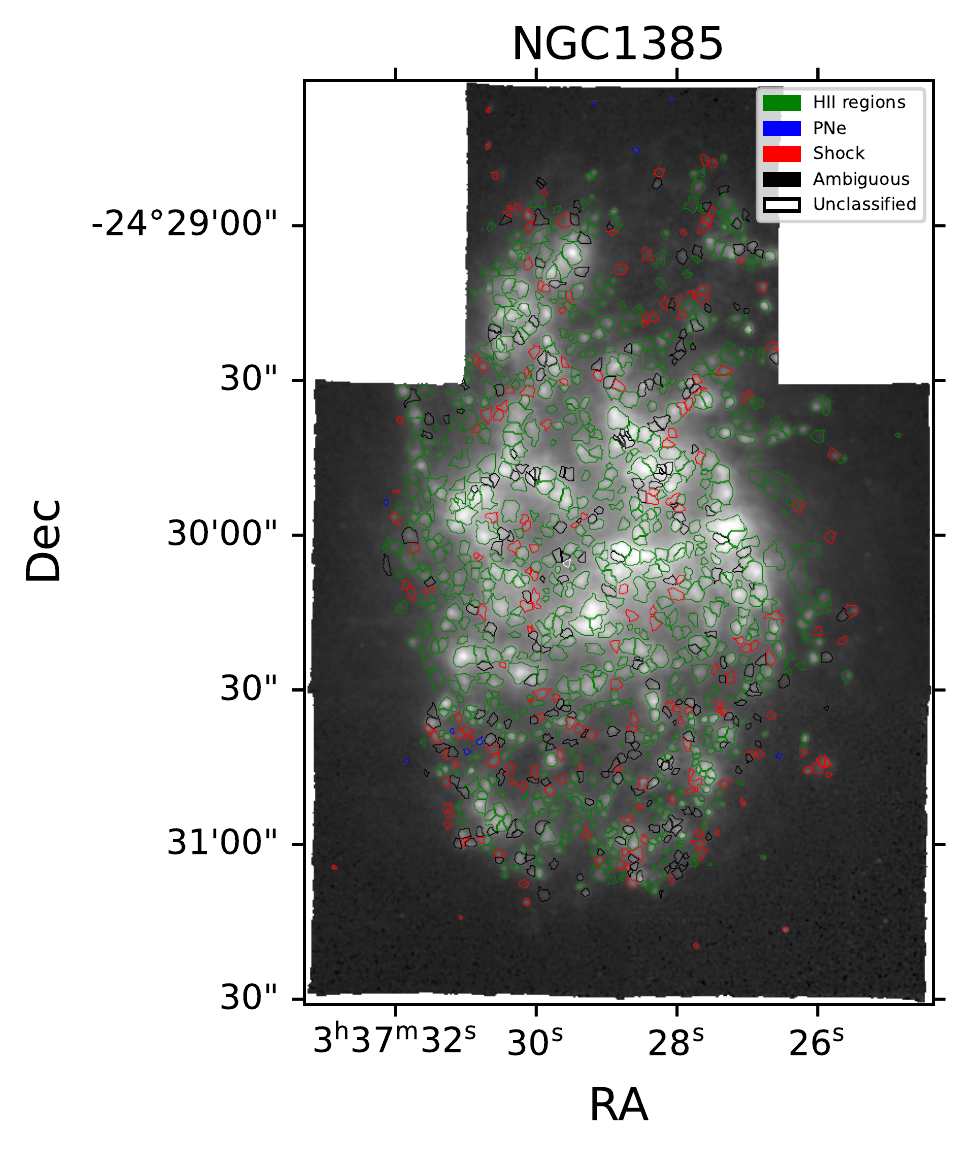}
\caption{Boundaries of the nebulae detected in \ngc1385 superimposed on the detection map for the galaxy. The colour of the contour represents the classification of the nebula according to our model-comparison-based algorithm.}
\label{fig:NGC1385_map}
\end{figure*}

\begin{figure*}
\centering
\includegraphics[width=0.94\textwidth]{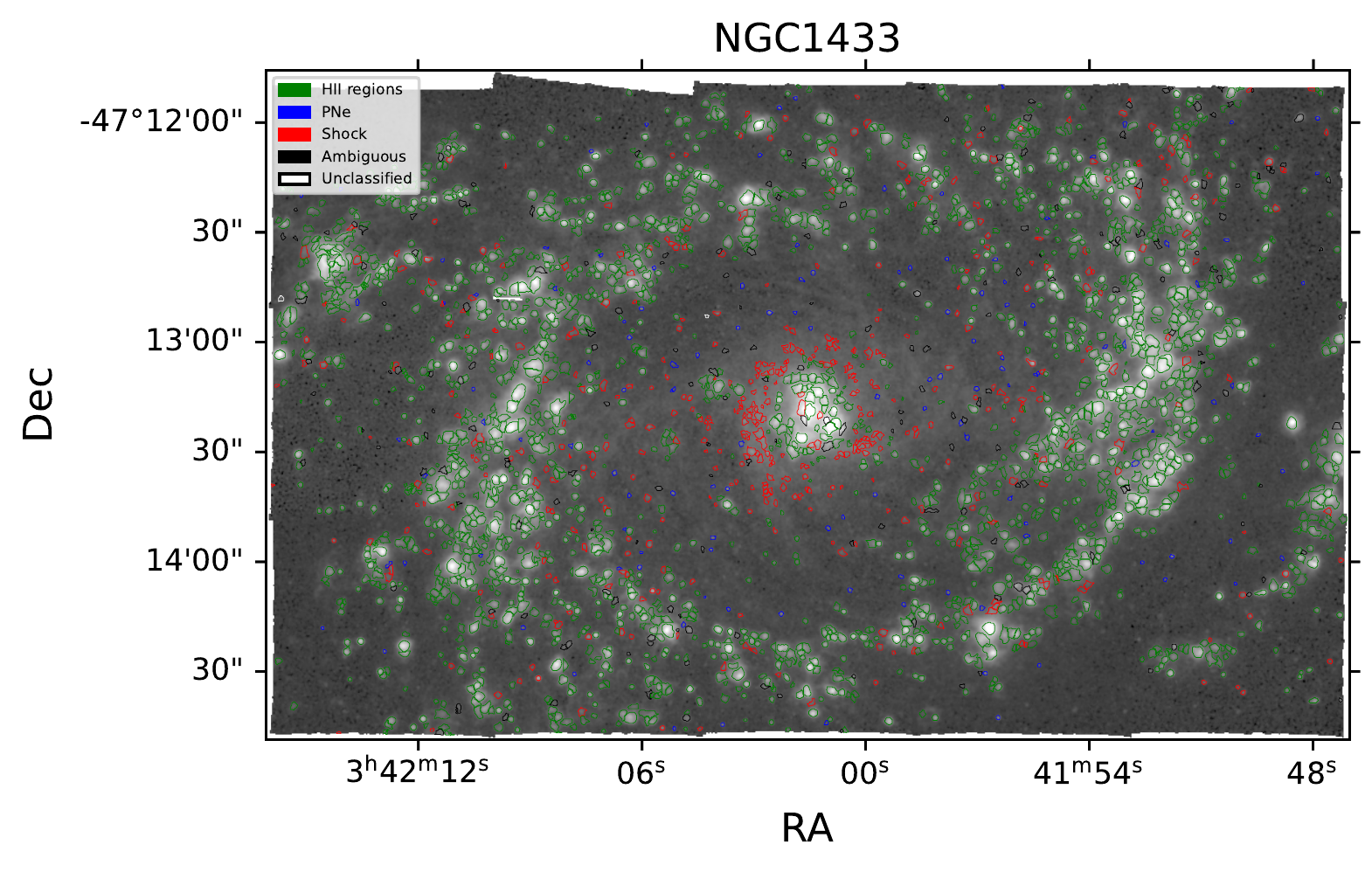}
\caption{Boundaries of the nebulae detected in \ngc1433 superimposed on the detection map for the galaxy. The colour of the contour represents the classification of the nebula according to our model-comparison-based algorithm.}
\label{fig:NGC1433_map}
\end{figure*}

\begin{figure*}
\centering
\includegraphics[width=0.94\textwidth]{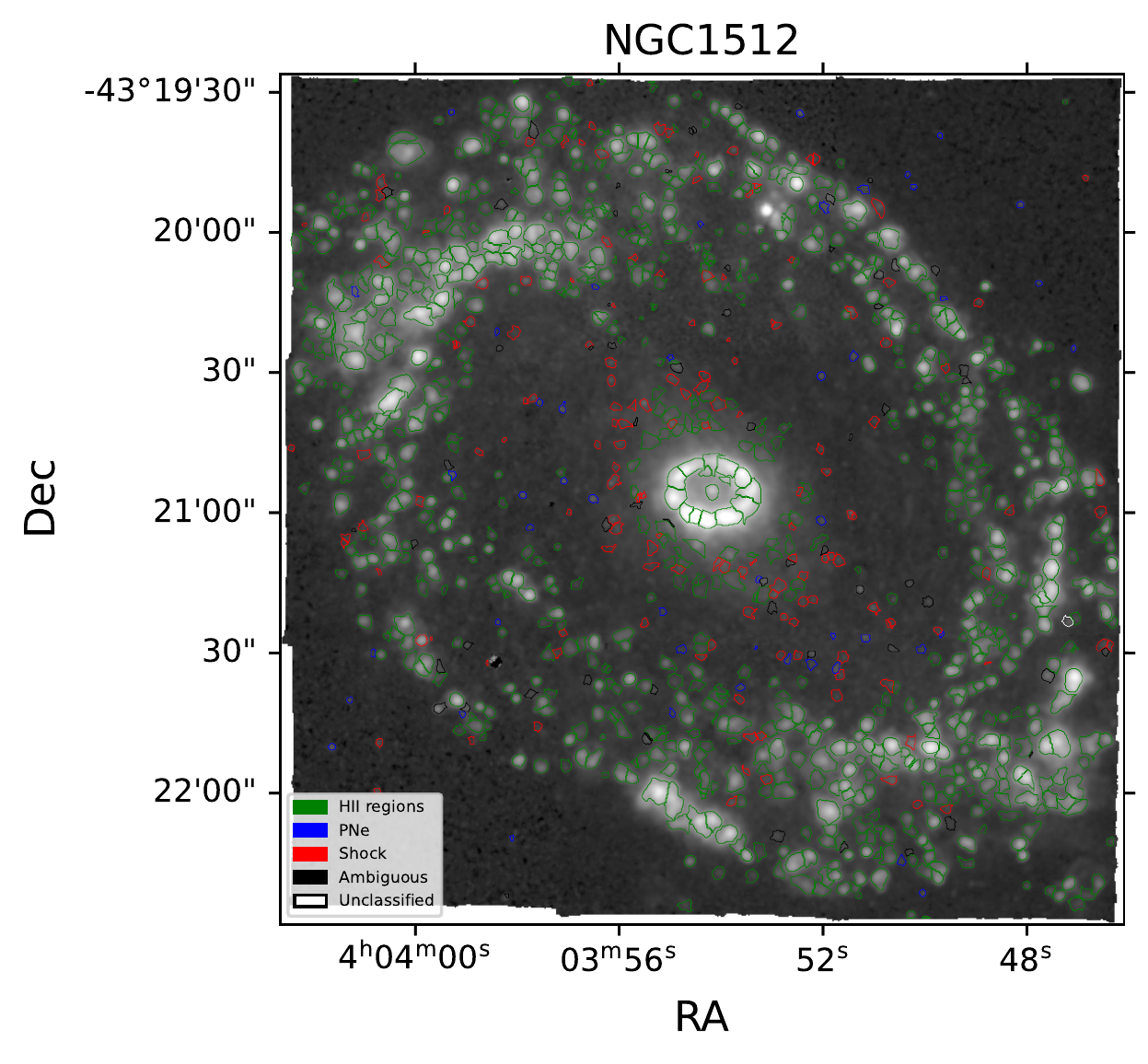}
\caption{Boundaries of the nebulae detected in \ngc1512 superimposed on the detection map for the galaxy. The colour of the contour represents the classification of the nebula according to our model-comparison-based algorithm.}
\label{fig:NGC1512_map}
\end{figure*}

\begin{figure*}
\centering
\includegraphics[width=0.94\textwidth]{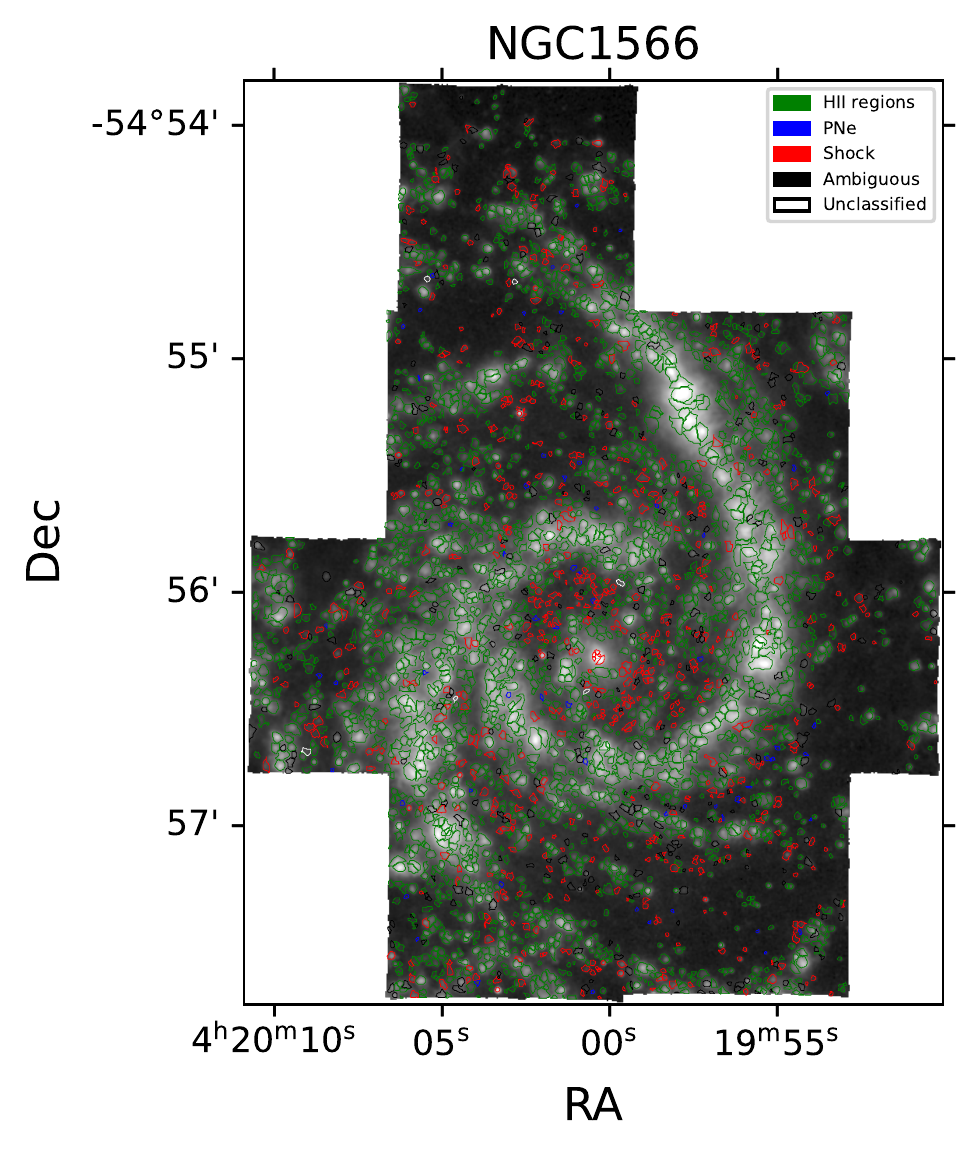}
\caption{Boundaries of the nebulae detected in \ngc1566 superimposed on the detection map for the galaxy. The colour of the contour represents the classification of the nebula according to our model-comparison-based algorithm.}
\label{fig:NGC1566_map}
\end{figure*}

\begin{figure*}
\centering
\includegraphics[width=0.94\textwidth]{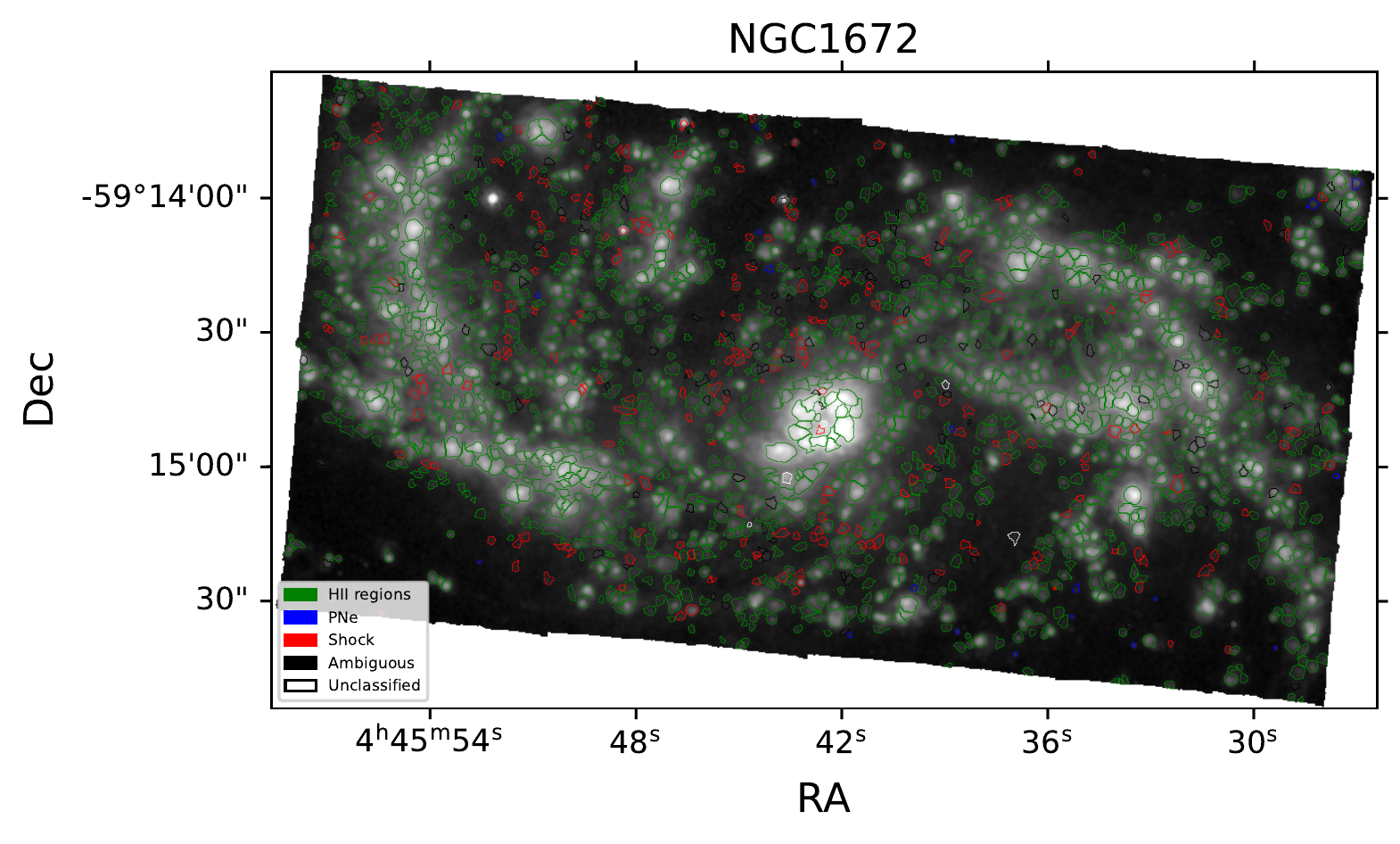}
\caption{Boundaries of the nebulae detected in \ngc1672 superimposed on the detection map for the galaxy. The colour of the contour represents the classification of the nebula according to our model-comparison-based algorithm.}
\label{fig:NGC1672_map}
\end{figure*}

\begin{figure*}
\centering
\includegraphics[width=0.94\textwidth]{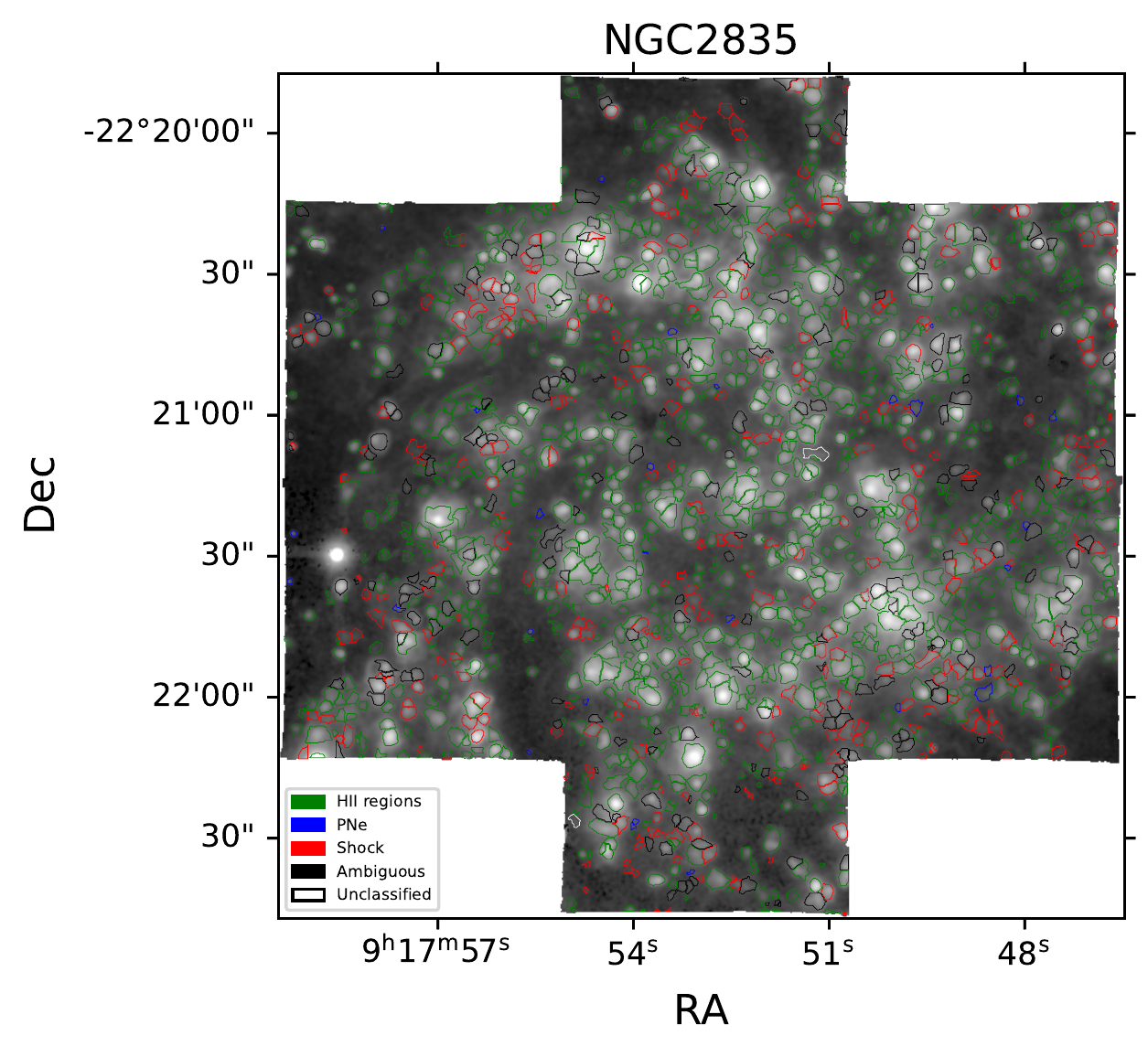}
\caption{Boundaries of the nebulae detected in \ngc2835 superimposed on the detection map for the galaxy. The colour of the contour represents the classification of the nebula according to our model-comparison-based algorithm.}
\label{fig:NGC2835_map}
\end{figure*}

\begin{figure*}
\centering
\includegraphics[width=0.94\textwidth]{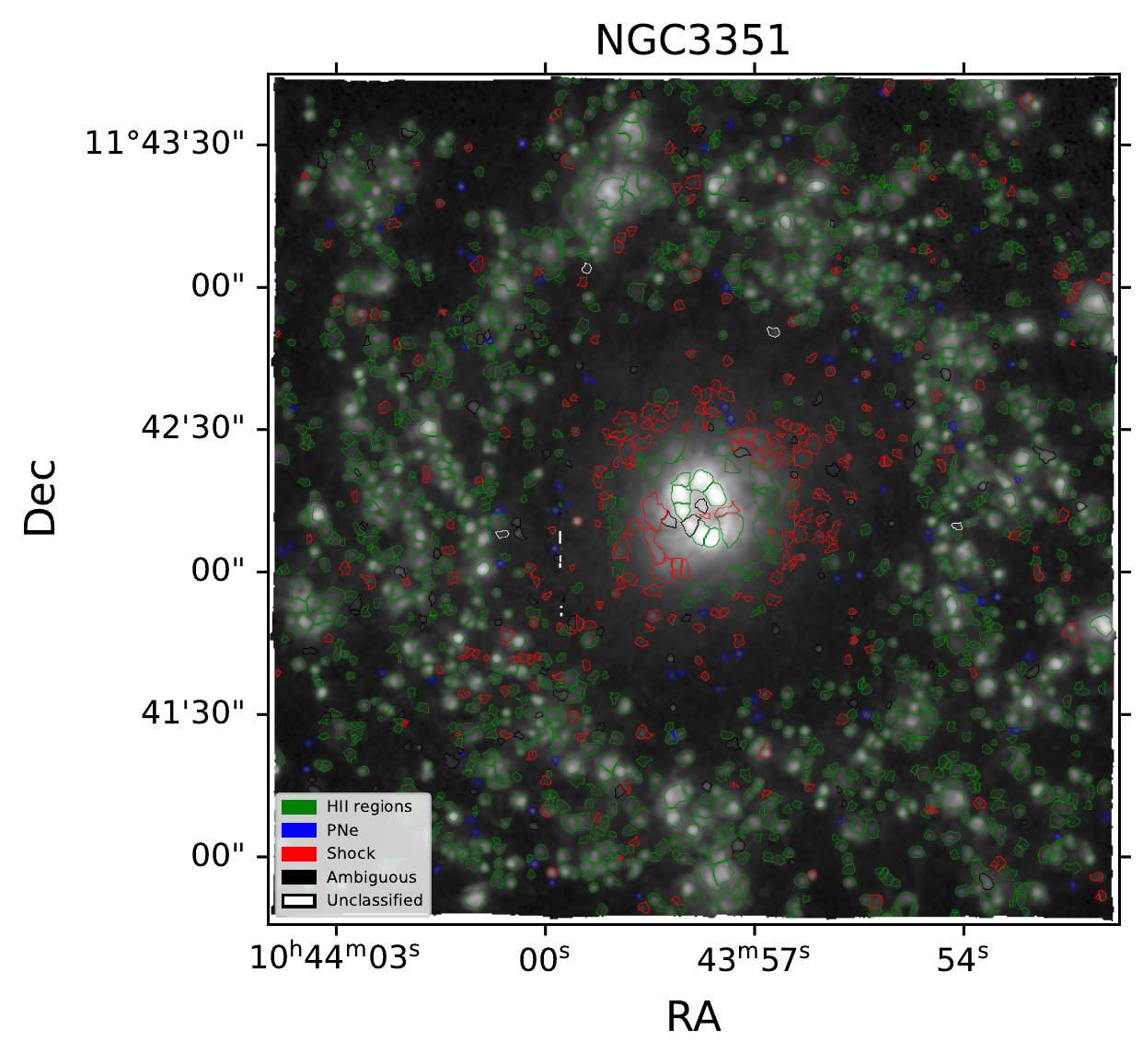}
\caption{Boundaries of the nebulae detected in \ngc3351 superimposed on the detection map for the galaxy. The colour of the contour represents the classification of the nebula according to our model-comparison-based algorithm.}
\label{fig:NGC3351_map}
\end{figure*}

\begin{figure*}
\centering
\includegraphics[width=0.94\textwidth]{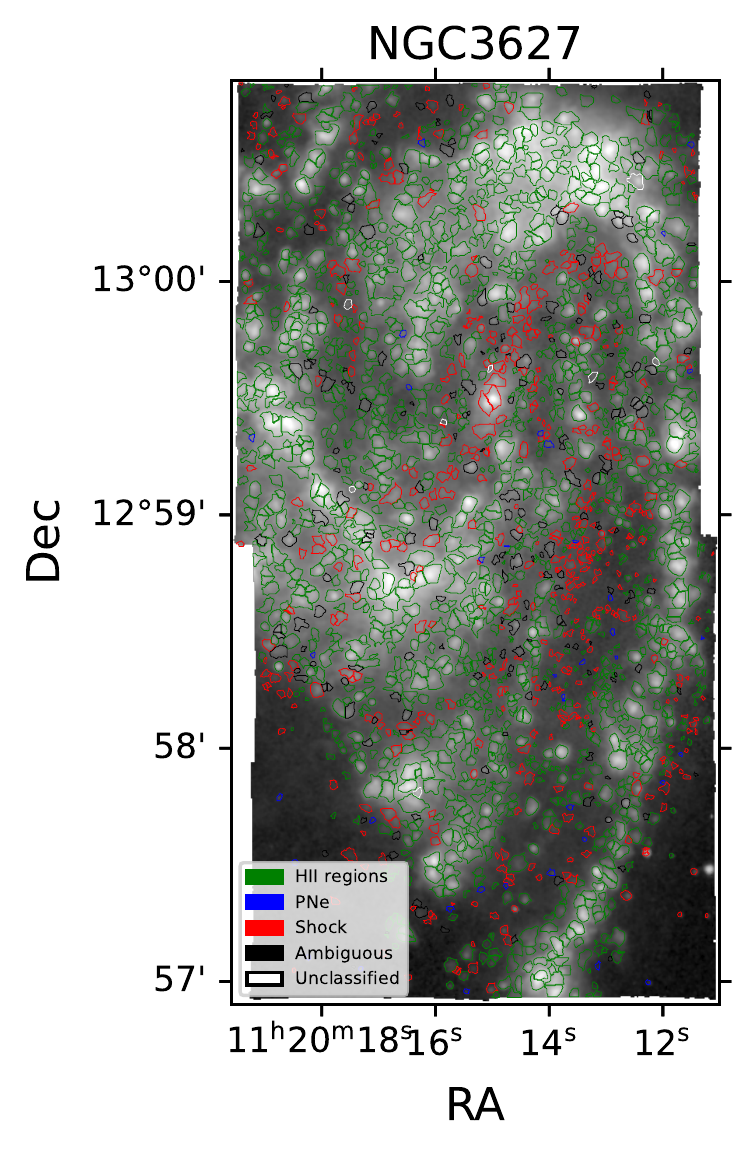}
\caption{Boundaries of the nebulae detected in \ngc3627 superimposed on the detection map for the galaxy. The colour of the contour represents the classification of the nebula according to our model-comparison-based algorithm.}
\label{fig:NGC3627_map}
\end{figure*}

\begin{figure*}
\centering
\includegraphics[width=0.94\textwidth]{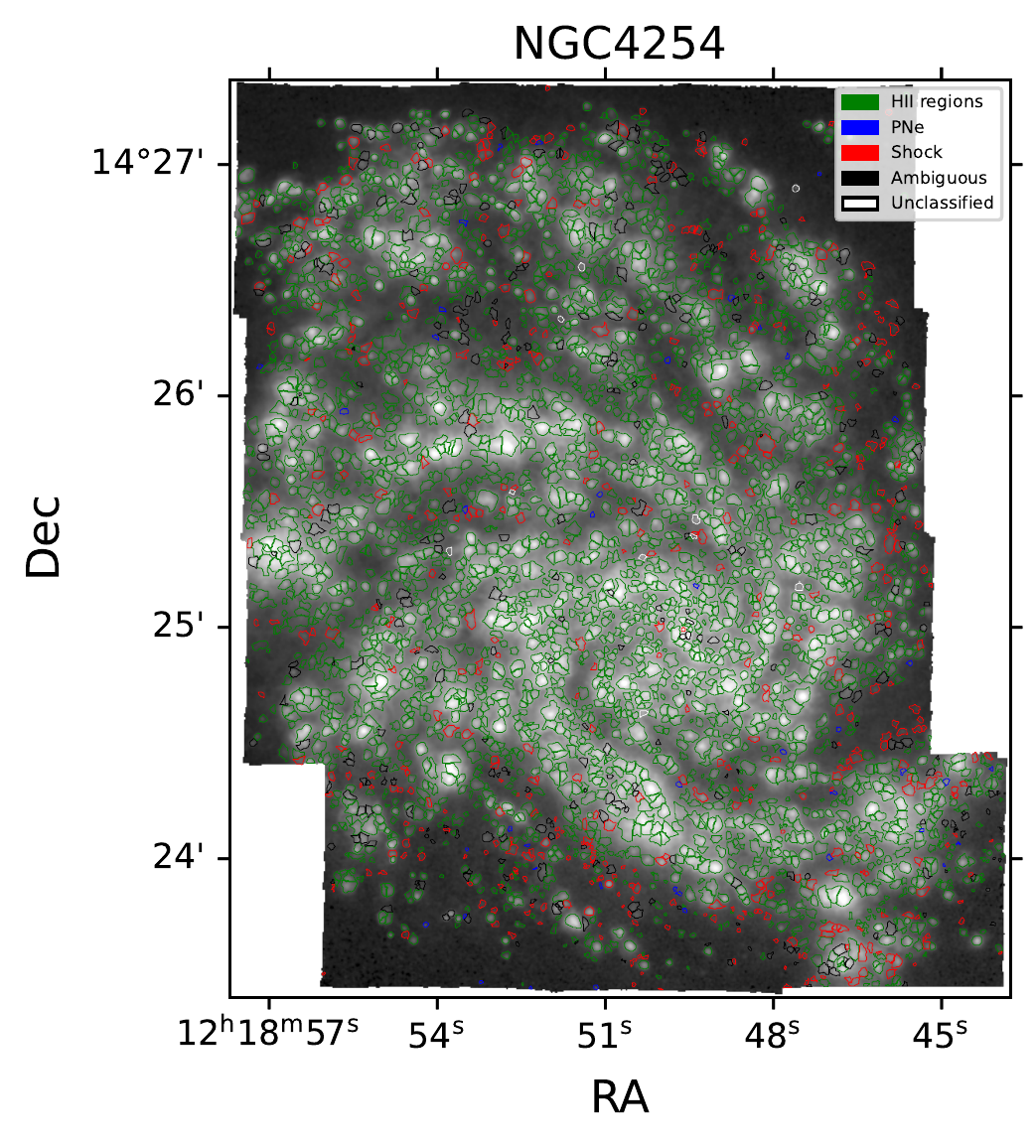}
\caption{Boundaries of the nebulae detected in \ngc4254 superimposed on the detection map for the galaxy. The colour of the contour represents the classification of the nebula according to our model-comparison-based algorithm.}
\label{fig:NGC4254_map}
\end{figure*}

\begin{figure*}
\centering
\includegraphics[width=0.94\textwidth]{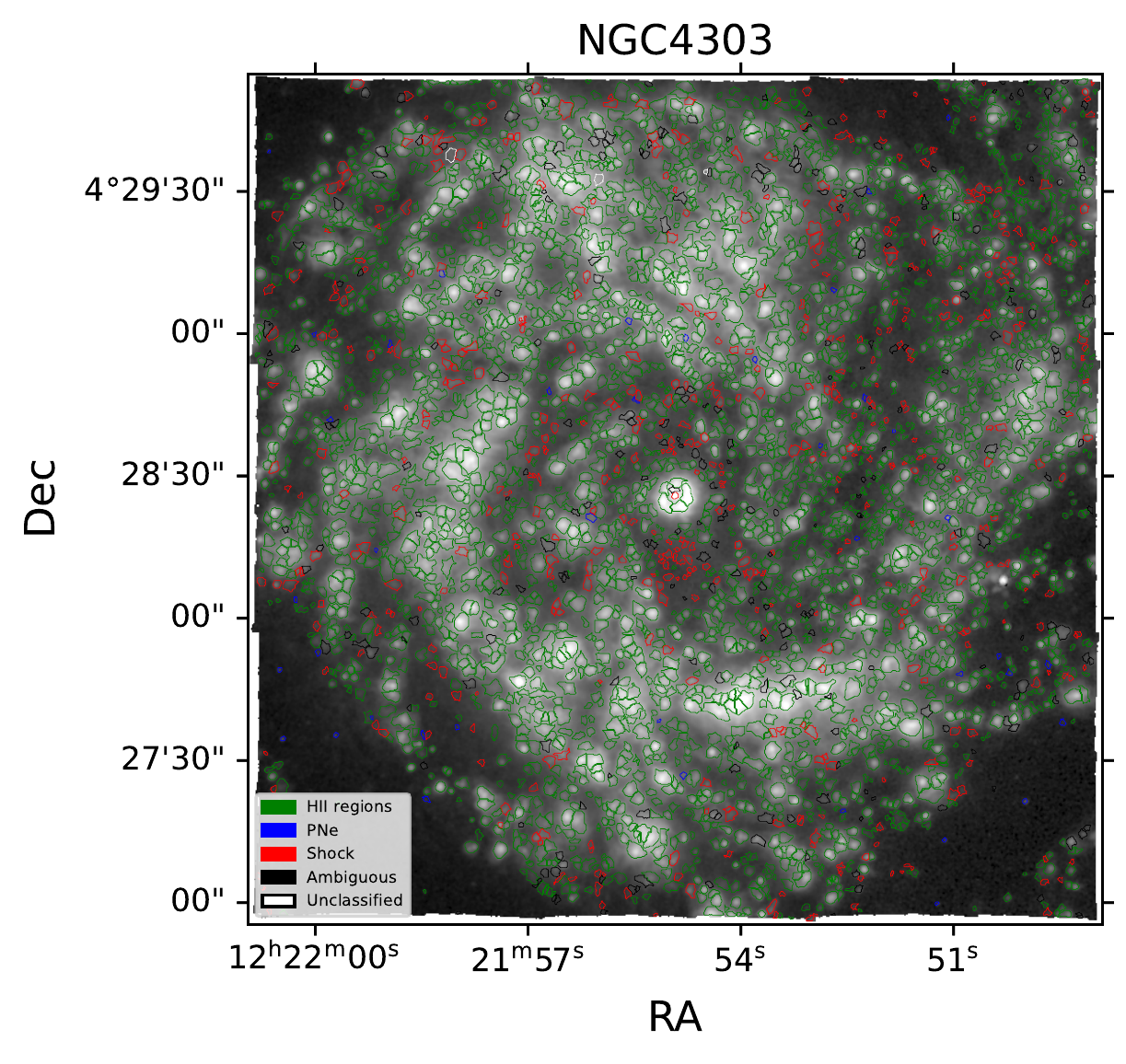}
\caption{Boundaries of the nebulae detected in \ngc4303 superimposed on the detection map for the galaxy. The colour of the contour represents the classification of the nebula according to our model-comparison-based algorithm.}
\label{fig:NGC4303_map}
\end{figure*}

\begin{figure*}
\centering
\includegraphics[width=0.94\textwidth]{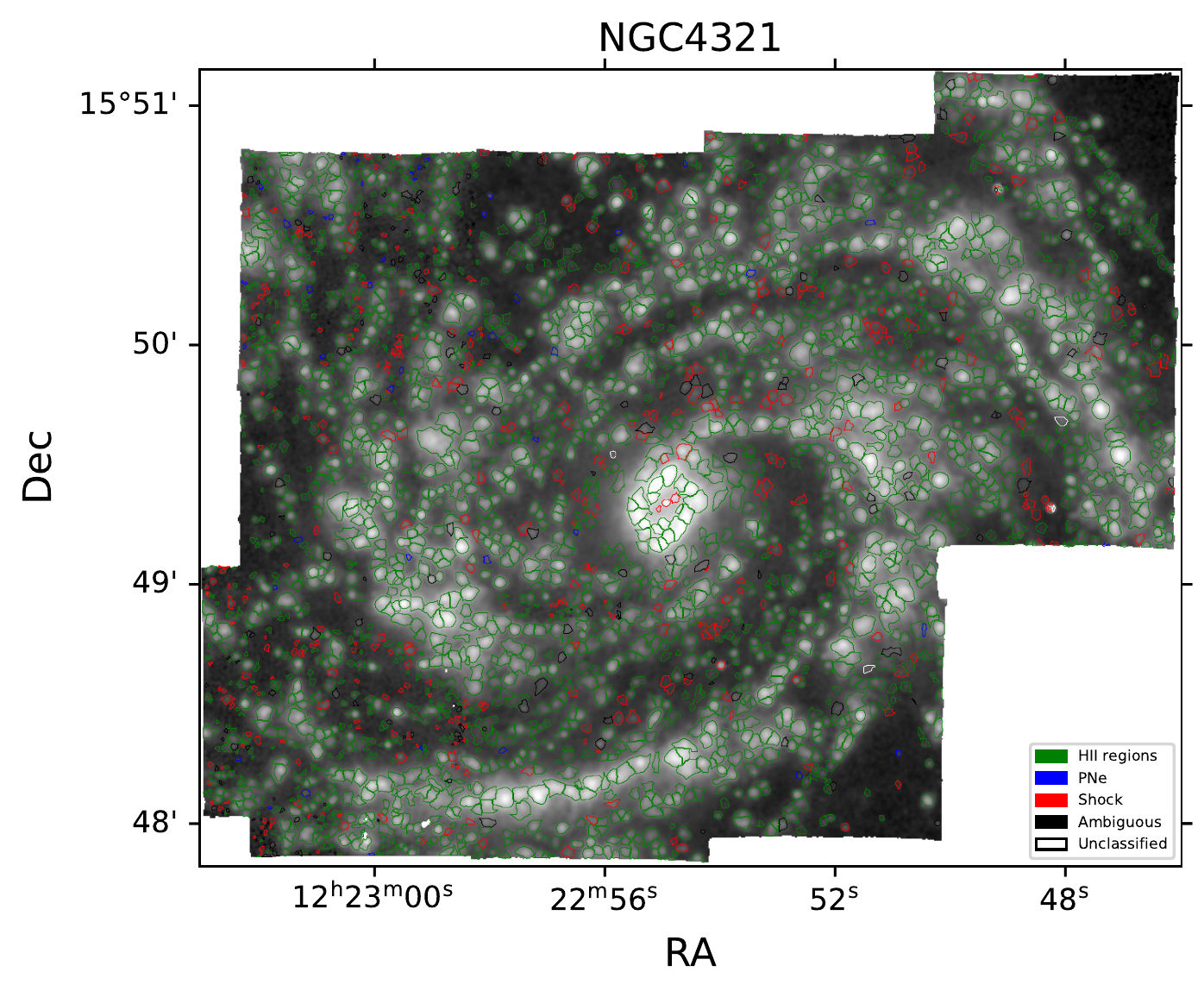}
\caption{Boundaries of the nebulae detected in \ngc4321 superimposed on the detection map for the galaxy. The colour of the contour represents the classification of the nebula according to our model-comparison-based algorithm.}
\label{fig:NGC4321_map}
\end{figure*}

\begin{figure*}
\centering
\includegraphics[width=0.94\textwidth]{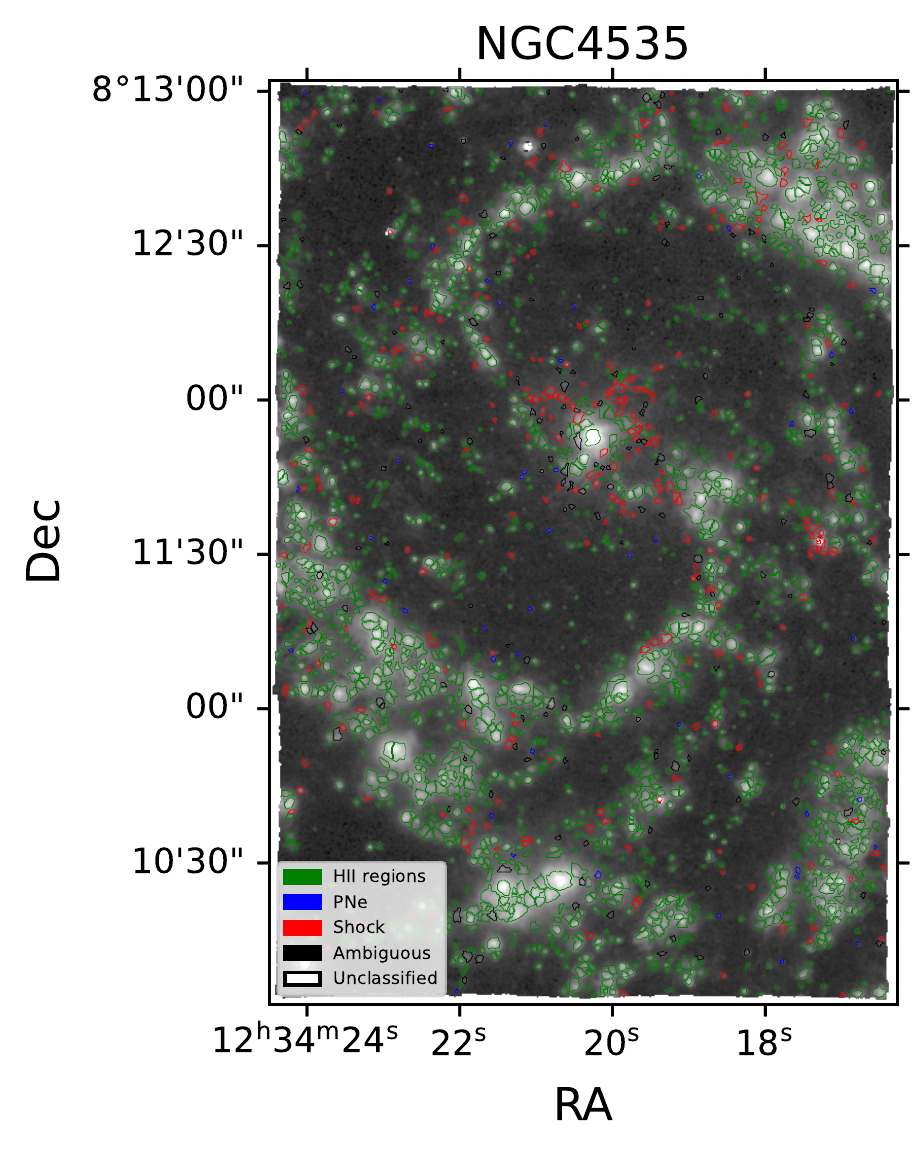}
\caption{Boundaries of the nebulae detected in \ngc4535 superimposed on the detection map for the galaxy. The colour of the contour represents the classification of the nebula according to our model-comparison-based algorithm.}
\label{fig:NGC4535_map}
\end{figure*}

\begin{figure*}
\centering
\includegraphics[width=0.94\textwidth]{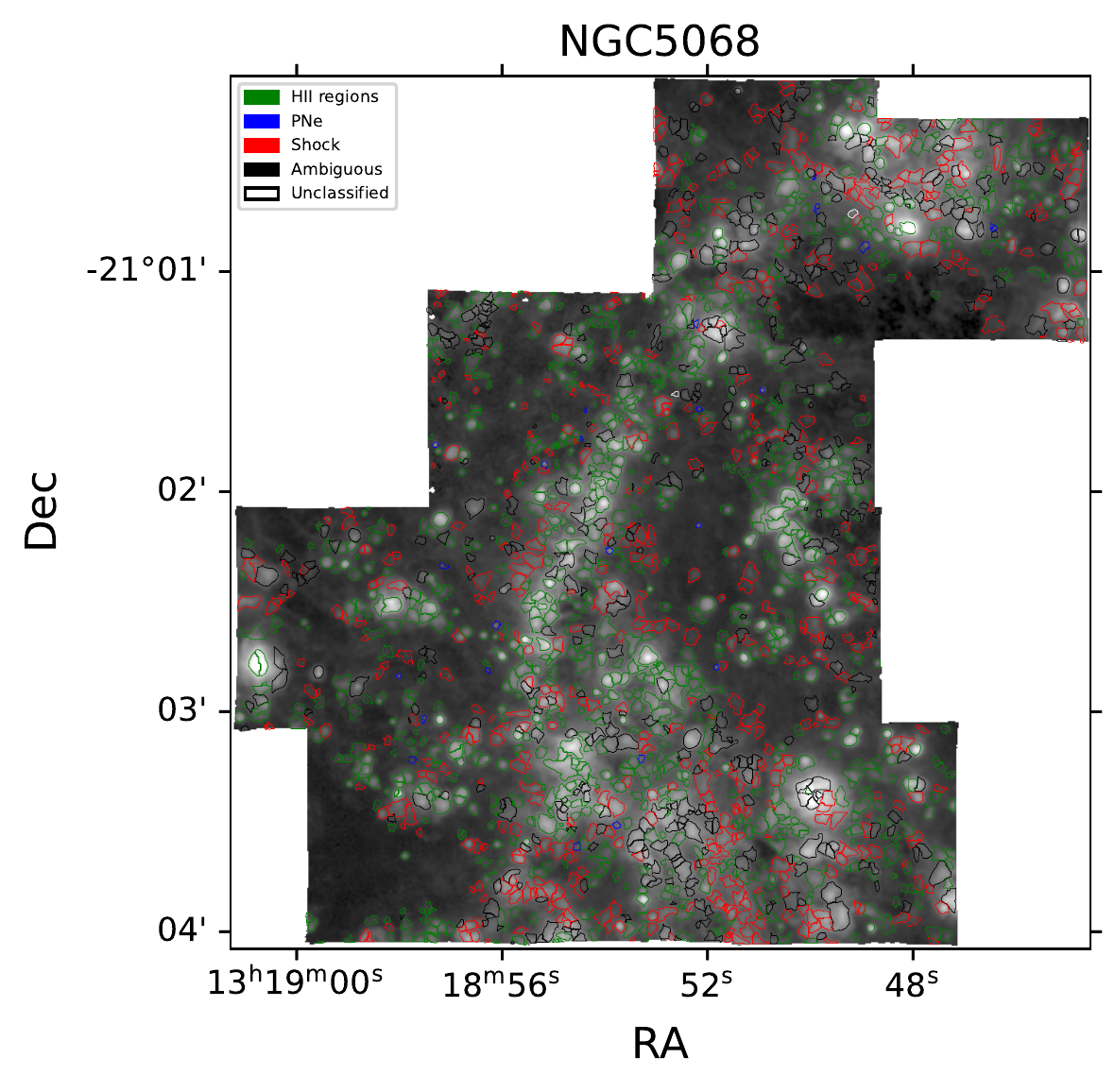}
\caption{Boundaries of the nebulae detected in \ngc5068 superimposed on the detection map for the galaxy. The colour of the contour represents the classification of the nebula according to our model-comparison-based algorithm.}
\label{fig:NGC5068_map}
\end{figure*}

\begin{figure*}
\centering
\includegraphics[width=0.94\textwidth]{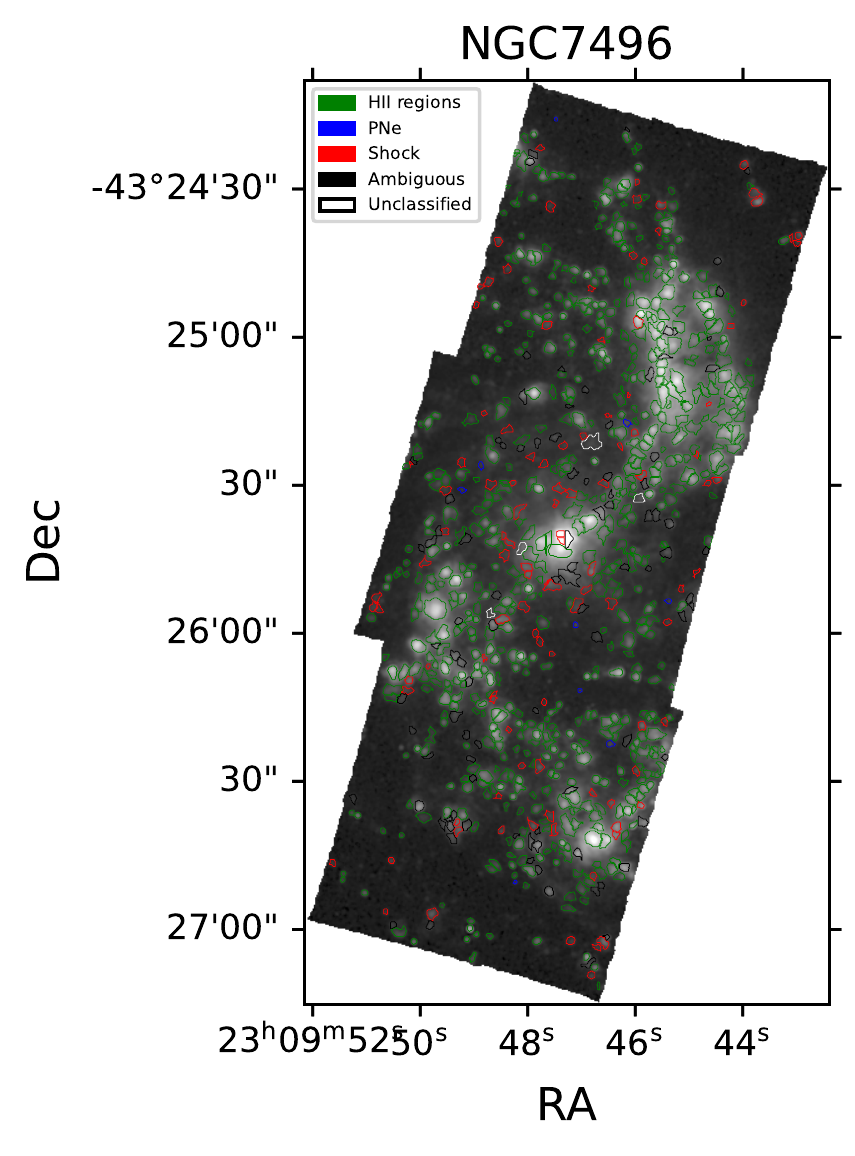}
\caption{Boundaries of the nebulae detected in \ngc7496 superimposed on the detection map for the galaxy. The colour of the contour represents the classification of the nebula according to our model-comparison-based algorithm.}
\label{fig:NGC7496_map}
\end{figure*}

\end{appendix}

\end{document}